\pdfoutput=1
\pdfmapfile{-mpfonts.map}
\documentclass[a4paper,runningheads]{llncs}
\usepackage{graphicx}
\newif\iflongversion
\longversiontrue
\longversionfalse
\newif\ifcompact
\compacttrue
\compactfalse
\usepackage{setup}

\begin{document}
\title{Programming with Union, Intersection,\\ and Negation
Types}
%
%
\author{Giuseppe Castagna\vspace{-2mm}
}
\authorrunning{G. Castagna}
%
\institute{CNRS - Université Paris Cité\vspace{-5mm}
}

%
\maketitle              
\begin{abstract}
\blfootnote{Based on joint work with Pietro Abate, Véronique
Benzaken, Alain Frisch, Hyeonseung Im,
Victor Lanvin, Micka\"el Laurent,  Sergue\"{\i} Lenglet, Matthew Lutze,
Kim Nguyen, Luca Padovani, Tommaso Petrucciani, and Zhiwu Xu.

\smallskip
\scriptsize\noindent\emph{This is a preprint of the following chapter: \emph{G.\ Castagna, Programming
with Union, Intersection, and Negation
Types, published in \emph{The French School of Programming}, edited by B.\ Meyer, 2023, Springer}. Reproduced with
permission.}}
In this essay I present the advantages and, I dare say, the beauty of
programming in a language with set-theoretic types, that is, types
that include union, intersection, and negation type connectives. I
show by several examples how set-\break theoretic types are necessary
to type some common programming patterns, but also how they play a key
role in typing several language constructs---from branching and
pattern matching to function overloading and type-cases---very
precisely.

\quad I start by presenting the theory of types known as semantic subtyping
and extend it to include polymorphic types. Next, I discuss the design
of languages that use these types. I start by defining a theoretical
framework that covers all the examples given in the first part of the
presentation. Since the system of the framework cannot be effectively
implemented, I then describe three effective restrictions of this
system: $(i)$~a polymorphic language with explicitly-typed functions,
$(ii)$ an implicitly typed polymorphic language \emph{à la}
Hindley-Milner, and $(iii)$ a monomorphic language that, by
implementing classic union-elimination, precisely reconstructs
intersection types for functions and implements a very general form of
occurrence typing.

\quad I conclude the presentation with a short overview of other aspects of
these languages, such as pattern matching, gradual typing, and
denotational semantics.



\end{abstract}
\section{Introduction}
\label{sec:intro}
\lstset{basicstyle=\linespread{0.7}\ttfamily\color{darkblue}}
\lstset{literate=
               {->}{$\to$}{1}
               {'a}{$\alpha$}{1}
               {forall}{$\forall$}{1}
               {neg}{$\neg$}{1}}
In this essay we present the use of set-theoretic types in programming
languages and
outline their theory. Set theoretic types include union types
$t_1\vee t_2$, intersection types $t_1\wedge t_2$, and negation types
$\neg t$. In strict languages it is sensible to interpret a type as the
set of values that have that type (e.g., \Bool{} is interpreted
as the set containing the values \code{true} and \code{false}). Under
this assumption, then, $t_1\vee t_2$ is the set of values that are
either of type $t_1$ \emph{or} of type $t_2$;  $t_1\wedge t_2$ is the set of values that are
both of type $t_1$ \emph{and} of type $t_2$; $\neg t$ is the set of all
values that are \emph{not} of type $t$. 
Set-theoretic types are polymorphic when they include type variables
(that we range over by Greek letters, $\alpha$, $\beta$,\ldots).

To give an idea of the kind of programming that set-theoretic types
enable and that we describe in this
article, consider the classic \emph{recursive flatten} function that
transforms arbitrarily nested lists in the list of their elements. In a
ML-like language with pattern matching it can be defined as
simply as\vspace{-2mm}
\begin{alltt}\color{darkblue}\morecompact
    let rec flatten = function
       |  [] -> []
       |  h::t -> (flatten h)@(flatten t)
       |  x -> [x]
\end{alltt}
The function \code{flatten} returns the empty list \code{[\,]} when its argument is an empty
list; if its argument is a non-empty list, then it flattens the argument's
head \code{h} and tail \code{t} and returns the concatenation
(denoted by \code{\texttt{@}}) of the results; if its argument
is not a list (i.e., the first two patterns do not match), then \code{flatten} returns the list containing just the argument.

The \code{flatten} function is completely polymorphic: it can be applied to
any argument and, if lists are finite,  always
terminates. Although its semantics is easy to understand, giving a simple
and general polymorphic type to this function (i.e., a type that,
without complex metaprogramming constructions,
allows the function to be applied to every well-typed argument)
defies all existing programming languages~\cite{gre19} with a single exception: CDuce~\cite{cduce}. This is
because CDuce is a language that uses a complete set of
set-theoretic type connectives and we need all of them (union,
intersection, and negation) to define \type{Tree($\alpha$)}, the type of
nested lists whose elements are of type \type{$\alpha$}:
\begin{alltt}\negskip
     type \type{Tree(\(\alpha\))} = \type{(\(\alpha\)\setminus{}List(Any)) | List(Tree(\(\alpha\)))}\negskip
\end{alltt}
in this type definition ``\lstinline!|!'' denotes a union,
``\lstinline!\!'' difference (i.e., intersection with the negation:
$t_1\setminus{}t_2\eqdef t_1\wedge\neg t_2$), \type{List($t$)} is the type of lists of elements of type
$t$, and \type{Any} is the type of all values, so
that \type{List(Any)} is the type of any list.\footnote{%
We mainly use ``\type{\&}'',
``\type{|}'', and ``\type{\setminus}'' in code snippets for
intersections, unions, and differences and reserve ``$\vee$'' ,
``$\wedge$'',
and ``$\neg$'' for formal types.}
In words,  \type{Tree($\alpha$)} is the type of nested lists whose
leaves (i.e., the elements that are not lists) have type
\type{$\alpha$}. Thus it is either a leaf or a list of
\type{Tree($\alpha$)}.
Then, it just  suffices to annotate \code{flatten} with the right type
\\[1mm]
\hspace*{3em}\code{let rec flatten: \type{Tree($\alpha$)$\to$List($\alpha$)} = function
...}
\\[1mm]
for the  definition to type-check in CDuce. In other terms, in CDuce
the above definition of \code{flatten} is of type  
\type{$\forall\alpha$.Tree($\alpha$)\To{}List($\alpha$)}.
The important point is that whatever the type of the argument
of \code{flatten} is, the application is always well-typed: if the
argument is not a list, then \type{$\alpha$} is instantiated to the
type of the argument; if it is a list, then it is also a nested
list, and \type{$\alpha$} is instantiated with the union of the types
of the non-list elements of this nested list. In other
terms, \code{flatten} can be applied to expressions of any type and
the type inferred for such an application is \type{List($t$)} where
the type $t$ is the union of the types of all the leaves of the
argument, a non-list argument being itself a leaf. For instance, the
type statically deduced for the application\\[1mm]
\hspace*{3em}\code{flatten [3 "r" [4 [true 5]] ["quo" [[false] "stop"]]]}
\\[1mm]
is 
\type{List(Int|Bool|String)}.\footnote{%
CDuce syntax is actually slightly different. The
valid CDuce code for our example is:\\
\code{\hspace*{1em}type Tree(\textquotesingle{a}) = (\textquotesingle{a}\setminus[Any*]) | [ (Tree(\textquotesingle{a}))* ]}\\
\code{\hspace*{1em}let flatten ( (Tree(\textquotesingle{a})) -> [ \textquotesingle{a}* ] )}\\[-.6mm]
\code{\hspace*{3em}  |  [] -> []}\\[-.6mm]
\code{\hspace*{3em}  |  [h;t] -> (flatten h)@(flatten t)}\\[-.6mm]
\code{\hspace*{3em}  |  x -> [x]}\\
and the type deduced by CDuce for the application is more precise
than the above since it is: \code{[ (Bool |
3--5 | \textquotesingle{o}\textquotesingle--\textquotesingle{u}\textquotesingle)* ]} (``\texttt{--}'' is for intervals
and \code{[$t$*]} for lists of $t$ elements).}

The overall type inference system is quite expressive: it types more
expressions or gives more precise types (but worse error messages) than
typical core-ML systems.  However, such a deduction is possible only
because the function \lstinline!flatten! is explicitly typed: fail to
specify the type annotation \type{Tree($\alpha$)$\to$List($\alpha$)}
and \code{flatten} will be rejected by all existing type-checking
systems.

That current type-systems cannot infer a type as
sophisticated as the type of \code{flatten} without an explicit
annotation is not surprising, since its definition combines the full palette of set-theoretic connectives (union, intersection, \emph{and} negation) and
recursive types.  However, an important limitation of current
programming languages is that none of them is able to infer intersection
types for functions without explicit annotations. So while any ML-like
language can deduce for
\\[1mm]
\hspace*{3em}\code{let not\_ = fun x -> if x then false else true}
\\[1mm]
the type \type{Bool\(\to\)Bool}, current languages with intersection
types cannot deduce for the same function the more precise
type \type{(true\(\to\)false)\&(false\(\to\)true)} (where \type{true}
and \type{false} denote the singleton types containing the respective
values) without being instructed
to do so by an explicit type annotation. The latter type is an
intersection of types, meaning that \code{not\_} has both
type  \type{true\(\to\)false} \emph{and}
type  \type{false\(\to\)true}. The intersection type is  more
precise than the type \type{Bool\(\to\)Bool}: it states that when \code{not\_} is
applied to an expression of type \type{true}, the result is not only a
Boolean but actually \code{false}, and likewise for arguments of
type \type{false}. As we show later on, this degree of 
``precision'' between two types is
formally defined
since \type{(true\(\to\)false)\&(false\(\to\)true)} is a strict subtype
of  \type{Bool\(\to\)Bool}: every function of the former type is also
of the latter type, but not viceversa. 
Actually, if we adopt for if-then-else a semantics similar to the one in
JavaScript, that is, we consider every value different from false to be
``truthy'' (i.e., equivalent to true), then an even better intersection type for \code{not\_} would
be \type{(\(\neg\)false\(\to\)false)\&(false\(\to\)true)} which completely
specifies the behavior of the function since the function \code{not\_} above returns \code{false}
for every argument that is not \code{false} (i.e., for ``truthy''
values such as \code{42}). The
more precise is a type the fewer functions it types, the most precise
type being one that,
as \type{(\(\neg\)false\(\to\)false)\&(false\(\to\)true)}  completely
defines the semantics of a  function.%
\footnote{For the sake of
precision, JavaScript considers being ``truthy'' every value different
from the eight specific
``falsy'' values
(\code{false}, \code{""}, \code{0}, \code{-0}, \code{0n}, \code{undefined}, \code{null},
and \code{NaN}). Both types can be defined by using union and negation:\\[1mm]
\begin{tabular}{@{\quad}l}\color{darkblue}\tt
   type \type{ Falsy = false | "" | 0 | -0 | 0n | undefined | null | NaN}\\\color{darkblue}\tt
   type \type{Truthy = \(\neg\)Falsy}\\
\end{tabular}\\[1mm]
and the type to be deduced for \code{not\_} would then be \type{(Truthy\(\to\)false)\&(Falsy\(\to\)true)}.}

We will discuss recent systems by~\citet{CLLN20,CLNL22} that are able to
deduce the most precise intersection type for the definition
of \code{not\_} even without any annotation. This inference is obtained by
considering the conditional in the definition of \code{not\_} akin to
a type-case that
tests whether \code{x} is of type \type{\(\neg\)false} or not. The body
of \code{not\_} is then analyzed separately under the hypotheses
that \code{x} has type  \type{$\neg$false} and \type{$\neg\neg$false}
(i.e., \type{false}),
yielding the corresponding intersection type. This is performed also for multiple
arguments, allowing the cited systems to deduce for\negskip
\begin{alltt}\color{darkblue}
    let and\_ = fun x y ->
        if x then (if y then false else true) else false
\end{alltt}
the following type:

\centerline{\type{(false\To Any\To
false) \& ($\neg$false\,\To\,(($\neg$false\To{true})\&(false\To{false})))}.}

\noindent
This type completely specifies the semantics of \code{and\_}: if the first
argument is \code{false}, then the result will be \code{false} for
a second argument of any type; if the first argument is not \code{false},
then the result will be \code{true} for a second argument not \code{false},
and \code{false} otherwise.  It is important to notice that the
analysis performed in~\cite{CLLN20,CLNL22} is type-theoretic rather
than syntactic: the arrows forming the intersection type of a function
are not determined by a syntactic recognition of type-cases, but are
inferred from the types
involved in the definition of the function. To illustrate the
advantages of a type-based approach over a syntactic one it suffices to
consider the following definition of \code{or\_} that combines the
previous \code{not\_} and \code{and\_} definitions according to De
Morgan's laws:\\[1mm]
\hspace*{3em}\code{let or\_ = fun x y -> not\_ (and\_ (not\_ x) (not\_ y))}
\\[1mm]
The type \type{($\neg$false\To
Any\To true)\,\&\,(false\;\To\;(($\neg$false\To{true})\&(false\To{false})))}
is deduced for this definition despite that no branching appears in it. For the same
reasons we could equivalently define the previous \code{and\_} function
using a double call to \code{not\_} so that a  second argument that is
not false
yields true:\\[1mm]
\hspace*{3em}\code{let and\_ = fun x y -> if x then not\_ (not\_ y) else
false}
\\[1mm]
and obtain the same type as for the previous definition of \code{and\_}.

The ultimate goal of the research we present in this article is to define a
programming language whose type-inference subsumes ML-core
type-inference, that can also deduce intersections of arrows types for
implicitly-typed functions such as
 \code{not\_}\,, \code{and\_}\,,
and \code{or\_}\,, and where the
programmer would be obliged to specify type annotations only in
particular cases, such as for \code{flatten}. Unfortunately, while
there exist systems that provide some of these features, it is not currently 
possible to have all of them simultaneously in a unique language, as we discuss in
Section~\ref{sec:languages}.

\paragraph{Roadmap.} This article aims at giving a rather
comprehensive---though, high-level---presentation of the current
status of the research on set-theoretic types and semantic subtyping and it was written
with a sequential reading in mind. However, other reading paths are possible:
here we provide a roadmap through
the remaining sections of this presentation and suggest two such reading paths.

Section~\ref{sec:motivations} provides some general examples and motivations for using set-theoretic
types. In particular, it shows the use of set-theoretic types to type programming features and
idioms, such as pattern matching, occurrence typing, and function overloading.

Section~\ref{sec:types} is devoted to types. It starts by showing some
of the limitations of syntactic
approaches for union and intersection types
(\S\ref{sec:syntactic}),
limitations that justify the use of a semantic approach. Next, we describe the semantic subtyping
approach: we interpret types as sets, and use this interpretation to
define the subtyping relation which, in turn, characterizes union and
intersection types; the difficult part of the definition is the
interpretation
of function types and recursive types (\S\ref{sec:semsub}). Finally, we
extend the semantic subtyping approach to polymorphic types by adding type
variables to types (\S\ref{sec:polysemsub}).

Section~\ref{sec:languages} is devoted to languages. We present
various languages that use set-theoretic types. We start with a
theoretical language that can express and type all the examples given in this work
(\S\ref{sec:theoretical}). The theoretical
language is powerful but its typing is not effective, therefore we
describe three practical systems that partially implement this
theoretical language, each implementation being the result of a
certain number of choices and trade-offs that we will discuss.  The
first language is an explicitly typed version of the theoretical
language (\S\ref{sec:cduce})---in practice, it is the functional core of
polymorphic CDuce---and, as such, it gives up type
reconstruction. The second language is implicitly typed,  its type system
performs type reconstruction, but it cannot infer intersections of arrow types
(\S\ref{sec:implicit}): it gives up function overloading. The third one
uses techniques developed for \emph{occurrence typing} to perform
reconstruction for intersection types (of arrow types, too) but only for types without type
variables (\S\ref{sec:occtyping}): it gives up parametric
polymorphism.

Section~\ref{sec:features} explores three further aspects of this
research: pattern matching (\S\ref{sec:pm}), where we show how
set-theoretic types enable precise typing for pattern-matching
expressions and how they can be used to define properties such as exhaustiveness
and redundancy;  gradual typing
(\S\ref{sec:gradual}), where we show that by using set-theoretic type
connectives we can give a surprising characterization of gradual types
(i.e., types that use an unknown or dynamic type identifier to
establish a boundary between static and dynamic typing) in terms of
non-gradual ones; denotational semantics
(\S\ref{sec:semantics}), where we show that the interpretation for
types introduced in \S\ref{sec:semsub} to define semantic subtyping can be adapted to give a set-theoretic
denotational semantics to the language CDuce.

The practical-oriented reader will start with the general motivations
of Section~\ref{sec:motivations} and then proceed with the first part of
Section~\ref{sec:types} (stopping at the end of
\S\ref{sec:syntactic}) to have an idea of the limitations of
current syntactic approaches. The reader will then proceed to
Section~\ref{sec:languages} till the end of \S\ref{sec:cduce} but
taking care of skipping the more theoretic subsections,
namely, \S\ref{sec:negation} and \S\ref{sec:feasibility}.  Finally,
the practical-oriented reader may also be interested in exploring \S\ref{sec:pm} which shows
the subtleties of using set-theoretic types to type pattern-matching expressions,
and  \S\ref{sec:gradual} which gives an \emph{aperçu} of gradual
typing and shows how this technique benefits from the integration of
set-theoretic types.

The theoretical-oriented reader can skim through
Section~\ref{sec:motivations}, then jump
directly to \S\ref{sec:semsub}  which sets the groundwork for the
semantic subtyping approach, and next proceed with
\S\ref{sec:polysemsub} for polymorphic types. The reader will then continue with
Section~\ref{sec:languages}, first by studying the theoretical language
of \S\ref{sec:theoretical} and then choosing the next topic(s)
according to her/his particular interests: \S\ref{sec:cduce} for a more
programming-oriented presentation; \S\ref{sec:implicit} for type
reconstruction by constraint generation and solving;
\S\ref{sec:occtyping} for a more cutting edge approach to union
types and their elimination rule. The reader will conclude with \S\ref{sec:semantics} on
the denotational semantics of CDuce.

Both reading paths are summarized below, the former in red and the latter in blue.

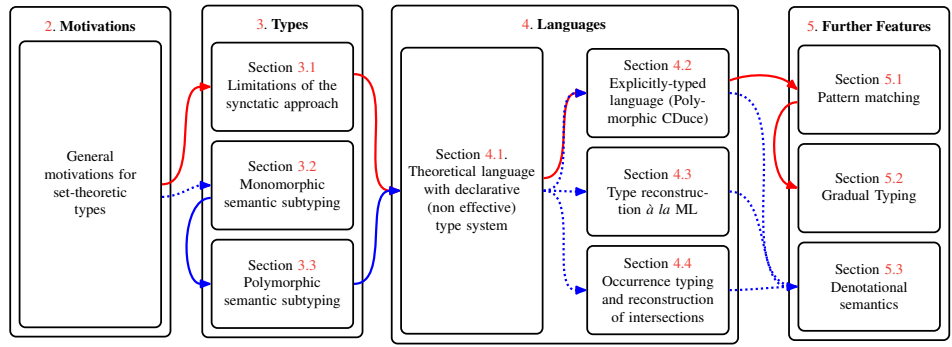
\usetikzlibrary{arrows,positioning,fit,arrows.meta} 
\tikzset{
    >=stealth',
    punkt/.style={
           rectangle,
           rounded corners,
           draw=black, very thick,
           text width=8.5em,
           minimum height=2em,
           text centered},
    pil/.style={
           ->,
           thick,
           shorten <=2pt,
           shorten >=2pt,}
}

\smallskip

\noindent
\resizebox{\textwidth}{!}{%
\begin{tikzpicture}[node distance=1cm, auto,looseness=0.9]

 \node (group0) {\ref{sec:motivations}. \bf Motivations};

\node[punkt, inner sep=5pt,below=0.1cm of group0,minimum height=6.8cm]  (general){General\\motivations for set-theoretic\\ types};

 \node[punkt, fit=(group0) (general), inner sep=2mm] (all) {};

 \node[right=2.6 of group0] (group1) {\ref{sec:types}. \bf Types};

\node[punkt, inner sep=5pt,below=0.1cm of group1,minimum height=2.1cm]   (3-1) {Section~\ref{sec:syntactic} \\ Limitations of the synctatic approach};
 \node[punkt, inner sep=5pt,below=0.2cm of 3-1,minimum height=2.1cm] (3-2) {Section~\ref{sec:semsub}\\ Monomorphic semantic subtyping};
 \node[punkt, inner sep=5pt,below=0.2cm of 3-2,minimum height=2.1cm] (3-3) {Section~\ref{sec:polysemsub} \\Polymorphic semantic subtyping};
 
 \node[punkt, fit=(group1) (3-1) (3-2)(3-3), inner sep=2mm] (all) {};

 \node[right=4.8cm of group1] (group2) {\ref{sec:languages}. \bf Languages};

\node[punkt, inner sep=5pt,below=0.2cm of group2, minimum height=6.8cm] (4-1) at (9,-.3){Section~\ref{sec:theoretical}. Theoretical language with declarative (non effective) type system};
 \node[punkt, inner sep=5pt,right=1.7 of 4-1,minimum height=2.1cm]
 (4-2) at (10,-1.6) {Section~\ref{sec:cduce}\\Explicitly-typed
 language (Polymorphic CDuce)};
 \node[punkt, inner sep=5pt,below=0.2cm of 4-2,minimum height=2.1cm] (4-3) {Section~\ref{sec:implicit}\\Type reconstruction \emph{à la} ML};
 \node[punkt, inner sep=5pt,below=0.2cm of 4-3,minimum height=2.1cm] (4-4) {Section~\ref{sec:occtyping}\\ Occurrence typing and reconstruction of intersections};

 \node[punkt, fit=(group2) (4-1)(4-2) (4-3) (4-4) , inner sep=2mm] (all2) {};
 
 \node[right=4.5cm of group2] (group3) {\ref{sec:features}. \bf Further Features};

 \node[punkt, inner sep=5pt,below=0.2cm of group3,minimum height=2.1cm] (5-1) {Section~\ref{sec:pm}\\Pattern matching};
 \node[punkt, inner sep=5pt,below=0.2cm of 5-1,minimum height=2.1cm] (5-2)  {Section~\ref{sec:gradual}\\Gradual Typing};
 \node[punkt, inner sep=5pt,below=0.2cm of 5-2,minimum height=2.1cm] (5-3)  {Section~\ref{sec:semantics}\\Denotational semantics};

 \node[punkt, fit=(group3) (5-1) (5-2)(5-3), inner sep=2mm] (all3) {};

\draw [line width=1.5pt,red,-{Latex[round]}] (general.east) to[out=0,in=180] (3-1.west);
\draw [line width=1.5pt,red,-{Latex[round]}] (3-1.east)+(0,0.3) to[out=0,in=180,above=9mm] (4-1.west);
\draw [line width=1.5pt,red,-{Latex[round]}] (4-1.east)+(0,0.3) to[out=0,in=180] (4-2.west);
\draw [line width=1.5pt,red,-{Latex[round]}] (4-2.east)+(0,0.3) to[out=20,in=140] (5-1.west);
 \draw [line width=1.5pt,red,-{Latex[round]}] (5-1.west)+(0,-0.3) to[out=180,in=160] (5-2.west);
 \draw [line width=1.5pt,blue,dotted,-{Latex[round]}] (general.east)+(0,-0.3) to[out=0,in=180] (3-2.west);
  \draw [line width=1.5pt,blue,-{Latex[round]}] (3-2.west)+(0,-0.3) to[out=180,in=180] (3-3.west);
 \draw [line width=1.5pt,blue,-{Latex[round]}] (3-3.east) to[out=0,in=180] (4-1.west);
 \draw [line width=1.5pt,blue,dotted,-{Latex[round]}] (4-1.east) to[out=0,in=180] (4-2.west);
 \draw [line width=1.5pt,blue,dotted,-{Latex[round]}] (4-1.east) to[out=0,in=180] (4-3.west);
 \draw [line width=1.5pt,blue,dotted,-{Latex[round]}] (4-1.east) to[out=0,in=180] (4-4.west);
 \draw [line width=1.5pt,blue,dotted]
                (4-2.east) to[out=0,in=180] (5-3.west) ;
 \draw [line width=1.5pt,blue,dotted,-{Latex[round]}] (4-3.east) to[out=0,in=180] (5-3.west);
 \draw [line width=1.5pt,blue,dotted]
                (4-4.east) to[out=0,in=180] (5-3.west);
\end{tikzpicture}\negskip
}

\section{Motivations}
\label{sec:motivations}
In the previous section we gave few specific examples of use of
polymorphic set-theoretic types. One of the key features of
these types that makes them versatile is that they encompass all the
three main forms of polymorphism, namely:
\begin{description}
  \item[\rm\textit{Parametric polymorphism:}]
    which describes code that can act uniformly on any type,
    using type variables that can be instantiated with the appropriate type
    (e.g., typing the identity function as
    $\forall\alpha. \alpha \to \alpha$). In this article we consider
    only the so-called \emph{prenex} or \emph{second-class}
    polymorphism (in the sense of~\cite{harperbook06}) where variable
    quantification cannot appear below type constructors or type connectives.
    
  \item[\rm\it Ad-hoc polymorphism:]
    which allows code that can act on more than one type,
    possibly with different behavior in each case,
    as in function overloading
    (e.g., allowing \code{+} to have
    both types \type{$ \Int \times \Int \to \Int $}
    and \type{{String}$\times${String}$\to${String}},
    corresponding to different implementations).
  \item[\rm\it Subtype polymorphism:]
    which creates a hierarchy of more or less precise types for the same code
    allowing it to be used wherever any of these type is expected (e.g., typing $ \code{3} $ as both $ \Int $ and \type{Real},
    with \type{\Int$\leq$Real}).
\end{description}
In this section we reframe polymorphic set-theoretic types in a more general
setting showing how these types allow us
to type
several features and idioms of programming languages effectively.
We illustrate this with some examples.

\paragraph{Union types:}

The simplest use cases for union types include branching constructs.
In a language with union types,
we can type precisely conditionals that return results of different types:
for instance, \code{if $ e $ then 3 else true} has type \type{$ \Int \lor \Bool $}
(provided that $ e $ has type  \Bool).
Without union types, it could have an approximated type (e.g., a top type)
or be ill-typed.
Similarly, we can use union types
for structures like lists that mix different types: we already saw 
an example of this in the previous section when an application of \code{flatten}
returned the list  \code{[3 "r" 4 true 5 "quo" false "stop"]} of type \type{List(\Int|\Bool|{String})}.

This makes union types invaluable
to design type systems for existing untyped languages:
witness for example
their inclusion in Typed Racket \cite{THF08} which allows the
incremental addition of statically-checked type annotations on a
dialect of Scheme
and in TypeScript \cite{TypeScript} and Flow \cite{Flow}
which extend JavaScript with static type checking.

\paragraph{Function overloading:}

We can use intersection types
to assign more than one type to an expression.
This is particularly relevant for functions.
We have already seen it in the previous section for the functions \code{not\_}\,, \code{and\_}\,,
and \code{or\_}. But even the
simple identity function
can be typed as $ (\Int \To \Int) \land (\Bool \To \Bool) $:
this means it has both types $ \Int \To \Int $ and $ \Bool \To \Bool $,
because it maps integers to integers and Booleans to
Booleans.\iflongversion
\else\ \fi
This type describes a uniform behavior over two different argument
types (the function uniformly maps an argument into itself
independently from the argument's type),
which can also be described using parametric polymorphism.
However,
intersection types let us express \emph{ad-hoc} polymorphism (i.e.,
function overloading)
if coupled with some mechanism
that allows functions to test the type of their arguments. 
For example, let  $\tcase{e}{t}{e_1}{e_2}$ be the type-case expression that first evaluates $e$ to a
value $v$ and continues as  $e_1$ if $v$ is of type $t$, and as
$e_2$ otherwise. The function $ \lambda x. \tcase{x}\Int{(x + 1)}{\lnot x}$
checks whether its argument $ x $ is an $ \Int $
and returns the successor of $ x $ in that case, and the Boolean negation
(hereinafter denoted by ``$\neg$'') of $x$ otherwise.
The function can be applied to integers, returning their successor,
and to Booleans, returning their negation.
This behavior can be described by the same type
$ (\Int \to \Int) \land (\Bool \to \Bool) $ used for the identity function,
but does not correspond to parametric behavior.

A function of type $ (t_1 \to t_1') \land (t_2 \to t_2') $
can be safely applied  to any argument of type $ t_1 \lor t_2 $,
since it is defined on both $ t_1 $ and $ t_2 $.
We know that the result will always have type $ t_1' \lor t_2' $.
However, if we know the type of the argument more precisely,
we can predict the type of the result more precisely:
for example, if the argument is of type $ t_1 $,
then the result will be of type $ t_1' $. So the intersection type of
the function  $ \lambda x. \tcase{x}\Int{(x + 1)}{\lnot x}$ allows us
to deduce that its application to an integer will return an integer. 

We said that the type $ (\Int \To \Int) \land (\Bool \To \Bool) $
can be assigned to the identity function
and expresses parametric behavior.
In this respect,
we can see intersection types as a finitary form of parametric polymorphism;
however, they are not constrained to represent uniform behavior,
as our other example illustrates.
Conversely,
we could see a polymorphic type (or type scheme)
$ \forall{\alpha. \alpha \to \alpha} $
as an infinite intersection (intuitively, $
  \bigwedge_{t \in \textsf{Types}} t \to t
$, where \textsf{Types}  is the set of all types),
but infinite intersections do not actually exist in our types.

\paragraph{Occurrence typing:}

\emph{Occurrence typing} or \emph{flow typing}
\cite{THF10,Pearce2013,Chaudhuri2017} is a typing technique pioneered by Typed Racket
that uses the information provided by a type test to specialize the
type of some variables in the branches of a conditional.
For example, if $ x $ is of type $ \Int \lor \Bool $,
then to type the expression $ \tcase{x}\Int{e_1}{e_2} $
we can assume that the occurrences of $ x $ in $ e_1 $ have type $ \Int $
and those in $ e_2 $ have type $ \Bool $,
because the first branch will only be reached if $ x $ is an $ \Int $
and the second if it is not an $ \Int $
(and is therefore a $ \Bool $).
Intersection and negation types are useful to describe this type discipline.
If we test $x$ for the type $ \Int $ as in our example,
then we can assign to $ x $
the type $ \Int $ if the test succeeds
and $ \lnot \Int $ if it fails.
Using intersections, we can add this information to what we already knew,
so the type of $ x $
is $ (\Int \lor \Bool) \land \Int $ (which is equivalent to $ \Int $)
in the first branch
and $ (\Int \lor \Bool) \land \lnot \Int $ (which is equivalent to $ \Bool $)
in the second branch.
We already implicitly used this technique when, earlier in this
section, we said that $ \lambda x. \tcase{x}\Int{(x + 1)}{\lnot x}$ is
of type  $ (\Int \to \Int) \land (\Bool \to \Bool) $ since we must assume that
$x$ is of type $\Int$ to type $x+1$ and that it is of type \Bool{} to
type $\lnot x$: we took into account the result of the type-test.

This method of refining types according to conditionals
is important in type systems for dynamic languages
and in those that enforce null safety:
some examples include
Ceylon \cite{Ceylon}, Dart~\cite{googledart}, Flow, 
Kotlin \cite{Kotlin}, Typed Racket, TypeScript,
and Whiley \cite{Pearce2013whiley}.
In particular, Ceylon relies on intersection types
\cite{Ceylon,Muehlboeck2018}
and Whiley on both intersection and negation types
\cite{Pearce2013}.

This same method is at the basis of the systems by~\citet{CLLN20,CLNL22}
we cited in the introduction as the sole capable of inferring
intersection of arrow types for functions without needing explicit type
annotations. These systems use the characteristics  of
set-theoretic types, as outlined above, to implement and generalize occurrence typing and
decide how to split the type analysis to deduce intersection types for
function expressions, as we detail in Section~\ref{sec:occtyping}.

\paragraph{Encoding disjoint union types:}

Disjoint union types (also known as variant or sum types)
are an important feature of functional programming languages.
They can be encoded using union types and product (or record, or object) types.
It is also useful to have \emph{singleton types}, that is,
types that correspond to a single value as we already saw with the two types \type{true} and \type{false} for the respective constants,
both subtypes of the Boolean type
(which we can then see as equivalent to the union \type{$ \type{true} \lor \type{false} $}).

For instance, consider this example in Flow.%
  \footnote{%
    This example is copied verbatim from the documentation of Flow,
    available at \url{https://flow.org/en/docs/types/unions}.}
\begin{align*}
  & \textsf{type Success = \{ success: true, value: boolean \}}
  \\[-1.2mm]
  & \textsf{type Failed  = \{ success: false, error: string \}}
  \\[-1.2mm]
  & \textsf{type Response = Success \texttt{|} Failed}
  \\[1mm]
  & \textsf{function handleResponse(response: Response) \{}
  \\[-1.2mm]
    & \quad \textsf{if (response.success) \{
      var value: boolean = response.value \}}
  \\[-1.2mm]
    & \quad \textsf{else \{
      var error: string = response.error \}}
  \\[-2mm]
  & \textsf{\}}
\end{align*}
The type \K{Response} is the union (denoted by {\texttt{|}})
of two object types: both have a Boolean field \K{success},
but the types state that \K{success} must be
\K{true} for objects of type \K{Success}
and \K{false} for objects of type \K{Failure}.
An analogous type could be declared in OCaml as
\code{type response = Success of bool | Failed of string}.
Occurrence typing is used to distinguish the two cases,
like pattern matching could do in ML:
if \textsf{response.success} is true,
then \K{response} must be of type \K{Success};
if it is false, \K{response} must be of type \K{Failure}.

\paragraph{Encoding bounded polymorphism:}

Using union and intersection types,
we can encode bounded polymorphism without adding specific syntax
for the bounds in quantifications.
For example, a type scheme with bounded polymorphism is
$ \forall(\alpha \leq t). \alpha \to \alpha $:
it describes functions that can be applied to arguments
of any subtype of $ t $
and that return a result of the same type as the argument.
Using intersection types, we can write this type scheme as
$ \forall{\alpha. (\alpha \land t) \to (\alpha \land t)} $,
writing the bound on the occurrences of the type variable
and not on the quantifier: as the previous type scheme it accepts only
arguments of a type smaller than $t$ and returns results of the same
type.\footnote{Of course, the syntax $\forall(\alpha \leq t). \alpha
\to \alpha$ is likely to be clearer to a programmer and should be
privileged. The point is that set-theoretic types provide all is needed to account for bounded polymorphism without the
need to add new machinery or rules for this sort of typing.} 
Analogously, we can use union types to represent lower bounds:
in general, a bound $ t' \leq \alpha \leq t $ on a type variable
can be eliminated by replacing every occurrence of $ \alpha $ in the type
with $ t'\lor (\alpha \land t)$, yielding bounded quantifications of
the form $\forall( t' {\leq} \alpha {\leq} t). t''$. Notice however that
the form of bounded polymorphism we obtain by this encoding is limited, insofar
as two bounded types may be in subtyping relation only if they have
the same bounds,\footnote{This is necessary only for bounded variables
that occur in the type both in covariant and in contravariant
positions. Notice however that variables that do not satisfy this
property can be easily eliminated by replacing \type{Any} for all
covariantly-only occurring variables, and \type{$\neg$Any} for all
contravariantly-only occurring ones.}
yielding a second-class polymorphism more akin to
\textsf{Fun}~\cite{Fun} (where $\forall (\alpha{\leq} s_1).t_1\leq
\forall (\alpha{\leq} s_2).t_2$ is possible only for $s_1=s_2$) than to
$F_{<:}$~\cite{Fsub} (which allows $\forall (\alpha{\leq} s_1).t_1\leq
\forall (\alpha{\leq} s_2).t_2$ even for $s_1\neq s_2$, typically
$s_2\leq s_1$).

As a concrete example, consider again the \code{flatten} function of
the introduction. We can give this function a type slightly more
precise than the one in the introduction by using
the annotation
\type{Tree($\alpha$)$\to$List(($\alpha$\setminus{}List(Any)))} which
states that the elements of the resulting list cannot be themselves lists: the
list is flat. With such an annotation the current version of
polymorphic CDuce deduces for \code{flatten} the (equivalent) type
\type{$\forall\alpha$.Tree($\alpha$\setminus{}List(Any)))$\to$List(($\alpha$\setminus{}List(Any)))}. Since
$t_1\setminus\,t_2=t_1{\land}\neg t_2$, this corresponds to the bounded
quantification
\type{$\forall(\alpha\leq\neg$List(Any)$)$.Tree($\alpha$)$\to$List($\alpha$)}
stating that \type{$\alpha$} can be instantiated with any type that is
not a list (though the domain \type{Tree($\alpha$)} can still match any type).

\paragraph{Typing pattern matching:}

Pattern matching is widely used in functional programming.
However, using pattern matching in ML-like languages,
we can write functions that cannot be given an exact domain in the type system.
For instance, the OCaml code\\[1mm]
\hspace*{3em}\code{let f = function 0 -> true | 1 -> false}
\\[1mm]
defines a function that can only be applied to the integers \K{0} and \K{1},
but OCaml infers the unsafe type $ \K{int} \to \K{bool} $
(albeit with a warning that pattern matching is not exhaustive).
The precise domain cannot be expressed in OCaml.
Using set-theoretic types and singleton types,
we can express it precisely as $ \K{0} \lor \K{1} $.
Furthermore, we can use the inference of intersection of arrows we outlined
in the introduction, which for the function \code{f} gives the type
\type{(0\To{true})\&(1\To{false})} which completely defines the
semantics of \code f.

More generally, set-theoretic types are a key ingredient to achieve a
precise typing of pattern matching. For instance, in a language as
CDuce the set of values that match a given pattern form a type (see Section~\ref{sec:pm}). This
can be used to precisely type a single branch of pattern matching
since the set of values processed by a given branch are all the values
in the type of the matched expression \emph{minus} (set-theoretic
difference) the \emph{union} (set-theoretic union) of all the values
matched by the preceding branches, \emph{intersected} (set-theoretic
intersection) with the values matched by the pattern of the branch at
issue. We will give all the details about it in Section~\ref{sec:pm} but
in this essay we
already met several examples of application of this technique.  For instance, in the
definition of \code{flatten} in the introduction, the first pattern \code{[]}
captures the empty list, that is the singleton type \type{[]}; the
second pattern \code{h::t} captures all the non-empty lists, that is the
type \type{List(Any)\setminus []}; the third pattern \code{x} captures
all values, that is the type \type{Any}. From that CDuce deduces that
the variable \code{x} in the third branch will capture any value that is not captured by
the two previous patterns, that is \type{Any} minus
\type{[]$\lor($List(Any)\setminus{[]}$)$}$=$\type{List(Any)} (i.e.,
the type captured by the first pattern union the type captured by the second pattern) and deduces
that the list returned by the branch cannot contain other lists:
this is the key mechanism that allows CDuce to type \code{flatten}
also when it is annotated with the more precise type
\type{Tree($\alpha$)$\to$List(($\alpha$\setminus{}List(Any)))}.

\paragraph{Negation Types:} We have already seen several applications
of negation types. In the examples we gave we mostly used type
differences, since they better fit a usage for programming, but this is
completely equivalent since negation and differences can encode each
other (i.e., $t_1\setminus{}t_2=t_1\wedge\neg t_2$ and $\neg t =
\type{Any}\setminus{}t$). As a matter of fact, type difference is
pervasive in all programming languages that use union types. However
the vast majority of these languages hide type difference to the
programmer and use it only as a meta-operation on types, implemented in
the type-checker which uses it to produce precise types or analyze the
flow of values in pattern matching. For instance, to 
type the type-case expression $ \tcase{x}\Int{e_1}{e_2} $ where $ x $ has type $
\Int \lor \Bool $, a type checker such as the one for Flow would assume that the occurrences of $ x $ in $ e_2 $
have type $\Bool$, since this type is the result of $ (\Int \lor
\Bool) \setminus \Int $. But to compute this result it would use an internal
type-difference operator without exposing it to the programmer: the
programmer can write its types by using unions, intersections, but not
differences.

Nevertheless, first-class difference (or negation) types are useful to
type several programming patterns and idioms. We already seen this
with the \code{flatten} function, whose type critically relies on the
use of difference types to define the type of nested lists. But much
simpler examples exist: consider for instance a function
$\lambda x. \tcase{x}{\Int}{(x + \K{1})}x$.
It can act on arguments of any type,
computing the successor of integers and returning any other argument unchanged.
Using intersection and difference types, plus parametric polymorphism,
we can type it as $
  \forall{\alpha}.
    (\Int \to \Int) \land
    (\alpha \setminus \Int \to \alpha \setminus \Int)
$, which expresses the function's behavior fairly precisely and that corresponds to the
    bounded quantification $
    \forall(\alpha\leq\neg\Int).
    (\Int \to \Int) \land
    (\alpha \to \alpha)$: the function returns integer results for integer
    arguments and returns $\alpha$ results for $\alpha$ arguments that
    are not integers.

Although the example $\lambda x. \tcase{x}{\Int}{(x + \K{1})}x$ is not
very enthralling, it yields a type that is extremely useful in practice
since it precisely types functions defined by pattern matching with a
last default case that returns the argument. In 
\cite[Appendix~A]{polyduce2} the reader can find the detailed presentation of a couple of compelling examples
of standard functions (on binary trees and SOAP envelopes) whose
typing is only possible or can be improved thanks to 
set-theoretic types that use differences as in the example above.
In particular, \cite{polyduce2} shows how
to type the function to insert a new node in a red-black tree (one of
the most popular implementation of self-balancing binary search tree, due to~\citet{GS78}).
The types used in the definition given in  \cite{polyduce2} enforce three out of the four invariants of red-black trees,%
  \footnote{%
    Specifically,
    that the root of the tree is black, that the leaves of the tree are black,
    and that no red node has a red child;
    the missing invariant is that every path from the root to a leaf
    should contain the same number of black nodes.\label{foot1}}
requiring only the addition of type annotations to the code
and no other change to a standard implementation
due to Okasaki \cite{okasaki_1998,okasaki99} to which the reader can
refer for more details. The core of Okasaki's definition
is the \code{balance} function which is defined (in our ML-like syntax)
as follows:
\begin{alltt}\color{darkblue}\morecompact
  type \type{RBTree(\(\alpha\))} = \type{Leaf | ( (Red|Black), \(\alpha\), RBTree(\(\alpha\)), RBTree(\(\alpha\)))}

  let balance = function
    | (Black, z, (Red, y, (Red, x, a, b), c), d)
    | (Black, z, (Red, x, a, (Red, y, b, c)), d)
    | (Black, x, a, (Red, z, (Red, y, b, c), d))
    | (Black, x, a, (Red, y, b, (Red, z, c, d))) ->
           (Red, y, (Black, x, a, b), (Black, z, c, d))
    | x -> x
\end{alltt}
which is of type \type{ RBTree(\(\alpha\))\To{}RBTree(\(\alpha\))}.
In the definition of   \type{RBTree(\(\alpha\))} nothing distinguishes a red-black tree from a vanilla
binary tree with some red or black tags. If we want to enforce some of the
invariants of red-black trees (cf.\ Footnote~\ref{foot1}) we must
modify the type definition as follows
\begin{alltt}\color{darkblue}\morecompact
  type \type{RBTree(\(\alpha\))} = \type{BTree(\(\alpha\)) | RTree(\(\alpha\))}
  type \type{ BTree(\(\alpha\))} = \type{( Black, \(\alpha\), RBTree(\(\alpha\)), RBTree(\(\alpha\))) | Leaf  }
  type \type{ RTree(\(\alpha\))} = \type{( Red, \(\alpha\), BTree(\(\alpha\)), BTree(\(\alpha\)))}
\end{alltt}
However, with these definitions the \code{insert} function of binary
trees no longer type-checks. But it is just the matter of giving a
precise type to \code{balance}, since it suffices to add the following type
annotation:\\[1mm]
\centerline{\type{(\,Unbalanced($\alpha$)\;\To\;{RTree($\alpha$)})\;\,\&\;\,(\,$\beta$\setminus{Unbalanced($\alpha$)}\;\To\;$\beta$\setminus{Unbalanced($\alpha$)}\,)}
}\\[1mm]
where
\begin{alltt}\color{darkblue}\morecompact
  type  \type{WrongTree(\(\alpha\))} = \type{(Red, \(\alpha\), RTree(\(\alpha\)), BTree(\(\alpha\)))}
                     \,| \type{(Red, \(\alpha\), BTree(\(\alpha\)), RTree(\(\alpha\)))}\medskip
  type \type{Unbalanced(\(\alpha\))} = \type{(Black, \(\alpha\), WrongTree(\(\alpha\)), RBTree(\(\alpha\)))}
                     \,| \type{(Black, \(\alpha\), RBTree(\(\alpha\)), WrongTree(\(\alpha\)))}
\end{alltt}
The two type definitions state that a wrong tree is a red tree with a
black child and that an unbalanced tree is a black tree with a wrong
child. The annotation describes the semantics of
\code{balance}: it transforms an unbalanced tree into a red tree and
leaves any other argument unchanged. We recognize in this type the
pattern of our simpler example (the same pattern appears also in some other
parts of the red-black tree implementation). It is then possible to
deduce for the insertion function for red-black trees the type
\type{BTree(\(\alpha\))\(\,\to\alpha\to\,\){BTree}(\(\alpha\))}.  For the valid CDuce code with an
explanation of subtler typing details, the reader can refer to Appendix
A of~\cite{polyduce2}.

\section{Types}
\label{sec:types}

We have seen in the previous section that set-theoretic types play a
key role in typing several language constructs---from branching and
pattern matching to function overlo\-ading---very precisely. However,  we have glossed over exactly how a type checker should treat them.
It is essential to define
a suitable notion of \emph{subtyping} on these types.
The informal description we have given
suggests that certain properties should hold.
In particular, we expect union and intersection types
to satisfy commutative and distributive properties of Boolean algebras.
Moreover, we expect, for example,\\[1mm]\centerline{$
  (\Int \to \Int) \land (\Bool \to \Bool)
  \leq
  (\Int \lor \Bool) \to (\Int \lor \Bool)$
}\\[1mm]
to hold, so that the typing of functions with type-cases works as we sketched.
To model occurrence typing, we want
$ (\Int \lor \Bool) \land \Int $ to be equivalent to $ \Int $
and $ (\Int \lor \Bool) \land \lnot \Int $ to be equivalent to $ \Bool $.



\subsection{Limitations of syntactic systems}\label{sec:syntactic}
Arguably, it is intuitive to view types and subtyping
in terms of sets and set inclusion,
especially to describe set-theoretic types.%
  \footnote{%
    For instance,
    this model is used to explain subtyping in the online documentation of Flow
    at \url{https://flow.org/en/docs/lang/subtypes}.}
We can see a type as the set of the values of that type
in the language we consider.
Then, we expect $ t_1 $ to be a subtype of $ t_2 $
if every value of type $ t_1 $ is also of type $ t_2 $,
that is, if the set of values denoted by $ t_1 $
is included in the set denoted by $ t_2 $.
In this view, union and intersection types
correspond naturally to union and intersections of sets;
negation corresponds to complementation with respect to the set of all values.

However, most systems reason on subtyping without using this
interpretation of types as set of values. These systems (we call them,
syntactic systems) rather use rules
that are sound but not complete with respect to this model:
that is, they do not allow $ t_1 \leq t_2 $
in some cases in which every value of type $ t_1 $
is in fact a value of type $ t_2 $.
Incompleteness is not necessarily a problem,
but it can result in unintuitive behavior.
We show two examples below.

\paragraph{Lack of distributivity:}
Consider this code in Flow.%
  \footnote{%
    Adapted from the StackOverflow question at
    \url{https://stackoverflow.com/questions/44635326}.}
\begin{align*}
  & \textsf{type A = \{ a: number \}} \\[-1.2mm]
  & \textsf{type B = \{ kind: "b", b: number \}} \\[-1.2mm]
  & \textsf{type C = \{ kind: "c", c: number \}} \\[1mm]
  & \textsf{type T = (A \& B) \texttt{|} (A \& C)} \\[-1.2mm]
  & \textsf{function f\,(x: T) \{
      return (x.kind \texttt{===} "b") ? x.b : x.c \}
    }
\end{align*}
The first three lines declare three object types;
in \K{B} and \K{C}, \textsf{"b"} and \textsf{"c"} are the singleton types
of the corresponding strings.
The type \K{T} is defined as the union of two intersection types,
namely, \K{A\&B} (the type of objects with a fields \K{a} and \K{b} of
type \K{number} and a field \K{kind} of type \K{"b"}) and  \K{A\&C} (the type of objects with a fields \K{a} and \K{c} of
type \K{number} and a field \K{kind} of type \K{"c"}).

The function \K{f} is well typed:
as in \K{handleResponse} before,
occurrence typing recognizes that
\K{x} is of type \textsf{A \& B} in the branch \textsf{x.b}
and of type \textsf{A \& C} in the branch \textsf{x.c}.
However, if we replace the definition of \K{T}
to be \textsf{type T = A \& (B \texttt{|} C)},
the code is rejected by the type checker of Flow.
Occurrence typing does not work
because \K{T} is no longer explicitly a union type.
Flow considers \textsf{(A \& B) \texttt{|} (A \& C)} to be
a subtype of \textsf{A \& (B \texttt{|} C)}:
indeed, this can be proven just by assuming
that unions and intersections are respectively joins and meets for subtyping.
But in Flow's type system, subtyping does not hold in the other direction,
because Flow does not consider distributivity.

\paragraph{Union and product types:}

Apart from distributivity laws,
we could also expect interaction between union and intersection types
and various type constructors.
Consider product types;
we might expect the two types
$ (t_1 \times t) \lor (t_2 \times t) $ and $ (t_1 \lor t_2) \times t $
to be equivalent (i.e., each one subtype of the other one):
intuitively, both of them describe the pairs
whose first component is either in $ t_1 $ or in $ t_2 $
and whose second component is in $ t $.
But this reasoning is not always reflected in the behavior of type checkers.

For example, consider this code in Typed Racket
(similar examples can be written in Flow or TypeScript).
\begin{align*}
  & \textsf{%
      (define-type U-of-Pair (U (Pair Integer Boolean) (Pair String Boolean)))}
  \\[-1.2mm]
  & \textsf{(define-type Pair-of-U (Pair (U Integer String) Boolean))}
  \\[1mm]
  & \textsf{(define f (lambda ([x : U-of-Pair]) x))}
  \\[-1.2mm]
  & \textsf{(define x (ann (cons 3 \#f) Pair-of-U))}
  \\[-1.2mm]
  & \textsf{(f x)}
\end{align*}
We define two type abbreviations.
In Typed Racket, \K{U} denotes a union type and \K{Pair} a product type,
so \textsf{U-of-Pair} is $
  (\K{Integer} \times \K{Boolean}) \lor (\K{String} \times \K{Boolean})
$, and \textsf{Pair-of-U} is $
  (\K{Integer} \lor \K{String}) \times \K{Boolean}
$.
The two types are not considered equivalent.
To show it, we define a function \K{f} whose domain is \textsf{U-of-Pair}
(for simplicity, we take the identity function)
and try to apply it to an argument \K{x} of type \textsf{Pair-of-U};
to define \K{x}, we use an explicit type annotation (\K{ann})
to mark the pair \textsf{(cons 3 \#f)} as having type \textsf{Pair-of-U}.
The application is rejected.
If we exchange the two type annotations, instead, it is accepted:
the type checker considers \textsf{U-of-Pair} a subtype of \textsf{Pair-of-U},
but not the reverse.

\subsection{Semantic subtyping}\label{sec:semsub}

In a nutshell we have to define the subtyping relation  so that the types satisfy all
the commutative and distributive laws we expect from their set-theoretic
interpretation. But a ``syntactic'' definition of
subtyping---i.e., a definition given by a set of deduction rules---is
hard to devise since, as shown by the previous examples, it may
yield a definition that is sound but not complete. To obviate this problem we
follow the \emph{semantic subtyping} approach~\cite{FCB02,FCB08}. In this approach  subtyping is defined
by giving an interpretation $ \Inter{\cdot} $ of types as sets
and defining $ t_1 \leq t_2 $
as the inclusion of the interpretations, that is, $ t_1 \leq t_2 $
is defined as $ \Inter{t_1} \subseteq \Inter{t_2} $.
Intuitively, we can see $ \Inter{t} $
as the set of values that inhabit $ t $ in the language.
By interpreting union, intersection, and negation
as the corresponding operations on sets
and by giving appropriate interpretations to the other constructors,
we ensure that
subtyping will satisfy all expected commutative and distributive laws.

Formally, we proceed as follows. We first fix two countable sets:
a set $ \Constants $ of \emph{language constants}
(ranged over by $ c $)
and a set $ \BasicTypes $ of \emph{basic types}
(ranged over by $ b $).
For example, we can take constants to be Booleans and integers: $
  \Constants = \Set{\K{true}, \K{false}, \K{0}, \K{1}, \K{-1}, \dots}
$.
$ \BasicTypes $ might then contain $ \Bool $ and $ \Int $;
however, we also assume that, for every constant $ c $,
there is a ``singleton'' basic type which corresponds to that constant alone
(for example, a type for $ \K{true} $, which will be a subtype of $ \Bool $).
We assume that a function $
  \ConstantsInBasicType : \BasicTypes \to \Pset(\Constants)
$ assigns to each basic type the set of constants of that type
and that a function $
  \BasicTypeOfConstant{(\cdot)} : \Constants \to \BasicTypes
$ assigns to each constant $ c $ a basic type $ \BasicTypeOfConstant{c} $
such that $ \ConstantsInBasicType(\BasicTypeOfConstant{c}) = \Set{c} $.

\begin{definition}[Types]\label{def:types}
  The set $ \Types $ of \emph{types} is the set of terms $ t $
  coinductively produced by the following grammar\\
  \centerline{\(
    t ::= b \mid t \times t \mid t \to t \mid
    t \lor t \mid \lnot t \mid \Empty
  \)}\\[1mm]
  and which satisfy two additional constraints: $(1)$ \emph{regularity}:
          the term must have a finite number of different sub-terms; $(2)$ \emph{contractivity}:
          every infinite branch must contain an infinite number
          of occurrences of the product or arrow type constructors.
\end{definition}
We use the abbreviations
$
  t_1 \land t_2 \eqdef \lnot (\lnot t_1 \lor \lnot t_2)
$, $
  t_1 \setminus t_2 \eqdef t_1 \land (\lnot t_2)
$, and $
  \Any \eqdef \lnot \Empty
$ (in particular, $\Any$ corresponds to the type~\ANY{} we used in the examples
  of Section~\ref{sec:motivations}).
We refer to $ b $, $ \times $, and $ \to $ as \emph{type constructors},
and to $ \lor $, $ \lnot $, $ \land $, and $ \setminus $
as \emph{type connectives.} As customary, connectives have priority over constructors and negation has the highest priority---e.g., $\neg s{\vee}t\to u{\wedge}v$ denotes  $((\neg s){\vee}t)\to (u{\wedge}v)$.

Coinduction accounts for recursive types and it is coupled with a
contractivity condition which excludes infinite terms
that do not have a meaningful interpretation as types or sets of values:
for instance, the trees satisfying the equations
$ t = t \lor t $
(which gives no information on which values are in it)
or $ t = \lnot t $ (which cannot represent any set of values).
Contractivity also gives an induction principle on $\Types$ that allows us to apply the induction hypothesis
below type connectives (union and negation),
but not below type constructors (product and arrow).
As a consequence of contractivity,
types cannot contain infinite unions or intersections.
The regularity condition is necessary
only to ensure the decidability of the subtyping relation.

In the semantic subtyping approach
we give an interpretation of types as sets;
this interpretation is used to define the subtyping relation
in terms of set containment.
We want to see a type as the set of the values that have that type in
a given language.
However, this set of values cannot be used directly to define the interpretation,
because of a problem of circularity.
Indeed, in a higher-order language,
values include well-typed $\lambda$-abstractions;
hence to know which values inhabit a type
we need to have already defined the type system (to type $\lambda$-abstractions),
which depends on the subtyping relation,
which in turn depends on the interpretation of types.
To break this circularity,
types are actually interpreted as subsets of a set $ \Domain $,
an \emph{interpretation domain},
which is not the set of values,
though it corresponds to it intuitively
(in \cite{FCB08}, a correspondence is also shown formally:
we return to this point in Section~\ref{sec:closing}).
We use the following domain.

\begin{definition}[Interpretation domain]\label{def:domain}
The \emph{interpretation domain} $ \Domain $ is the set of finite terms $ d $
produced inductively by the following grammar\vspace{-2mm}
\begin{align*}
  d & ::=  c \mid (d, d) \mid \Set{(d, \domega), \dots, (d, \domega)}
    &
    \domega & ::= d \mid \Omega\\[-7mm]
\end{align*}
where $ c $ ranges over the set $ \Constants $ of constants
and where $ \Omega $ is such that $ \Omega \notin \Domain $.
\end{definition}

The elements of $ \Domain $ correspond, intuitively,
to the results of the evaluation of expressions.
These can be constants or pairs of results,
so we include both in $ \Domain $. Also, in a higher-order language,
the result of a computation can be a function which are represented in this model by finite relations
of the form $ \Set{(d_1, \domega_1), \dots, (d_n, \domega_n)} $,
where $ \Omega $ (which is not in $ \Domain $)
can appear in second components to signify
that the function fails (i.e., evaluation is stuck) on the corresponding input.

The restriction to \emph{finite} relations
is standard in semantic subtyping and it is one of
its subtler aspects (see~\cite{CF05} for a detailed explanation of
this aspect). In principle, given some mathematical domain
$\Domain$, we would like to interpret $t_1\to t_2$ as the set of functions
from $\Inter{t_1}$ to $\Inter{t_2}$. For instance if we consider
functions as binary relations, then $\Inter{t_1\to t_2}$ could be the
set
$
\{\ f\subseteq\Domain^2\mid \text{for all }(d_1,d_2){\in}f , \text{ if } d_1{\in}\Inter{t_1}\text{ then } d_2{\in}\Inter{t_2}\ \}
$
or, compactly, \raisebox{-.5pt}[1em]{$\Pset(\overline{\Inter{t_1}{\times}\overline{\Inter{t_2}}})$},  where the $\overline S$  denotes the complement of the set $S$ within the appropriate universe (in words, these are the sets of pairs in which it is \emph{not} true that the first projection belongs to $\Inter{t_1}$ and the second does not belong to $\Inter{t_2}$).
But here the problem is not circularity but cardinality, since this would require $\Domain$ to contain $\Pset(\Domain^2)$, which is impossible. The
solution  given by~\cite{FCB08} relies on the observation that in
order to use types in a programming language we do not need to know
what types are, but just how they are related (by subtyping). In other
terms, we do not require the interpretation of an arrow
type to be \emph{exactly} the set of all functions of that type. We just
require that this interpretation induces the same subtyping relation
as interpreting an arrow type with this set would yield. That is, the
interpretation must satisfy the (weaker) property\\[1mm]\centerline{
$
\Inter{s_1{\to}s_2}\subseteq\Inter{t_1{\to}t_2} \iff  \Pset({\overline{\Inter{s_1}{\times}\overline{\Inter{s_2}}}})\subseteq\Pset({\overline{\Inter{t_1}{\times}\overline{\Inter{t_2}}}})
$.}\\[1mm]
If we interpret $t_1{\to}t_2$ as the set
$\PsetFin({\overline{\Inter{t_1}{\times}\overline{\Inter{t_2}}}})$
(where $ \PsetFin $ denotes the restriction of the powerset to finite
subsets), then this property holds. 

The above explains why we use a domain $\Domain$ with finite
relations and define the interpretation $ \Inter{t} $ of a type $ t $
so that it satisfies the following equalities,
where $ \Domain_\Omega = \Domain \cup \Set{\Omega} $:
\begin{align*}
  \Inter{t_1 \lor t_2} & = \Inter{t_1} \cup \Inter{t_2} &
  \Inter{\lnot t} & = \Domain \setminus \Inter{t} &
  \Inter{\Empty} & = \emptyset
  \\
  \Inter{b} & = \ConstantsInBasicType(b) &
  \Inter{t_1 \times t_2} & = \Inter{t_1} \times \Inter{t_2} \\
  \Inter{t_1 \to t_2} & =
    \Set{R \in \PsetFin(\Domain \times \Domain_\Omega) \mid
      \forall (d, \domega) \in R. \: d \in \Inter{t_1} \implies \domega \in \Inter{t_2}}
    \span\span\span\span 
\end{align*}
This interpretation is reminiscent of a common practice in denotational
semantics that consists in interpreting functions as the set of their finite
approximations: we will discuss this relation more in
Section~\ref{sec:semantics}. A consequence of this
interpretation is that the type $\Empty\to\Any$ contains all the
(well-typed) functions: it will play an important role in
Section~\ref{sec:languages}. The interpretation also explains the need
of the constant $\Omega$: this constant is used to ensure that $\Any\to\Any$ is not a
supertype of all function types: in a domain without $\Omega$ (i.e.,
where the last of the equalities above would use $d$ instead of $\domega$) every well-typed function could be subsumed to  $\Any\to\Any$
and, therefore, every application could be given the type $\Any$,
independently from the types of the function and of its argument; thanks to $\Omega$
instead $\Any\to\Any$ contains only the functions whose domain is
exactly $\Any$, since a function with domain, say, \Int{}, could map
non-integer elements to $\Omega$, thus excluding it from $\Any\to\Any$
(since $\Omega\not\in\Any$): see Section 4.2 of~\cite{FCB08} for
details.

We cannot take the equations above
directly as an inductive definition of $ \Inter{\cdot} $
because types are not defined inductively but coinductively.
Therefore we give the following definition,
which validates these equalities
and which uses the aforementioned induction principle on types
and structural induction on $ \Domain $.

\begin{definition}[Set-theoretic interpretation of types]\label{def:interpretation-of-types}
We define a binary predicate \hbox{$ (\domega : t) $}
(``the element $ \domega $ belongs to the type $ t $''),
where $ \domega \in \Domain \cup \Set{\Omega} $ and $ t \in \Types $,
by induction on the pair $ (\domega, t) $ ordered lexicographically.
The predicate is defined as:
\bgroup\setlength{\jot}{0ex}
\begin{align*}
  (c : b) & = c \in \ConstantsInBasicType(b) \\
  ((d_1, d_2) : t_1 \times t_2 ) & =
    (d_1 : t_1) \mathrel{\mathsf{and}} (d_2 : t_2) \\
  (\Set{(d_1, \domega_1), \dots, (d_n, \domega_n)} : t_1 \to t_2) & =
    \forall i \in \Set{1, \dots, n} . \:
    \mathsf{if} \: (d_i : t_1) \mathrel{\mathsf{then}} (\domega_i : t_2)\hspace*{-7mm} \\
  (d : t_1 \lor t_2) & = (d : t_1) \mathrel{\mathsf{or}} (d : t_2) \\
  (d : \lnot t) & = \mathsf{not} \: (d : t) \\
  (\domega : t) & = \mathsf{false} & \text{otherwise}
\end{align*}\egroup
We define the \emph{set-theoretic interpretation}
$ \Inter{\cdot} : \Types \to \Pset(\Domain) $
as $ \Inter{t} = \Set{d \in \Domain \mid (d : t)} $.
\end{definition}
Notice that $\Omega\not\in\Inter t$, for any type $t$. Finally,
we define the subtyping preorder and its associated equivalence relation
as:

\begin{definition}[Subtyping]\label{def:subtyping}
  We define the \emph{subtyping} relation $ \leq $
  and the \emph{subtyping equivalence} relation $ \simeq $
  as
  \(
    t_1 \leq t_2 \iffdef \Inter{t_1} \subseteq \Inter{t_2}\) and   
  \(t_1 \simeq t_2 \iffdef (t_1 \leq t_2) \mathrel{\mathsf{and}} (t_2 \leq t_1)
    \: .
  \)
\end{definition}
The subtyping relation is decidable. Deciding whether $t_1$ is a
subtype of $t_2$ is equivalent to deciding whether $t_1\wedge\neg t_2$
is the empty type, insofar as  $t_1\leq
t_2 \iff \Inter{t_1}\subseteq\Inter{t_2}\iff \Inter{t_1}\cap(\Domain\setminus\Inter{t_2})\subseteq\varnothing\iff\Inter{t_1\wedge\neg
t_2}\subseteq\varnothing \iff t_1\wedge\neg t_2\leq\Empty$. A
detailed description of the subtyping algorithm and of the data
structures used to implement it efficiently can be found in~\cite{Cas20}.

\subsection{Polymorphic Extension}\label{sec:polysemsub}

The examples we gave at the beginning of this article used polymorphic
types. Syntactically, this means extending the grammar of types with type variables
drawn from a countable set $\TypeVars$ ranged over by $\alpha$:
\begin{equation}\label{eq:polytypes}
    t ::= b \mid t \times t \mid t \to t \mid
    t \lor t \mid \lnot t \mid \Empty\mid\alpha
\end{equation}
However, extending the semantic subtyping approach to define a
subtyping relation on these types is not straightforward and has been a longstanding open
problem. The reason is explained by \citet{HFC09} who
point out that the naive solution of defining subtyping of
two polymorphic types as equivalent to the subtyping of all their
ground instances
yields a subtyping relation that is both untreatable and
counterintuitive. They demonstrate this by defining the following
problematic example:\\[1mm]
\centerline{$t\times\alpha\leq(t\times\neg t)\lor(\alpha\times
  t)$}\\[1mm]
One could expect this judgment not to hold, because the type variable $\alpha$ appears in
unrelated positions in the two types (in the second component on the left of a product, in the first one on
the right). According to the naive definition, instead, the judgment holds if and only
if $t$ is a singleton type.

The solution to this problem was found
by~\citet{CX11} who argue that one should consider only 
interpretations of types where judgments such as the above do not hold. This should ensure that
subtyping on type variables behaves closer to the expectations for parametric
polymorphism, so that a type variable can occur on the right-hand side of
a subtyping judgment only if it occurs in a corresponding position on the
left-hand side. To that end~\citet{CX11}
propose \emph{convexity} as a general property of interpretations
that avoid pathological behavior such as the example above. We leave the interested reader to
refer to~\cite{CX11} for details on the convexity property and its
interpretation. Here we present a very simple way to extend the
 interpretation of Definition~\ref{def:interpretation-of-types}
into a convex interpretation for polymorphic types. The idea, due to
\citet{Gesbert2011,Gesbert2015}, is to consider
the domain $\Domain$ of Definition~\ref{def:domain} in which all elements are
labeled by a finite set of type variables\vspace{-1.5mm}
\begin{align*}
  d & ::=  c^L \mid (d, d)^L \mid \Set{(d, \domega), \dots, (d, \domega)}^L
    &
    \domega & ::= d \mid \Omega\\[-6.5mm]
\end{align*}
with $L\in\PsetFin(\TypeVars)$, and interpret a type variable
$\alpha$ by the set of all elements that are labeled by $\alpha$, that
is $\Inter{\alpha}=\Set{ d\mid \alpha\in\tags(d)}$ (where we define
$\tags(c^L)=\tags((d, d' )^L)=\tags(\Set{(d_1, \domega_1), \dots,
  (d_n, \domega_n)}^L)=L$). The interpretation of all other types
disregards labels (e.g., the interpretation of \Int{}
is the set of all integer constants labeled by any set of
variables). It is straightforward to modify
Definition~\ref{def:interpretation-of-types} to validate the equality
$\Inter{\alpha}=\Set{ d\mid \alpha\in\tags(d)}$: it suffices to
use the new domain and just add the  clause\\[1mm]
\centerline{$(d:\alpha)\quad=\quad\alpha\in\tags(d)$}
\\[1mm]
No further modification is needed (apart from adding labels in the
first clauses) and Definition~\ref{def:subtyping}
is still valid.\footnote{The reason why the interpretation thus
obtained is convex is that every type is interpreted into an infinite
set (even singletons types, since, e.g., $\Inter{\type{true}} =
\Set{\textsf{true}^L\mid L\in\PsetFin(\TypeVars)}$). See~\citet{CX11} to
see how this implies convexity.}

While the interpretation of type variables is not very intuitive, it
is easy to check that it yields a subtyping relation that has all the
sought properties. First and foremost, according to this interpretation
a type is empty if and only if all its instances are empty. In
particular, as expected, the interpretation of a type variable
$\alpha$ is never empty (it contains all the elements tagged by
$\alpha$) insofar as $\alpha$ could be instantiated into a non-empty
type. Also, the interpretation of a variable is contained in the
interpretation of another variable if and only if the two variables
are the same.\footnote{It is important to avoid confusion between
(sub)type equivalence and unification. There is a fundamental
difference between being the same type (i.e., denoting the same set of
values) and to be unifiable: two variables can always be unified, but
if they are not the same, then it is not safe to use an expression
whose type is one variable where an expression whose type is a different variables is expected. For instance,
\code{fun(x:\(\alpha\)→\(\alpha\)\,,\,y:\(\beta\))\,=\,xy}
is not well typed (since the programmer wrote these variables, then they
are considered monomorphic and it is unsound to use a \(\beta\)
expression where an \(\alpha\) expression is expected)
while \code{fun(x:\(\alpha\)→\(\alpha\)\,,\,y:\(\alpha\))\,=\,xy} is
well typed, and so is \code{fun(x,y)\,=\,xy}. To type the latter the
type system assigns the type \code{\(\alpha\)} to \code{x}
and \code{\(\beta\)} to \code{y} and then \emph{unifies} \code{\(\alpha\)}
with \code{\(\beta\)→\(\gamma\)} which yields  the
type \code{((\(\beta\)→\(\gamma\))×\(\beta\))→\(\gamma\)} of the function.}

Finally, $\alpha\land t$ is empty if and only if $t$ is
empty since, otherwise, we could obtain a non-empty type by
instantiating $\alpha$ with $t$. For instance, since
$\Inter{\alpha}=\Set{ d\mid \alpha\in\tags(d)}$ and 
$\Inter\Int=\Set{n^L\mid n\in\mathbb{Z}}$, then
$\Inter{\alpha\land\Int} = \Set{n^L\mid n\in\mathbb{Z},\alpha\in L}$: we see that
$\alpha\land\Int$ is not empty since it contains at least
$\mathtt{42}^{\{\alpha\}}$. Likewise, 
since $\Inter{\alpha\land\Int}$ contains both $\mathtt{42}^{\{\alpha\}}$ and
$\mathtt{42}^{\{\alpha,\beta\}}$, then neither
$\alpha\land\Int\leq\beta$ nor $\alpha\land\Int\leq\lnot\beta$ hold,
the former because $\mathtt{42}^{\{\alpha\}}\not\in\Inter\beta$ the
latter because
$\mathtt{42}^{\{\alpha,\beta\}}\not\in\Inter{\lnot\beta}$ (by
definition $\Inter{\lnot\beta}=\Set{d\mid \beta\not\in\tags(d)}$). 
The subtyping relation is again decidable
(see~\cite{CX11} for a detailed description of the algorithm) and, although it is not evident from the
interpretation, the subtyping relation is preserved by type
substitutions, a property needed to ensure soundness for
polymorphic type systems.

\section{Languages}
\label{sec:languages}
The natural candidate languages for the types we presented in the previous
section are $\lambda$-calculi (functional languages) with at least pairs and type-cases. Intuitively,
we need a $\lambda$-calculus because we have arrow types, we need
pairs to inhabit product types, while type-cases are needed to
define overloaded functions and thus inhabit any intersection of
arrow types. We will see in Section~\ref{sec:cduce}
(cf.~Section~\ref{sec:closing} in particular) that the correspondence
between set-theoretic types and a language satisfying these criteria can be formally stated.

We start by describing a generic theoretical language that covers all
the features we outlined in the motivation section. While
theoretically interesting the language will not be effective: it is a
language so generic and expressive that defining a reasonably complete
type-inference algorithm seems very hard. We will then discuss some
trade-offs and define three effective (sub-)systems for which type
inference is possible, but each of which will be able to capture only
a part of the examples of Section~\ref{sec:motivations}.

\subsection{Theoretical framework}\label{sec:theoretical}

The expressions and values of our theoretical language\ are defined as
follows:
\begin{equation}\label{slgrammar}
  \begin{array}{lrclr}
    \textbf{Expressions}\quad &e &::=& c\mid x\mid\lambda x.e
    \mid e e\mid (e,e)\mid \pi_i e \mid 
    \tcase{e}{t}{e}{e}\\[.3mm]
    \textbf{Values} & v  &::=& c\mid\lambda x.e\mid (v,v)
  \end{array}
\end{equation}
Expressions are an untyped $\lambda$-calculus with
constants $ c $, pairs $ (e, e) $, pair projections $ \pi_i{e} $, and
type-cases $ \tcase{e}t e e $. 
A type-case $ \tcase{e_0}t{e_1}{e_2} $ is a dynamic type test that
first evaluates $ e_0 $
and, then, if $ e_0 $ reduces to a value $ v $
evaluates $ e_1 $ if $ v $ has type $ t $ or $ e_2 $ otherwise.
Formally, the reduction semantics is that of a call-by-value
pure $\lambda$-calculus with pairs and type-cases.
The reduction is given by the
following notions of reductions (where  $e\subs x{v}$ denotes the
capture avoiding substitution of $v$ for $x$ in $e$)
\[\qquad\begin{array}{rcl}
  (\lambda x.e)v &\reduces& e\subs x{v}
\\[1mm]
  \pi_1 (v_1,v_2) &\reduces& v_1
\\
  \pi_2 (v_1,v_2) &\reduces& v_2
\end{array}
\qquad\qquad\qquad\qquad
\begin{array}{rcll}
  \tcase{v}{t}{e_1}{e_2} &\reduces& e_1\qquad\text{if } v\in t
  \\[1mm]
  \tcase{v}{t}{e_1}{e_2} &\reduces& e_2\qquad\text{if } v\not\in t  
\end{array}\]
together with the context rules that implement a leftmost outermost
reduction strategy, that is, $E[e]\reduces
E[e']$ if $e\reduces e'$ where the evaluation contexts $E[\cdot]$ are
defined as $E ::=  [\,]\mid v E \mid E e \mid
   (v,E) \mid (E,e) \mid \pi_i E \mid 
    \tcase{E}{t}{e}{e}$. In the reduction rules we used the
 notation $v\in t$ to indicate that the value $v$ has type $t$. Here,
 this corresponds to deducing the judgment $\varnothing\vdash v: t$ using the rules given in
 Figure~\ref{fig:declarative}, rules
  that form the type-system of our
 language; but we will see that for the three system variations
 we present later on, the relation $v\in t$ can be defined without
 resorting to the type-system: this is an important property since we
 do not want to call the type-inference algorithm to decide at run-time the branching
 of a type-case. 

\ifcompact
\input{fig-decl-typing-less-condensed}
\else
\begin{figure}\vspace{-2mm}
\begin{mathpar}
    \Infer[Const]
    { }
    {\Gamma\vdash c:\basic{c}}
    { }
    \and
    \Infer[Var]
  { }
  {\Gamma \vdash x: \Gamma(x)}
  {x\in\dom\Gamma}
 \\
  \Infer[$\to$I]
    {\Gamma,x:t_1\vdash e:t_2}
    {\Gamma\vdash\lambda x.e: \arrow{t_1}{t_2}}
    { }
  \qquad 
  \Infer[$\to$E]
  {
    \Gamma \vdash e_1: \arrow {t_1}{t_2}\quad
    \Gamma \vdash e_2: t_1
  }
  { \Gamma \vdash {e_1}{e_2}: t_2 }
  { }
  \\
  \Infer[$\times$I]
  {\Gamma \vdash e_1:t_1 \and \Gamma \vdash e_2:t_2}
  {\Gamma \vdash (e_1,e_2):\pair {t_1} {t_2}}
  { }
 \qquad
  \Infer[$\times$E$_1$]
  {\Gamma \vdash e:\pair{t_1}{t_2}}
  {\Gamma \vdash \pi_1 e:t_1}
  { } \quad
  \Infer[$\times$E$_2$]
  {\Gamma \vdash e:\pair{t_1}{t_2}}
  {\Gamma \vdash \pi_2 e:t_2}
  { }
\\
\hspace*{-.5mm}\Infer[$\wedge$]
  {\Gamma \vdash e:t_1 \ \ \ \Gamma \vdash e:t_2}
  {\Gamma \vdash e: {t_1}\wedge {t_2}}
  { }
\
\Infer[$\vee$]
    {     \Gamma \vdash e':t_1{\vee} t_2\quad 
      \Gamma\!, x:t_1\vdash e:t\quad \Gamma\!, x:t_2\vdash e:t
    }  
    {
    \Gamma\vdash e\subs x {e'}  : t
    }
    { 
    }\  
   \Infer[$\leq$]
    { \Gamma \vdash e:t\quad t{\leq} t' }
    { \Gamma \vdash e: t' }
    { }
\\
  \Infer[$\Empty$]
  {
    \Gamma \vdash e:\Empty
  }
  {\Gamma\vdash \tcase {e} t {e_1}{e_2}: \Empty}
  { }
\qquad
  \Infer[$\in_1$]
  {
    \Gamma \vdash e:t \and \Gamma\vdash e_1: t_1
  }
  {\Gamma\vdash \tcase {e} t {e_1}{e_2}: t_1}
  { }
  \quad
  \Infer[$\in_2$]
  {
    \Gamma \vdash e:\neg t \and \Gamma\vdash e_2: t_2
  }
  {\Gamma\vdash \tcase {e} t {e_1}{e_2}: t_2}
  { }
\end{mathpar}
\caption{Declarative type system\label{fig:declarative}}\vspace{-4mm}
\end{figure}

\fi
The rules in the first three rows of Figure~\ref{fig:declarative} do
not deserve any special comment: they are the standard typing rules
for a simply-typed $\lambda$-calculus with pairs where, as customary,
$\Gamma$ ranges over type environments, that is, finite mappings from
variables to types, $\Gamma, x:t$ denotes the extension of the
environment $\Gamma$ with the mapping $x\mapsto t$ provided that
$x\not\in\dom{\Gamma}$ (we will use $\varnothing$ to denote the empty type environment).

The rules in the fourth row are also standard. The first
rule \Rule{$\land$} is
the classic introduction rule for intersection: it states that if an
expression has two types, then it has their intersection, too. The
second rule \Rule{$\lor$} is the classic
union elimination rule as it was first introduced
by MacQueen et al.~\cite{MacQueen1986}: it states that given some expression (here,  $e\subs x {e'}$) with a subexpression $e'$ of type
$t_1{\vee}t_2$, if we can give to this expression the type $t$ both under the
hypothesis that $e'$ produces a result of type $t_1$ and under the
hypothesis that $e'$ produces a result in $t_2$, then we can safely
give this expression type $t$ (to satisfy type soundness, this rule needs a further restriction when used with polymorphic types: see~\cite{CLN24}). The last rule of the row is the
\emph{subsumption} rule \Rule{$\leq$} that states that if an expression has some
type $t$, then it has all  super-types of $t$, too. Together, the rules
in the first four lines of Figure~\ref{fig:declarative} form a well-known type-system, since they are the
same rules as those in Definition 3.5 of the classic work on union and
intersection types by Barbanera, Dezani, and
de'Liguoro~\cite{BDD95}. Although the rules are textually the same as in~\cite{BDD95},
there is an important difference between the system in~\cite{BDD95}
and the one in Figure~\ref{fig:declarative}, namely, that our types
are a strict extension of
those of~\cite{BDD95} since we also have recursive types,
negation types, and the empty type. As a consequence our subsumption
rule uses the subtyping relation of Definition~\ref{def:subtyping}
which is more general than the one in~\cite{BDD95}  of which it
is a conservative extension (cf.~\cite{itrs02}).

Finally, the last three rules are specific to systems with
set-theoretic types and type-cases. They are rather new (they were first introduced
in~\cite{CLNL22}) and provide a natural and nifty way to type
type-case expressions. The first rule states that if the tested
expression $e$ has the empty type (i.e., if $e$ diverges, that is, it can
only produce a value that is in the empty set), then so has the
whole type-case expression.  The second rule states that if $e$ can
only produce a result in $t$, then the type of $\tcase {e} t
{e_1}{e_2}$ is the type of $e_1$. The third rule states that if $e$
can only produce a result in $\neg t$, then the type of $\tcase {e} t
{e_1}{e_2}$ is the type of $e_2$: since the negation type $\neg t$ is
interpreted set-theoretically as the set of all values that are
\emph{not} of type $t$, this means that, in that case, $e$ can only
produce a result \emph{not} of type $t$.
The reader may wonder how we type a type-case expression  $\tcase
{e} t {e_1}{e_2}$ when the
tested expression $e$ is neither of type $t$ nor of type $\neg t$. As a
matter of fact, a type-case is interesting only if we cannot statically
determine whether it will succeed or fail.  For instance, when
discussing occurrence typing, we informally described how to type the
expression $\tcase x\Int{(x+1)}{\lnot x}$ when $x$ is of
type \type{\Int$\lor$\Bool}, that is, in that case, when $x$ is neither of type
$\Int$ nor of type \type{$\neg$\Int}. Here is where the union
elimination rule \Rule{$\lor$} shows its full potential. Even though the
expression $e$ tested in $\tcase
{e} t {e_1}{e_2}$ has some type $s$ that is neither contained in
(i.e., subtype of) $t$ nor in $\neg t$, we can use intersection and
negation to split $s$ into
the union of two types that have this property, since $s \simeq (s\wedge
t)\vee(s\wedge\neg t)$.
We can thus apply the union rule and check the
type-case under the hypothesis that the tested expression has type $(s\wedge
t)$ and under the hypothesis that it has type $(s\wedge
\neg t)$. For instance, for  $\tcase x\Int{(x+1)}{\lnot x}$ we check
the type-case under the hypothesis that $x$ has type \Int{} (i.e., \type{(\Int$\lor$\Bool)$\land$\Int})
 and deduce the type \Int,  and under the hypothesis that $x$ has type \Bool{} (i.e., \type{(\Int$\lor$\Bool)$\land\neg$\Int})
 and deduce the type \Bool{} which, by subsumption, gives for the whole
 expression the type \type{\Int$\lor$\Bool}.

A final important remark is that the deduction system in
Figure~\ref{fig:declarative} is defined modulo
$\alpha$-conversion. This is crucial in systems with union
types since the rule \Rule{$\vee$} breaks the $\alpha$-invariance
property (see Section 2.4 in~\cite{CLNL22} and Discussion 12.5 in \cite{HS08}).

\subsubsection{On deriving negation types.}\label{sec:negation}

The language and the typing rules we just defined are expressive
enough to cover all the examples we described in the first two
sections. However, the rules of Figure~\ref{fig:declarative} are yet
not enough to cover the whole palette of application of set-theoretic
types. The reason is that in the current system the only way to derive
for an expression a negation type is to use the subsumption rule. For
instance, we can deduce $\K{42}:\lnot\Bool$ by subsumption, since
$\K{42}:\Int$ and $\Int\leq\lnot\Bool$ (since all integer constants
are contained in the set of values that are not Booleans). But while subsumption is
sufficient for values formed only by constants,\footnote{These are
 either constants or possibly nested pairs of constants.  All these values
 have a smallest type deduced by the rules of
 Figure~\ref{fig:declarative} and this smallest type
 is \emph{indivisible} (i.e., its only proper subtype is the empty
 type: cf.~\cite{CX11}). An indivisible type acts like a point in the
 set-theory of types: it is
 either contained in a type or in its negation, and so are their
 values. This is the reason why
 subsumption suffices to determine the negation types of this sort of values:
 they are all the types that do not contain the minimal type of the
 value at issue.  Functions, in general, do not have a smallest
 type.\label{fn:negationconst}}
it is not enough for values with functional components.  For
example, consider the successor function $\lambda x.(x+1)$. This
function has type $\Int\to\Int$ but \emph{not} type $\Bool\to\Bool$:
it maps integers to integers but when applied to a Boolean it does not
return a Boolean. Therefore, one would expect the type system to
deduce for the successor function the
type \type{$\lnot(\Bool\to\Bool)$}. However, in this case subsumption
is of no use since \Int\To\Int{} is \emph{not} a subtype
of \type{$\neg(\Bool\to\Bool)$}, and rightly so since it is easy to
find a value that is of the former type but not of the latter one: for
instance, the identity function $\lambda x.x$ is a function that has
type $\Int\to\Int$ but---since it is also of
type \Bool\To\Bool---is \emph{not} of
type \type{$\neg(\Bool\to\Bool)$}.

Intuitively, we would like the type-system to deduce for an expression
$e$ the type $\neg t$ whenever $(i)$ $e$ is typable with some type
$t'$ and $(ii)$ it is not possible to deduce the type
$t$ for it. In a sense we would like to have a rule such as the pseudo-rule
\Rulem{\lnot} here below on the left:
\begin{mathpar}
   \Infer[$\lnot$]
  {\Gamma\vdash e:t'\and\Gamma \not\vdash e:t}
  {\Gamma \vdash e:\lnot t}
  { }
  \and
   \Infer[$\lnot^{(\to)}$]
  {\Gamma\vdash \lambda x.e:t'\and\Gamma \not\vdash \lambda x.e:t\to t''}
  {\Gamma \vdash \lambda x.e:\lnot (t\to t'')}
  { }  
\end{mathpar}
This pseudo-rule, which puts in formulas what we explained in prose,
deduces negation types for a generic expression $e$. However, from a
practical perspective a less generic rule such
as \Rulem{\lnot^{(\to)}} above on the right would suffice:
as a matter of fact, deciding negation types is useful in practice to evaluate
type-cases and these are decided on values rather than generic
expressions. So from a practical viewpoint it suffices to
deduce negation types only for values rather than for all
expressions and, in particular, for $\lambda$-abstractions, since for
all the other values subsumption is enough (see Footnote~\ref{fn:negationconst}). So instead of deducing
generic negation types for generic expressions, it is enough to
deduce negated arrow types for $\lambda$-abstractions, yielding the less general
pseudo-rule \Rulem{\lnot^{(\to)}} above.

A different motivation for deducing negation types is that, for the
reasons we explain in Section~\ref{sec:problems}, few practical
systems implement the union elimination rule \Rulem{\lor} in its full
generality, insofar as deciding when \Rulem{\lor} is to be applied is
still an open problem (technically, this corresponds to determining an inversion lemma
for the \Rulem{\lor} rule). Now, in the absence of
a \Rulem{\lor} rule (e.g., in the systems in
Section~\ref{sec:cduce} and~\ref{sec:implicit}), the property of type preservation by
reduction (also known as the property of \emph{subject reduction}) requires the following property to hold:
\begin{equation}\label{prop:values}
  \textit
  {For every type $ t $ and well-typed value $ v $,
  either $ \varnothing \vdash v: t $
  or $ \varnothing \vdash v: \lnot t $ holds.}
\end{equation}
To illustrate why this is required, consider the expression $ \lambda{x. (x, x)} $
and the following typing derivation
(for some arbitrary type $ t $).
\[
  \TInfer[{[$\leq$]}]
    {
      \TInfer[{[$\land$]}]
      {
        \TInfer
        { \vdots }
        {
          \varnothing\vdash\lambda x.(x,x): t \to (t \times t)
        }
        {}
        \quad
        \TInfer
        { \vdots }
        {
         \varnothing\vdash\lambda x.(x,x): \lnot t \to (\lnot t \times \lnot t)
        }
        {}
      }
      {
         \varnothing\vdash\lambda x.(x,x):    ( t \to (t \times t) ) \land
          ( \lnot t \to (\lnot t \times \lnot t) )
      }
      {}
    }
    {
       \varnothing\vdash\lambda x.(x,x):  
        \Any \to ( (t \times t) \lor (\lnot t \times \lnot t) )
    }
    {}
\]
The subsumption rule can be applied because
\[
  ( t \to (t \times t) ) \land
  ( \lnot t \to (\lnot t \times \lnot t) )
  \leq
  \Any \to ( (t \times t) \lor (\lnot t \times \lnot t) )
  \: \text{:}
\]
in general, it holds that $
  (t_1' \to t_1) \land (t_2' \to t_2) \leq (t_1' \lor t_2') \to (t_1 \lor t_2)
$,
and $ t \lor \lnot t \simeq \Any $.
Now consider an arbitrary type $ t $ and a well-typed value $ v $.
Since $ v $ has type $ \Any $ by subsumption,
the application $ (\lambda{x. (x, x)}){v} $
can be typed as $
  (t \times t) \lor (\lnot t \times \lnot t)
$.
This application reduces to $ (v, v) $.
Therefore, either $ (v, v) $ has type $
  (t \times t) \lor (\lnot t \times \lnot t)
$ or subject reduction does not hold.
Since $ t \times t $ and $ \lnot t \times \lnot t $ are disjoint,
to derive the union type for $ v $
we need either the system to have the \Rulem{\lor} rule, or $ v $ to have either type $ t $ or type $ \lnot t $.
This illustrates the need for the property above which in particular
  requires to be able to derive negation types for functions other than
  by subsumption: e.g., since we cannot derive for $\lambda x.x$ the type
  $\Int\to\Bool$, then we must be able to derive for it the type $\neg(\Int\To\Bool)$.

However, both \Rulem{\neg} and \Rulem{\neg^{(\to)}}, untamed, do not make much sense (which is why we
called them pseudo-rules). First,
their definitions are circular since they depend on the very relation
they are defining. Furthermore, they
cannot be used in a deduction system since they would correspond to
non-monotone immediate-consequence operators for which a fix-point may
not exist and thus cannot be used to define the typing relation by
induction. Therefore, it is  necessary to put some tight-knit restrictions on the
inference of negation types. This requires a lot of care because the very
presence of negation types may yield to paradoxes, as it can be evinced
from considering
the recursive function\footnote{Thanks to recursive types it is easy
to define a polymorphic fix-point combinator and thus define recursive
functions: for every type
$t$ it is possible to define Curry's fix-point combinator $\Y t:
(t\To t)\To t$ as $\lambda f\col t\To t.\Delta^t\Delta^t$ where
$\Delta^t= \lambda x\col\mu X.X\To t. f(xx)$. Since our
calculus is strict, it is more interesting to define, for any type $s$
and $t$ the strict fix-point combinator $\Z {s,t}: ((s\To t)\,\To\,
s\To t) \,\To\, s\To t$ as $\lambda f\col (s\To t)\To\ s\To
t. \textup{E}^{s,t}\textup{E}^{s,t}$ where $\textup{E}^{s,t}=\lambda
x\col\mu X.X\To s\To t. f(\lambda v\col
s.xxv)$.\label{foot:fixpoint}} \code{let rec \(f\)
= \(\lambda{x}.\tcase{f}{\true\to\true}\false\true\)}; it is easy to
see that $f$ maps \true{} to \true{} if and only if it does not have type
$\true\to\true$.

As hard the inference of negation types is, it cannot be dismissed
lightheartedly, since the definition of the relation $v\in t$ depends
on it and so does the semantics of type-cases: if we perform a type
test such as $v\in\neg(\Bool\to\Bool)$, then we expect it to succeed at least
for some functional values (e.g., the successor function). In the second
part of this section we will show different solutions proposed in the
literature to infer negation types in a controlled way.

\subsubsection{On the feasibility of type-inference.}\label{sec:feasibility}
Defining 
inference of negation types is not the only problem to be solved before obtaining a
language usable in practice. The rules of
Figure~\ref{fig:declarative} are still a far cry from a practical system that can decide whether a program is well-typed or not. As customary, there are essentially two problems:
\begin{enumerate}[nosep]
  \item the rules are not \emph{syntax directed}: given a term, to type it we can try to
    apply some elimination/introduction rule, but also to apply the
    intersection rule \Rule{$\wedge$}, or the subsumption rule
    \Rule{$\leq$}, or the union rule \Rule{$\vee$}.
  \item some rules are \emph{non-analytic}:\footnote{We consider
  \emph{non-analytic} (or \emph{synthetic}) a rule in which
       the input (i.e., $\Gamma$ and $e$) of the judgment at the
       conclusion is not sufficient to determine the inputs of the judgments at
       the premises (cf.\ \cite{ML1994,types2019}).}
    if we use the \Rule{$\to$I} rule to type some
    $\lambda$-abstraction we do not know how to determine the type
    $t_1$ in the premise; if we use the \Rule{$\vee$} rule we
    know neither how to determine $e'$ nor how to determine the types
    $t_1$ and $t_2$ that split the type of $e'$.
\end{enumerate}
Notice that \Rulem{\vee} cumulates both problems. The problem that
some rules are not syntax directed can be already  solved in
this system for at least two of the three rules at issue: for the rules  \Rulem{\wedge}
and \Rulem{\leq} it is possible to eliminate them and refactor the use of
intersections and subtyping in the remaining rules. This essentially amounts to resorting to
some form of canonical derivations in which intersection \Rulem{\wedge} and
subsumption \Rulem{\leq} rules are used at specific places:
it can be proved (cf.~\cite{CLNL22}) that a typing judgment is provable
with the system of Figure~\ref{fig:declarative} if and only if
there exists a derivation for that typing judgment where $(i)$ subsumption is
  only used on the left premise of an application or a type-case rule,
  on the right premises of the union rule, and on the
  premise of a projection rule
  and $(ii)$~intersection is only used for expressions that are $\lambda$-abstractions, that is, all the premises of an intersection rule are the
  consequence of a \Rule{$\to$I}. This yields an equivalent system
  formed by the
  rules in Figure~\ref{fig:canonical},
\begin{figure}\vspace{-3mm}  
\begin{mathpar}
  \Infer[$\to$I$^{(\wedge)}$]
    {\forall i\in I\quad\Gamma,x:s_i\vdashC e:t_i}
    {\Gamma\vdashC\lambda x.e: \textstyle\bigwedge_{i\in I}\arrow{s_i}{t_i}}
    { }
  \qquad 
  \Infer[$\to$E$^{(\leq)}$]
  {
    \Gamma \vdashC e_1: t\leq \arrow {t_1}{t_2}\quad
    \Gamma \vdashC e_2: t_1
  }
  { \Gamma \vdashC {e_1}{e_2}: t_2 }
  { }
  \\
  \Infer[$\times$E$_1^{(\leq)}$]
  {\Gamma \vdashC e:t\leq\pair{t_1}{t_2}}
  {\Gamma \vdashC \pi_1 e:t_1}
  { } \quad
  \Infer[$\times$E$_2^{(\leq)}$]
  {\Gamma \vdashC e:t\leq\pair{t_1}{t_2}}
  {\Gamma \vdashC \pi_2 e:t_2}
  { }
  \\  
\Infer[$\vee^{(\leq)}$]
    {     \Gamma \vdash e':t_1\vee t_2 \and
      \Gamma, x:t_1\vdash e:s_1\leq t\and \Gamma, x:t_2\vdash
      e:s_2\leq t
    }  
    {
    \Gamma\vdash e\subs x {e'}  : t
    }{}
    \\
  \Infer[$\in_1^{(\leq)}$]
  {
    \Gamma \vdashC e:t_\circ\leq t \quad \Gamma\vdashC e_1: t_1
  }
  {\Gamma\vdashC \tcase {e} t {e_1}{e_2}: t_1}
  { }
  \and
  \Infer[$\in_2^{(\leq)}$]
  {
    \Gamma \vdashC e:t_\circ\leq\neg t \quad \Gamma\vdashC e_2: t_2
  }
  {\Gamma\vdashC \tcase {e} t {e_1}{e_2}: t_2}
  { }
\end{mathpar}\vspace{-2mm}
\caption{Canonical typing rules\label{fig:canonical}}\vspace{-5mm}
\end{figure}
plus the rules \Rule{Const}, \Rule{Var}, \Rulem{\times\textsc{I}},
and \Rulem{\Empty} of Figure~\ref{fig:declarative}, which do not
change. We improved the situation on the syntax-directed front since
we got rid of \Rulem{\land} and \Rulem{\leq}, but it looks as we
worsened the non-analytic front since now \emph{all} rules in
Figure~\ref{fig:canonical} are non-analytic. In particular, nothing
tells us how to determine the larger types in the subtyping relations
occurring at the premises of these rules. Actually, for the three
elimination rules \Rule{...E$^{(\leq)}$} in Figure~\ref{fig:canonical}
there exists a standard way to determine these larger types which
resorts to using some type operators defined by~\citet{FCB08}.  To
understand it, consider the rule \Rule{$\to$E} for applications in
Figure~\ref{fig:declarative}. It essentially does three things: $(i)$
it checks that the expression in the function position has a
functional type; $(ii)$ it checks that the argument is in the domain
of the function, and $(iii)$ it returns the type of the
application. In systems without set-theoretic types these operations
are straightforward: $(i)$ corresponds to checking that the expression
in the function position has an arrow type, $(ii)$ corresponds to checking that the argument is
in the domain of the arrow deduced for the function, and $(iii)$
corresponds to returning the codomain of that arrow. With
set-theoretic types things get more complicated, since in general the
type of a function is not always a single arrow, but it can be a union
of intersections of arrow types and their negations.\footnote{This
is the reason why, having eliminated in Figure~\ref{fig:canonical} the subsumption rule from the
system, we need in rule \Rule{$\to$E$^{(\leq)}$} to subsume the type
$t$ deduced for the function $e_1$ to an arrow type. Likewise for the
type $t$ of the expression $e$ in \Rule{$\times$E$_i$} which might be
different from a product.  } Checking that the expression in the
function position has a functional type is easy since it corresponds
to checking that it has a type subtype of
$\Empty{\to}\Any$, the type of all functions. Determining its domain and the type of the
application is more complicated and needs the operators $\dom{}$ and
$\circ$ defined as
\(\dom t \eqdef \max \{ u \mid t\leq u\to \Any\}\) and
\(\apply t s \eqdef\min \{ u \mid t\leq s\to u\}\).
In short, $\dom t$ is the largest domain of any single arrow that
subsumes $t$ while $\apply t s$ is the smallest codomain of any arrow type
that subsumes $t$ and has domain $s$. Thus the non-analytic
rule \Rule{$\to$E$^{(\leq)}$} in Figure~\ref{fig:canonical} can be replaced by its analytic
version \Rule{$\to$E$^{(\text{\tiny A})}$} below:
\begin{mathpar}
 \Infer[$\to$E$^{(\text{\tiny A})}$]
  {
    \Gamma \vdash e_1: {t_1}\quad
    \Gamma \vdash e_2: t_2
  }
  {\Gamma \vdash {e_1}{e_2}: t_1\circ t_2 }
  {\begin{array}{l}t_1\leq\Empty\to\Any\\[-1mm]t_2\leq\dom{t_1}\end{array}
  }
\and
\Infer[$\times$E$_i^{(\text{\tiny A})}$]
  {\Gamma \vdash e:t}
  {\Gamma \vdash \pi_i e:\bpi t}
  {t\leq\Any\times\Any}
\end{mathpar}
We need similar operators for projections since in the
rules \Rule{$\times$E$_i^{(\leq)}$} ($i=1,2$) the type $t$
of $e$ in $\pi_i e$ may not be a single product type but, say, a union
of products: all we know is that for the projection to be well-typed $t$ must be a subtype of
$\pair\Any\Any$. So let $t$ be a type such that $t\leq\pair\Any\Any$,
we define \(\bpl t\eqdef \min \{ u \mid t\leq \pair u\Any\}\) and
\(\bpr t\eqdef \min \{ u \mid t\leq \pair \Any u\}\) and replace each
non-analytic rule \Rule{$\times$E$_i^{(\leq)}$}  with the corresponding
analytic version \Rule{$\times$E$_i^{(\text{\tiny A})}$} above. All these type operators
can be effectively computed (see~\cite{FCB08}).

\subsubsection{Practical systems.}\label{sec:problems}

Although we showed how to handle some non-analytic and/or non-syntax-directed rules, filling the gap between our theoretical setting
and practical languages still requires to solve three non-trivial problems:
\begin{enumerate}[nosep]
\item how to infer the type of $\lambda$-abstractions since the
rule \Rulem{\to\textsc{I}^{(\land)}} in Figure~\ref{fig:canonical} not
only is non-analytic, but also states that we must be able to deduce 
for $\lambda$-abstractions intersection of arrows rather than just a
single arrow;
\item how to deduce negation types for expressions or, at least, how
to deduce negated arrow types for $\lambda$-abstractions.
\item how to infer the type of type-cases; this in particular implies
to tame the union-elimination rule \Rulem{\lor} which, in
its present formulation, is still too
generic: to use this rule to type some expression $e$ one has
to guess which subexpression $e'$ of $e$ to single out, which occurrences of
$e'$ in $e$ are to be tested by replacing them by $x$, and how to split
the type of this $e'$ in a union of types to be tested separately.
\end{enumerate}
These three problems  are tightly connected: the inference
of negation types essentially concerns the typing of functions and
it affects the semantics of type-cases; the typing of type-cases
must be based on their semantics, and since this typing is essentially
performed by the union elimination rule, the taming
of this rule cannot thus be disconnected from the semantics of
type-cases, ergo, from the deduction of negation types, ergo, from the
typing of $\lambda$-abstractions. 

Unfortunately, for these three problems there is not a
one-size-fits-all solution, yet. In the next three subsections we are going to present
three different ways to address these problems yielding to three
different practical systems, and discuss their advantages and
drawbacks. We can summarize these three solutions as
follows:
\begin{description}
\item[Core CDuce.] Everything is explicit: every function is
explicitly annotated with its type and the use of union elimination is
limited to type-cases and needs to explicitly specify a variable to capture
the tested value. In summary, no type-reconstruction, no general occurrence typing,
but intersection of arrows and negation types are inferred for the functions.

\item[Type Reconstruction.] Functions are implicitly typed, that is, they need no
annotation: their type is reconstructed but an intersection type can
be deduced only if the function is explicitly annotated with it. No
general occurrence typing and type-cases can test functional values only in a
limited form. Inference of negation types is not performed: it is only
used for proving type soundness.

\item[General Occurrence Typing.] Everything can be implicit, full use
of union elimination to implement occurrence typing, reconstruction of
intersections of arrow types for functions, but no polymorphism or
inference of negation types. The typing algorithm is sound but cannot be complete.
\end{description}

\subsection{Core CDuce}\label{sec:cduce}

The first solution for the three problems just described in Section~\ref{sec:problems}
was given with the definition of the calculus introduced
in~\cite{FCB02,FCB08} to study semantic subtyping, calculus which
constitutes the functional core of the programming language
CDuce~\cite{BCF03}. In this Core Calculus of CDuce (from now on, just
CDuce for short), the problem of inferring intersection types for
$\lambda$-abstractions is solved by annotating them with an
intersection type. Annotations solve also the problem of inferring negated
arrow types for $\lambda$-abstractions since, as we detail below, we can
deduce any negated arrow for a $\lambda$-abstraction as long as the
intersection of this type with the type annotating the function is not
empty.  Finally, for what concerns type-cases and union-elimination,
CDuce restricts union elimination to type-cases expressions whose syntax
is modified so that they specify the binding of the variable used in
the union rule.
    
Before detailing these technical choices it is important to understand
the reason that drove their definition. These choices were made to ``close the circle'':

\subsubsection{Closing the circle.}\label{sec:closing}
All this presentation long we spoke of types as sets of values. In
particular, we said that semantic subtyping consists in interpreting types
as sets of values and then defining one type to be  subtype of
another if and only if the interpretation of the one type is contained in the
interpretation of the other type.
Since a subtyping relation is a pre-order, then it immediately induces
the notions of least upper bound and greatest lower bound of a set of
types. It is then natural to use such notions---thus, the subtyping relation---to characterize,
respectively, union and intersection types. This property was used in
the context of XML processing languages by Hosoya, Pierce, and
Vouillon~\cite{hosoya00regular,hosoya01patter,hosoya01thesis,xduce_toit}:
by combining union types with product and recursive types they encoded XML typing systems such as DTDs or XML Schemas. The work of
Hosoya et al., however, had  an important limitation, since it could not
define the subtyping relation for functions types and, therefore, it
could not be used to type languages with higher order
functions. This impossibility resided in a circularity of the
definition we already hinted at in Section~\ref{sec:semsub}: to define
subtyping one needs to define the type of each value; for non
functional values this can be done by induction on their structure,
but with functional values---i.e., $\lambda$-abstractions---this
requires to type the bodies of the functions which, in turn, needs the very
subtyping relation one is defining.

The solution to this circularity problem was found
by Frisch et al.~\cite{FCB02,FCB08} and consisted of three steps: $(\Romanbar{I})$
interpret types as sets of elements of some domain $\Domain$ and
use this interpretation to define a subtyping relation;
$(\Romanbar{II})$  use the
subtyping relation just defined to type a specifically tailored functional language and in
particular its values; $(\Romanbar{III})$ show that if we interpret a type as the
set of values of this language that have that type, then this new
interpretation induces the same subtyping relation as the starting one
(which interprets types into subsets of the domain $\Domain$).

Step $(\Romanbar{I})$ yielded the definition of the interpretation we gave in
Definition~\ref{def:interpretation-of-types} resulting in the subtyping
relation of Definition~\ref{def:subtyping}. For step  $(\Romanbar{II})$
we need to define a language such that its set of values satisfies the property in
$(\Romanbar{III})$. For that, the language must provide enough values to
separate every pair of distinct types. In other terms, whenever two types
do not have the same interpretation in the denotational model, then
there must exist a value in the language that is in one type but not
in the other one. In
order to provide enough values to distinguish semantically different
types we need three ingredients, two of which are already present in
our system: $(i)$ the inference of intersection types for functions
$(ii)$ a type-case expression, and $(iii)$ a random choice
operator. We detail each of them in the next subsections.

\subsubsection{Inferring Intersection Types for Functions.}\label{inferringintersection}

The first problem we encounter is how to deduce intersection types for
functions. In particular, for every distinct pair of intersections of
arrows,  we want to be able to define a function that
distinguishes them (an
intersection of arrows is never empty since it contains at least the
function that diverges on all arguments). This is difficult
to do in practice unless  functions are explicitly annotated.
As a matter of fact, $\lambda x.x$ has type
$t\to t$ for every type $t$, and thus it has all possible finite
intersections of these types, thus providing an infinite search space
for an intersection type, without a best solution (since we do not
have infinite intersections). To address this
problem one could be tempted to annotate the
parameter of a function with the set of the domains of the
intersection type we want to deduce. In other terms,  one could try to explicitly list
the set of the types $s_i$ to be used by the
rule \Rulem{\to\textsc{I}^{(\land)}} so that, for instance,  the type
deduced for $\lambda x\col\{\Int,\Bool\}.x$ would be
$(\Int\To\Int)\wedge(\Bool\To\Bool)$. Still, this is not enough
because it does not avoid the paradoxes we presented at the end of
Section~\ref{sec:negation}. 
To solve these problems,
Frisch et al.~\cite{FCB08} annotate $\lambda$-abstractions with their
intersection types, thus providing also their return type(s). So the
identity function for integer and Booleans is written in 
the syntax of \cite{FCB08} as
$\lambda^{(\Int\To\Int)\wedge(\Bool\To\Bool)}x.x$ and the system
deduces for it the type $(\Int\To\Int)\wedge(\Bool\To\Bool)$. Using
the right annotations it is then easy to define a value that
distinguishes two  functional types that have different interpretations. 

The first modification to the system of Section~\ref{sec:theoretical} is then
to adopt for functions the syntax and typing rules of Frisch et al.~\cite{FCB08},
that is, we replace in \eqref{slgrammar} the production for $\lambda$-abstractions by
the production  $e \doublecoloneq  \lambda^{\wedge_{i{\in}I}s_i\to
t_i} x.e$ (where $I$ is finite)
and replace in Figure~\ref{fig:canonical}
the rule  \Rulem{\to\textsc{I}^{(\land)}} 
 by the following one:
\begin{mathpar}
\Infer[$\to$I$^{\textsc{\tiny (cduce)}}$]
{\forall i\in I\quad \Gamma, x:s_i\vdash e:t_i}
{\Gamma \vdash \lambda^{\wedge_{i{\in}I}s_i\to t_i} x. e: t\wedge t'}
{\begin{array}{l}
t=\wedge_{i{\in}I}(s_i\to t_i)\\[-1mm]
t'=\wedge_{j{\in}J}\neg(s'_j \to t'_j)\\[-1mm]
t\wedge t'\not\simeq\Empty
\end{array}
}
\end{mathpar}
This rule  (taken verbatim from~\cite{FCB08}) checks whether a $\lambda$-abstraction has all the arrow
types listed in its annotation $t$ and deduces for the  term this type $t$
intersected with an arbitrary finite number of negated arrow types. These
negated arrow types can be chosen freely provided that the type $
t\wedge t'$ remains non-empty. This rule ensures that given a function $\lambda^t x.e$ (where $t$
is an intersection type), for every type $t_1\to t_2$, either
$t_1\to t_2$ can be obtained by subsumption from $t$ or $\neg(t_1\to
t_2)$ can be added to the intersection $t$. In turn this ensures
that, for any function and any type $t$ either the function has type
$t$ or it has type $\neg t$ (see~\cite[Sections 3.3.2 and
3.3.3]{Pet19phd} for a thorough discussion on this rule). The consequences of this may look
surprising.  For
example, it allows the system to type $\lambda^{\Int\To\Int}x. x$ as
$(\Int\To\Int)\wedge\neg(\Bool\To\Bool)$ (notice the negation) even
though, disregarding its 
annotation, the function \emph{does} map Booleans to Booleans. But the
language is explicitly typed, and thus we cannot ignore the
annotations: indeed, the function does not have type
$\Bool\To\Bool$ insofar as its application to a Boolean does not
return another Boolean but an error $\Omega$. The point is that the
theory of semantic subtyping defined by~\cite{FCB08} gives expressions 
an \emph{intrinsic semantics} (in the sense of Reynolds~\cite{Rey03})
since the semantics of $\lambda$-abstractions depends on their explicit type
annotations. This aspect is apparent when one studies the denotational
semantics of CDuce (see Section~\ref{sec:semantics} and Lanvin's PhD dissertation~\cite[Chapter 11]{Lanvin21phd}): in particular, notice that
according to rule \Rule{$\to$I$^{\textsc{\tiny (cduce)}}$} we have $\lambda^{\Int{\to}\type{42}} x.{42}:\Int\To\type{42}$ while
$\lambda^{\Int{\to}\Int}
x.{42}:\neg(\Int\To\type{42})$ (notice the difference in the
annotations).  Therefore,
$\lambda^{\Int{\to}\Int} x.{42}$ and $\lambda^{\Int{\to}\type{42}}
x.{42}$ must have different denotations since they may yield different
results for a type-case on the type \Int\To\type{42}.

The purpose of the rule \Rule{$\to$I$^{\textsc{\tiny (cduce)}}$} is to ensure
that, given any function and any type $t$, either the function has
type $t$ or it has type $\neg t$. This property not only matches the view
of types as sets of values that underpins semantic subtyping, but also
it is necessary to ensure subject reduction, as we explained in Section~\ref{sec:negation} (see~\cite{FCB08} for
details).\footnote{Although by this rule it is possible to deduce infinitely
many distinct types for the same expression, the system still has a notion
of principality, obtained by the introduction of \emph{type schemes}:
see Section 6.12 in \cite{FCB08}.\label{fn:schemes}}

\subsubsection{Type-cases.}\label{typecase}

The second ingredient to obtain the system of Frisch et al.~\cite{FCB08} is the
modification of the syntax for type-case expressions by adding an
explicit binding. Formally, we replace the type-case expression
in~\eqref{slgrammar} by the following production:
\\[1mm]\centerline{\(e\doublecoloneq \ctcase x e t e e\)}\\[1mm]
The  expression $\ctcase x e t {e_1}{e_2}$ binds the value
produced by $e$ to the variable $x$, checks whether this value is of
type $t$, if so it reduces to $e_1$, otherwise it reduces to
$e_2$. Formally:
$$
\begin{array}{rcll}
 \hspace*{40mm}
  \ctcase x v t {e_1}{e_2} & \reduces &e_1\subs x v \hspace*{22mm}&\text{if } v\in t\\
  \ctcase x v t {e_1}{e_2} & \reduces &e_2\subs x v &\text{if }
  v\not\in t
\end{array}
$$
Since functions are explicitly annotated by their types, it is now
possible to define the relation $v\in t$ without using the
type-deduction system.\footnote{We have
  $v\in t\iffdef \exists\, s{\in}\typeof(v)\,.\,s\leq t$ where $\typeof(v)$ is
  inductively defined  as: $\typeof(c)\smalleqdef \{\BasicTypeOfConstant{c}\}$, $\typeof(\lambda^{\bigwedge_{i{\in}I}s_i\to t_i}
  x.e) \smalleqdef$ $\{t\mid t\simeq(\bigwedge_{i{\in}I}s_i\to
  t_i)\land(\bigwedge_{j{\in}J}\lnot(s'_j\to
  t'_j)), t\not\leq\Empty\}$,
  $\typeof((v_1,v_2))\smalleqdef\typeof(v_1)\times\typeof(v_2)$.\label{fn:typeof}}
It is easy to prove
  that for a well typed value $v$ and type $t$ that $v\in t$ is
  decidable (cf. Lemma 6.41 in ~\cite{FCB08})) and that we have $v\in
  t\iff\vdash v:t\iff \not\vdash v:\neg t \iff v\not\in\neg t$ (cf.\ Lemma 6.22 in~\cite{FCB08}).

Type-case expressions are needed to define full-fledged overloaded
functions as opposed to having just ``coherent overloading'' as found
in Forsythe~\cite{forsythe}. Indeed, the rule \Rule{$\to$I$^{\textsc{\tiny (cduce)}}$} we added in
the previous subsection, when it is not coupled with a type-case
expression, allows the system to type only a limited form
of \emph{ad hoc} polymorphism known as coherent overloading~\cite{Pierce92phd,forsythe}.  In
languages with coherent overloading, such as Forsythe or the system by~\citet{BDD95} (or our system
without type-case expressions), it is not
possible to distinguish $(s_1\To t_1)\wedge(s_2\To t_2)$ from
$(s_1{\vee}s_2)\to(t_1{\wedge} t_2)$, in the sense that they both type
exactly the same set of expressions.\footnote{It is not
possible to prove that the two types are equivalent in the system
of~\cite{forsythe} but this can be done for the system
of~\cite{BDD95}.} The equivalence (or indistinguishability) of
the two types above states that it is not possible to have a function
with two distinct behaviors chosen according to the type of the
argument: the behavior is the same for inputs of type $s_1$ or $s_2$
and the intersection of the arrow types is just a way to ``refine''
this behavior for specific cases. In the subtyping relation of
Definition~\ref{def:subtyping}, instead, the relation
\begin{eqnarray}\label{eq:inter}
s_1\vee s_2\to t_1\wedge t_2 & \leq & (s_1\to t_1)\wedge (s_2\to t_2)
\end{eqnarray}
is strict (i.e., the converse does not hold). Therefore, for the step $(\Romanbar{III})$ of~\cite{FCB08} to hold, the
language must provide a $\lambda$-abstraction that is in the
larger type but not in the smaller one, for instance because for some
argument in $s_1$ the  $\lambda$-abstraction returns a result that is in $t_1$ but not in
$t_2$. In general this may require the use of a type-case in the body of
the function, as for $\lambda^{(\Int\to\Bool)\land(\Bool\to\Int)} 
x.\ctcase y x \Int {(y\texttt{==}1)}{42}$ which is a function
that has type $(\Int\To\Bool)\wedge(\Bool\To\Int)$ but not
$\Int{\vee}\Bool\to\Int{\wedge}\Bool$: since
$\Int{\wedge}\Bool=\Empty$, then the second type contains only functions
that diverge on arguments in $\Int\lor\Bool$, which is not the case for
the function above. Thanks to the presence of type-cases we can thus distinguish these two
types by a value; without type-cases, the only functions in
$(\Int\To\Bool)\wedge(\Bool\To\Int)$ we could define would be
those that (disregarding their annotations) could be typed also by
\type{$\Int\lor\Bool\to\Empty$} and, thus, they would diverge on all their
arguments.

The typing rules for type-case expressions are, once
again, taken verbatim from~\cite{FCB08}
\begin{mathpar}
  \Infer[Case]
        {\Gamma\vdash e:t' \quad \Gamma, x:t'{\wedge} t\vdash e_1:s\quad  \Gamma, x:t'{\wedge}\neg t\vdash e_2:s}
        {\Gamma\vdash\ctcase x e t {e_1}{e_2}:s}{}
\qquad
  \Infer[Efq]
        {\quad}
        {\Gamma,x:\Empty\vdash e:t}{}        
\end{mathpar}
and they replace the rules \Rulem{\lor^{(\leq)}}, \Rulem{\in_1^{(\leq)}},
and \Rulem{\in_2^{(\leq)}} of Figure~\ref{fig:canonical} (thus solving
the last problem listed in Section~\ref{sec:problems}). 
The \Rule{Case} rule infers the
type $t'$ of the tested expression $e$, and then infers the types of
the branches by taking into account the outcome of the test. Namely, it
infers the type of $e_1$ under the hypothesis that $x$ is bound to a
value that was produced by $e$ (i.e., of type $t'$) and passed the
test (i.e., of type $t$): that is, a value of type $t{\wedge}t'$; it infers the type of $e_2$ under the
hypothesis that $x$ is bound to a value that was produced by $e$
(i.e., of type $t'$) and did \emph{not} pass the test (i.e., of type
$\neg t$). The reader will surely have recognized that the
rule \Rule{Case} is nothing but a specific instance of the
union-elimination rule \Rulem{\lor} for a type-case expression,
where the expression $e'$ of \Rulem{\lor} is the expression tested by
the type-case and the bind for the variable $x$ is explicitly given by
the syntax of the expression.
Finally, rule \Rule{Efq} (ex falso quodlibet) is used
for when in the rule \Rule{Case} either $t{\wedge} t'$ or $\neg t{\wedge}
t'$ is empty: this means that the corresponding branch cannot be
selected whatever the result of $e$ is and therefore, thanks
to \Rule{Efq} the branch is not typed (it is given any type, in
particular the type of the other branch). For more discussion on
the \Rule{Case} rule and its various implications, the reader can
refer to Section 3.3 of~\cite{FCB08} or Section 3.3 of~\cite{CF05}
(see also the related work section of~\cite{CLLN20}).

\subsubsection{Random choice.}\label{sec:choice}
The very last ingredient to obtain the system of Frisch et al.~\cite{FCB08} is the
addition of expressions for random choice.

Formally we add to the previous grammar the production
$e\doublecoloneq \choice(e,e)$. The semantics of this expression is just a random
choice of one of its arguments:
\begin{align*}
 \choice(e_1,e_2)  & \reduces e_1\\[-1mm]
 \choice(e_1,e_2)  & \reduces e_2
\end{align*}
The need for a choice operator can be evinced by considering the
interpretation of function spaces given in
Definition~\ref{def:interpretation-of-types}. Notice indeed that
functions are
interpreted as finite relations, but we do not require them to be
deterministic, that is, in a finite relation there may be two pairs
with the same first projection but different second projections. More
concretely, if $e_1:t_1$ and $e_2:t_2$ then $\choice(e_1,e_2)$ allows
us to define a value that separates the type $s\to t_1{\vee}\, t_2$ from the
type $(s\To t_1)\vee(s\To t_2)$ (in Definition~\ref{def:interpretation-of-types} the interpretation of the
latter type is strictly contained in the interpretation of the former
type), since $\lambda x.\choice(e_1,e_2)$ is a value in the first
type that it is not in the second type. This is formalized by the
following straightforward typing rule.
\begin{mathpar}
  \Infer[Choice]
        {\Gamma\vdash e_1:t_1\and\Gamma\vdash e_2:t_2}
        {\Gamma\vdash\choice(e_1,e_2):t_1\vee t_2}
        {}
\end{mathpar}
The complete deduction system for
Core CDuce is summarized in Figure~\ref{fig:core}.
\begin{figure}\vspace{-6mm}
\begin{mathpar}
    \Infer[Const]
    { }
    {\Gamma\vdash c:\basic{c}}
    { }
    \and
    \Infer[Var]
  { }
  {\Gamma \vdash x: \Gamma(x)}
  {x\in\dom\Gamma}
\\  
\Infer[$\to$I]
{\forall i\in I\quad \Gamma, x:s_i\vdash e:t_i}
{\Gamma \vdash \lambda^{\wedge_{i{\in}I}s_i\to t_i} x. e: t\wedge t'}
{\begin{array}{l}
t=\wedge_{i{\in}I}(s_i\to t_i)\\[-1mm]
t'=\wedge_{j{\in}J}\neg(s'_j \to t'_j)\\[-1mm]
t\wedge t'\not\simeq\Empty
\end{array}
}
\Infer[$\to$E]
  {
    \Gamma \vdash e_1: {t_1}\quad
    \Gamma \vdash e_2: t_2
  }
  {\Gamma \vdash {e_1}{e_2}: t_1\circ t_2 }
  {\begin{array}{l}t_1\leq\Empty\to\Any\\[-1mm]t_2\leq\dom{t_1}\end{array}
  }
\\
  \Infer[$\times$I]
  {\Gamma \vdash e_1:t_1 \and \Gamma \vdash e_2:t_2}
  {\Gamma \vdash (e_1,e_2):\pair {t_1} {t_2}}
  { }
\and
  \Infer[$\times$E$_i$]
  {\Gamma \vdash e:t}
  {\Gamma \vdash \pi_i e:\bpi t}
  {t\leq\Any\times\Any}
  \\
  \Infer[Case]
        {\Gamma\vdash e:t' \quad \Gamma, x:t{\wedge} t'\vdash e_1:s\quad  \Gamma, x:\neg t{\wedge} t'\vdash e_2:s}
        {\Gamma\vdash\ctcase x e t {e_1}{e_2}:s}{}
\qquad
  \Infer[Efq]
        {\quad}
        {\Gamma,x:\Empty\vdash e:t}{}
\\        
  \Infer[Choice]
        {\Gamma\vdash e_1:t_1\and\Gamma\vdash e_2:t_2}
        {\Gamma\vdash\choice(e_1,e_2):t_1\vee t_2}
        {}
\end{mathpar}
\caption{Algorithmic system for the core calculus of CDuce\label{fig:core}}
\end{figure}
It is
formed by the choice rule plus the rules \Rule{Case} and \Rule{Efq} of
Section~\ref{typecase}, the rule \Rulem{\to\textsc{I}^{(\land)}} of
Section \ref{inferringintersection}, the rules \Rule{$\to$E$^{(\text{\tiny
A})}$} and \Rule{$\times$E$_i^{(\text{\tiny A})}$} (for $i=1,2$) of
Section~\ref{sec:feasibility}, and the rules \Rule{Const}, \Rule{Var},
and \Rule{$\times$I}  of Figure~\ref{fig:declarative}. The resulting
deduction system is algorithmic: it is syntax-directed and formed by
analytic rules (with a small caveat for \Rule{$\to$I}, see Footnote~\ref{fn:schemes}).
The complete definition of the core calculus for CDuce
is summarized for the reader's convenience in
Appendix~\ref{app:cducecore}.  Finally, the resulting system has
enough points to distinguish all types that have different
interpretations.  In particular, the
value interpretation of types for this language, defined as
$\Inter{t}_{\mathcal V}=\{v\mid \varnothing\vdash v:t\}$, induces the same
subtyping relation as the interpretation $\Inter{\cdot}$ of
Definition~\ref{def:interpretation-of-types}: the circle is closed.

As a final remark, it is important to stress that all these types and
typing rules can also be used for languages that do not have all the
constructs described above and that, thus, do not ``close the
circle''. The only drawback is that the type system will not be used
at its full potential and result more restrictive than needed. For instance, if a language does not include
overloaded functions, then we could safely consider the
relation in~\eqref{eq:inter} to be an equivalence rather than a simple
containment and thus safely allow expressions of the type on the
right-hand side to be used where expressions of the type on the
left-hand side are
expected. This is forbidden by the current definition of the subtyping
relation which, for such a language, would thus be more restrictive
than it should be. Although this looks as a minor problem, it is however possible
to modify the definition of the subtyping relation to validate the
equality (and close the circle) as it is shown in Section~5.9 of
Frisch's PhD thesis~\cite{Frisch04phd}.

\subsubsection{Polymorphic language.}\label{sec:polyduce}

Hitherto, the system presented in this section is
mono\-morphic. Although we did not explicitly state it, the
meta-variable $t$ used so far ranged over the monomorphic types of
Definition~\ref{def:types}, which did not include type-variables. In
particular, Theorem 5.2 of \cite{FCB08} that states the equivalence of
type containment in the value interpretation and in the domain $\Domain$ and,
thus, ``closed the circle'', is valid only for monomorphic types: it
is not possible to give a value interpretation to polymorphic types
insofar as there is no value whose type is a type variable, even though
type-variables are \emph{not} empty types. Likewise, the property
that for every value $v$ and type $t$ either $v:t$ or $v:\neg t$ no
longer holds if type variables may occur in types: for instance, $42$
is neither of type $\alpha$ nor of type $\neg\alpha$.

Nevertheless, if we want  the function \code{flatten} in the
introduction to be applicable to any well-typed argument, then we need
to add polymorphic types to CDuce, since the monomorphic version of this function
requires a different implementation of \code{flatten} for each ground
instantiation of the type \type{Tree($\alpha$)\To List($\alpha$)}. A
similar argument holds for the function \code{balance} in
Section~\ref{sec:motivations}.

The extension of CDuce with polymorphic types is as conceptually simple as
its practical implementation is difficult.  To add polymorphism to
CDuce it suffices to take the grammar of the monomorphic expressions of CDuce as
is and use polymorphic types wherever monomorphic ones were used,
with a single exception: since the property that a value $v$ has
either type $t$ or type $\neg t$, no longer holds for every type $t$,
but just for closed types, then we restrict type-case expressions to
test only closed types, that is:\\[1mm]\centerline{$
\begin{array}{lrcl}
  \textbf{Types}& t & \doublecoloneq & b\mid t\times
  t\mid t\to t\mid t\lor t \mid \lnot t\mid \Empty \mid \alpha\\
  \textbf{Test Types}& \tau & \doublecoloneq & b\mid \tau\times
  \tau\mid \tau\to \tau\mid  \tau\lor  \tau \mid \lnot  \tau\mid \Empty\\
  \textbf{Expressions}& e & \doublecoloneq & c\mid
x \mid  \lambda^{\wedge_{i{\in}I}t_i\to t_i} x.e \mid ee\mid \pi_i
e\mid (e,e)\mid \ctcase x e \tau e e\mid \choice(e,e)\\
\end{array}$}\\[1mm]
To type these expressions all we need to do is to add a single typing
rule to account for the fact that if an expression has a polymorphic type, then
it has also all the instances of this type; and since there are
multiple instances of a type, then it has also their intersection. In
other terms we add to  Figure~\ref{fig:core} the following rule
\begin{mathpar}
  \Infer[Inst$^{(\land)}$]
     {\Gamma \vdash e:t}
     {\textstyle\Gamma \vdash e: \bigwedge_{i\in I}t\sigma_i}
     { }
\end{mathpar}
where $I$ is a finite set, $\sigma_i$'s denote type substitutions, that is, finite mappings
from variable to types, and $t\sigma_i$ is their application to a type
$t$. No other modification is necessary.

Thanks to these modifications it is now possible to define in CDuce, say, the
polymorphic identity function $\lambda^{\alpha\to\alpha}x.x$ which is
of type $\alpha\to\alpha$. By an application of
the \Rule{Inst$^{(\land)}$} rule we can deduce for it the type
$(\Int\To\Int)\land(\Bool\To\Bool)$ and, thanks to this deduction it
is possible to infer for the application
$(\lambda^{\alpha\to\alpha}x.x)(\choice(3,\true))$ the type
$\Int\lor\Bool$.\footnote{
As a side note, even if property~\eqref{prop:values} does not hold in
this system (e.g., \texttt{42} is neither of type $\alpha$ nor of type
$\neg\alpha$) this does not hinder the soundness of system since
subject reduction holds for all well typed terms with ground types, that
is for all ground instances of a program. This is enough to
prove the soundness of the system.}

Why then is  the practical implementation of this system so difficult? The reader will have
noticed that since we added \Rule{Inst$^{(\land)}$} to the deduction system in
Figure~\ref{fig:core}, then the system is no longer algorithmic. The new rule
is neither syntax-directed (it applies to a generic expression $e$)
nor analytic (it is not clear how to determine the set of type
substitutions $\{\sigma_i\}_{i\in I}$ to apply in the rule). The
latter, that is determining type substitutions, is the real challenge
for implementing polymorphic CDuce. We will not explain the details
about how to do it: this has needed two distinct papers (part
1~\cite{polyduce1} and part 2~\cite{polyduce2}) to which the reader
can refer for all details. Bottom line, all this complexity is hidden to the programmer: CDuce
does it for her.  Nevertheless, we want to outline some aspects that
can give the reader a flavor of the complexity that underlies this implementation.

A first ingredient that is necessary in all languages with implicit
parametric polymorphism (also known as \emph{prenex}
or \emph{second-order} polymorphism) is type unification: to type the application
of a (polymorphic) function of type $t\to t'$ to an argument of type
$t''$ one has to unify the type of the argument with the domain of the
function, viz., to find a type substitution $\sigma$ such that $t\sigma
= t''\sigma$. However, in a polymorphic language with subtyping this
may not be enough since, in general, we need a type-substitution that makes the type of
the argument \emph{a subtype} of the domain of the function. In other
terms we need to solve the \emph{type tallying
problem}~\cite{polyduce2}, that is, given two types $t$ and $t'$ find
all type substitutions $\sigma$ such that $t\sigma\leq t'\sigma$.  For
instance, consider the types we defined at the end of
Section~\ref{sec:motivations}: if we want to apply a function whose
domain is \type{RBTree($\alpha$)} (a red-black tree with generic
labels) to an argument of
type \type{RTree(\Int)} (a \emph{red} tree with integer labels), then we need the substitution
$\sigma=\{\type{\alpha}\mapsto\Int\}$ since
$(\type{RTree(\Int)})\sigma$ = \type{RTree(\Int)}$\leq(\type{BTree(\Int)\vee
RTree(\Int)})$ = (\type{RBTree(\(\alpha\))})$\sigma$ (notice that the
subtyping relation in the middle is strict). The type tallying
problem is decidable for the polymorphic types of
Section~\ref{sec:types}. In Castagna et al.~\cite[Appendix C]{polyduce2}
we defined an algorithm that returns a set of type-substitutions
that is sound and complete with respect to the tallying problem: every
substitution in the set is a solution, and every solution is an
instance of the substitutions in the set. The reason why the tallying
problem admits as solution a principal set of substitutions---rather
than a single principal substitution---is due to the presence of set-theoretic
types. For instance the problem of finding a substitution $\sigma$
such that
$(\alpha_1\times\alpha_2)\sigma\leq(\beta_1\times\beta_2)\sigma$
admits three incomparable solutions: $(i)$~$\{\alpha_1\mapsto\Empty\}$, $(ii)$~$\{\alpha_2\mapsto\Empty\}$, and
$(iii)$ $\{\alpha_1\mapsto\beta_1,\alpha_2\mapsto\beta_2\}$.

While the capacity of solving the type tallying problem is necessary to type the applications of polymorphic
functions, this capacity alone is not sufficient. The reason is that functions can
be typed not only by instantiating their types, but also by what is commonly
called \emph{expansion}: as stated by rule  \Rule{Inst$^{(\land)}$} an
expression, thus a function, can be typed by any intersection of
instantiations of its type. Consider the function:
\begin{alltt}\color{darkblue}
  let even : \type{(\Int\To\Bool)\,\&\,(\(\alpha\)\setminus\Int\To\(\alpha\)\setminus\Int)} =
    fun x -> (x\(\in\)\Int)\,?\,((x mod 2)==0)\,:\,x
\end{alltt}
or, in Core CDuce syntax, 
$\lambda^{(\Int\To\Bool)\land(\alpha\setminus\Int\To\alpha\setminus\Int)}x. \ctcase
y x \Int {((y\,\textsf{mod\,2})\texttt{==}0)}{y}$. The function is
polymorphic: if applied to an integer it returns a Boolean (i.e., whether
the argument is even or not), otherwise
it returns the argument. Notice that the type of this function is not
weird since it follows
the same pattern as the type of the \code{balance} function we defined in
Section~\ref{sec:motivations}. Next consider the classic \code{map} function:
\begin{alltt}\color{darkblue}\morecompact
  let rec map : \type{(\(\alpha\To\beta\))\To{List}(\(\alpha\))\To{List}(\(\beta\))} =
    fun f l -> match l with                                    \refstepcounter{equation}\text{\color{black}\rm(\theequation)}\label{map}
      |  [] -> []
      |  h::t ->  (f h)::(map f t)
\end{alltt}
and the partial application \code{map\,even} for which polymorphic
CDuce infers the  type
\begin{alltt}\color{darkblue}\morecompact
  map even : \type{( List(\Int) \To List(\Bool) ) \(\land\)
             ( List(\(\gamma\)\setminus\Int) \To (List(\(\gamma\)\setminus\Int)) \(\land\)}                  \refstepcounter{equation}\text{\color{black}\rm(\theequation)}\label{mapeven}
             \type{( List(\(\gamma\lor\)\Int)) \To List((\(\gamma\)\setminus\Int)\(\lor\)\Bool) )  }
\end{alltt}
stating that \code{map\,even} returns a function that when applied to a list
of integers it returns a list of Booleans; when applied to a list that
does not contain any integer, then it returns a list of the same type
(actually, the same list); and when it is applied to a list that may
contain some integers (e.g., a list of reals), then it returns a list
of the same type, without the integers but with some Booleans instead
(in the case of reals, a list with Booleans and with reals that are not
integers). The typing of \code{map\,even} shows that the sole tallying
is not sufficient to obtain such a precise type: the result is
obtained by inferring three different
instantiations\footnote{For \code{map\,even} we need to infer just two
instantiations, namely,
$\{{\alpha}\mapsto{(\gamma{\setminus}\Int)},{\beta}\mapsto{(\gamma{\setminus}\Int)}\}$
and
$\{{\alpha}\mapsto{(\gamma{\vee}\Int)},{\beta}\mapsto{(\gamma{\setminus}\Int)\vee\Bool}\}$. The
type in \eqref{mapeven} is redundant since the first type of the
intersection is an instance (e.g., for $\gamma{=}\Int$) of the
third. We included it just for the sake of the presentation.} of the
type of \code{map}, taking their intersection and tallying it with
the type of \code{even}. This is obtained by the CDuce type-checker by
trying different expansions of the types of the function and of the
argument,  implementing a dove-tail search. For a detailed
explanation the reader can refer to Castagna et al.~\cite{polyduce2}.

The language presented in this subsection is the core of the
polymorphic version of CDuce implemented in the development branch of
the language. It is possible to define in it the
functions \code{flatten}, \code{balance}, \code{map}, and \code{even} of
Sections~\ref{sec:intro}, \ref{sec:motivations} and here above as long as they are
explicitly typed: CDuce requires every function to be annotated
with its type. CDuce also performs occurrence typing, but it requires
the tested expression either to be a variable or to be explicitly
bound to a variable on which the union elimination
rule is applied.

In the next section we show how to get rid of the mandatory
annotations for functions (alas at the expense of inferring
intersection types for them), while in Section~\ref{sec:occtyping} we
present a language in which union elimination is implemented without
any restriction and intersection types for functions are inferred
without need of annotations (alas at the expense of polymorphism).

\textsf{let}$\mathsf{let}$

\subsection{An Implicitly-Typed Polymorphic Language with Set-Theoretic Types}\label{sec:implicit}

The polymorphic language of the previous section requires to
explicitly annotate every function with its type. While for top-level
functions this may be
often advisable and sometime necessary---e.g., for documentation purposes
or for exporting the functions in a library---, it is in general an annoying burden for
the programmer, especially for local functions many of which seldom require
to be documented with a precise type. Besides, determining the right
annotation may be mind-boggling if not impossible, even for very simple
functions: for instance, consider the function
$\lambda x.(\lambda y.(x,y))x$ which clearly has the type
$(\Int\To\Int\Times\Int)\land(\Bool\To\Bool\Times\Bool)$ since it
always returns a pair obtained by duplicating the function's
argument; as an exercise the
reader may try to annotate it without polymorphic types (and without
reading the solution in the footnote), so as to
deduce the above intersection type.\footnote{This is impossible since the type
to give to the local function $\lambda y.(x,y)$ depends on the
hypothesis on $x$: annotating the inner function with the above intersection type
would not work since when $x$ is of type \Int{}, then the local
function does not have type $\Bool\to\Bool\Times\Bool$ and similarly for the
case when $x$ is of type \Bool. The only solution is to annotate both
functions with the type $\alpha\to\alpha\Times\alpha$ and deduce the
intersection type by applying rule \Rule{Inst$^{(\land)}$}. See also
Section~\ref{sec:occtyping} which introduces more expressive
annotations that can type this example.}

To obviate these problems we studied how to type the
implicitly-typed language of grammar \eqref{slgrammar} in
Section~\ref{sec:theoretical}, whose $\lambda$-abstractions are not
annotated, using the polymorphic types of
Section~\ref{sec:polysemsub}. The first results of this study were
presented in~\citet{CPN16} whose system was later greatly
improved and superseded by Petrucciani's Ph.D.\
dissertation~\cite[see Part 2]{Pet19phd} on which the rest of this
subsection is largely based. In order to make type-inference for the
implicitly-typed $\lambda$-abstractions in \eqref{slgrammar} feasible,
we define a system that imposes several restrictions that are absent
from the explicitly-typed polymorphic CDuce we described in
Section~\ref{sec:polyduce}, namely:
\begin{enumerate}[topsep=0pt, partopsep=0pt]
  \item the system implements the so-called \emph{let-polymorphism},
    characteristic of languages of the ML-family or Hindley-Milner
    systems, according to
    which the type system can only instantiate the type of expressions\footnote{In practice, values, see the so called \emph{value restriction}
    suggested by~\citet{wright95}.} that are bound in a let construct. This
    contrasts with the system in Section~\ref{sec:polyduce} where the
    type of every expression can be instantiated.
  \item type-case expressions can test only types  that do not
    have any functional subcomponent other than $\Empty\to\Any$;
  \item the type-system does not infer negated arrow types for functions;
  \item the reconstruction algorithm does not infer intersection types for functions.
\end{enumerate}
Obviously, to implement let-polymorphism we need to extend the
grammar~\eqref{slgrammar} with a \K{let}-expression. The language thus has the
following definition:\\[1mm]
\centerline{$
\begin{array}{lrclr}
  \textbf{Test Types}& \tau & \doublecoloneq & b\mid \tau\times
  \tau\mid \Empty\to \Any\mid  \tau\lor  \tau \mid \lnot  \tau\mid \Empty\\
   \textbf{Expressions} &e &::=& c\mid x\mid\lambda x.e
    \mid e e\mid (e,e)\mid \pi_i e \mid 
    \tcase{e}{\tau}{e}{e} \mid \Let xee
\end{array}
$}\\[1mm]
As anticipated type-cases cannot test arbitrary types, since they use the restricted grammar
for test types $\tau$. There are two restrictions with respect to the types in~\eqref{eq:polytypes}: types
must be ground (as in Section~\ref{sec:polyduce}, $\alpha$ does not
appear in the definition of $\tau$) and the only arrow type that can appear
is $\Empty\To\Any$, that is, the type of all functions. This means that type-cases can
distinguish functions from non-functions but cannot distinguish, say,
the functions that have type $\Int\To\Int$ from those that do not. Type-cases of
this form have the same expressiveness as the type-testing primitives of
dynamic languages like JavaScript and Racket.
The definitions of values and of the reduction semantics rules
given in Section~\ref{sec:theoretical} do not change. To account for the new
\K{let}-expressions we add to these definitions the notion of reduction $\Let x  v e\reduces
e\subs x{v}$ together with the new evaluation context $\Let{x}Ee$.  As
for CDuce, the relation $v\in t$ (actually, $v\in\tau$) used in the
reduction semantics of type-cases can be defined independently from
the type system. Here the definition is even simpler than the one
given in Section~\ref{typecase} (cf. Footnote~\ref{fn:typeof}) since we have $v\in
t\iffdef\typeof(v)\leq t$ and $v\not\in t\iffdef\typeof(v)\leq\neg t$
where $\typeof(c)
\eqdef \BasicTypeOfConstant{c}$, $\typeof(\lambda
x.e) \eqdef$ $\Empty\To\Any$, and
$\typeof((v_1,v_2))\eqdef\typeof(v_1)\times\typeof(v_2)$. Note
that \typeof{} maps every $\lambda$-abstraction to
$\Empty\To\Any$. This approximation is allowed by the restriction on
test types in type-cases.

The most important changes with respect to the theoretical framework
of Section~\ref{sec:theoretical} are to be found in the type-system, defined in
Figure~\ref{fig:implicit} where,
\begin{figure}\vspace{-4mm}
  \begin{mathpar}
    \Infer[Const]
    { }
    {\Gamma\vdash c:\basic{c}}
    { }
    \and
    \Infer[Var]
  { }
  {\Gamma \vdash x: t\subst{\veca}{\vec t}}
  {\Gamma(x)=\forall\vec\alpha.t}
 \\
  \Infer[$\to$I]
    {\Gamma,x:t_1\vdash e:t_2}
    {\Gamma\vdash\lambda x.e: \arrow{t_1}{t_2}}
    { }
  \qquad 
  \Infer[$\to$E]
  {
    \Gamma \vdash e_1: \arrow {t_1}{t_2}\quad
    \Gamma \vdash e_2: t_1
  }
  { \Gamma \vdash {e_1}{e_2}: t_2 }
  { }
  \\
  \Infer[$\times$I]
  {\Gamma \vdash e_1:t_1 \and \Gamma \vdash e_2:t_2}
  {\Gamma \vdash (e_1,e_2):\pair {t_1} {t_2}}
  { }
 \qquad
  \Infer[$\times$E$_1$]
  {\Gamma \vdash e:\pair{t_1}{t_2}}
  {\Gamma \vdash \pi_1 e:t_1}
  { } \quad
  \Infer[$\times$E$_2$]
  {\Gamma \vdash e:\pair{t_1}{t_2}}
  {\Gamma \vdash \pi_2 e:t_2}
  { }
\\
  \Infer[Case]
        {\Gamma\vdash e:t' \quad \text{either }t'\leq\neg t\text{ or
          }\Gamma\vdash e_1:s\quad   \text{either }t'\leq t\text{ or }\Gamma\vdash e_2:s}
        {\Gamma\vdash\tcase e t {e_1}{e_2}:s}{}
\\
  \Infer[Let]
    {\Gamma\vdash e_1:t_1\and\Gamma,x:\forall\vec\alpha.t_1\vdash
      e_2:t}
    {\Gamma\vdash\Let x {e_1}{e_2}:t}
    {\vec\alpha\disjoint\Gamma}
\\
 \Infer[$\wedge$]
  {\Gamma \vdash e:t_1 \quad \Gamma \vdash e:t_2}
  {\Gamma \vdash e: {t_1}\wedge {t_2}}
  { }
  \qquad
  \Infer[$\leq$]
    { \Gamma \vdash e:t\quad t\leq t' }
    { \Gamma \vdash e: t' }
    { }
\end{mathpar}\vspace{-2mm}
\caption{Typing rule for let-polymorphism\label{fig:implicit}}\vspace{-4mm}
\end{figure}

as anticipated, $t$, $t'$, $t_1$, and $t_2$ range over
the polymorphic types defined in Section~\ref{sec:polysemsub} grammar~\eqref{eq:polytypes}.
The type system described in Figure~\ref{fig:implicit} is very similar to a standard Hindley-Milner
system: the differences are just the addition of subtyping and intersection
introduction, as well as a rule for type-cases.
As in Hindley-Milner type systems, we introduce a notion of \emph{type scheme}
separate from that of types. A type scheme, denoted by
$\forall\alpha_1,....,\alpha_n.t$ and abbreviated as
$\forall\vec\alpha.t$, binds the type variables $\alpha_1$,
...,$\alpha_n$ in $t$. We view types as a subset of type schemes identifying
$\forall\vec\alpha.t$ with $t$ if $\vec\alpha$ is empty. Type
environments map variables into type-schemes and we write
$\alpha\disjoint\Gamma$ for the property that $\alpha$ does not occur free in $\Gamma$
(type schemes are considered equivalent modulo $\alpha$-renaming of
the type variables).

If we compare the rules in Figure~\ref{fig:implicit} with those of the
theoretical framework in Figure~\ref{fig:declarative} we will notice
several differences. Foremost, the union elimination rule \Rulem{\lor}
is no longer present. Since this rule played a key role in typing
type-case expressions, then the three rules \Rulem{\Empty}, \Rulem{\in_1},
and \Rulem{\in_2} for type cases are replaced in
Figure~\ref{fig:implicit} by a single rule \Rule{Case} that skips the
typing of a branch when this is not selectable. The only other
difference is in the rule \Rule{Var}. This is classic in
Hindley-Milner system: type environments map variables into
type-schemes and \Rule{Var} instantiates these variables with a set of
types $\vec t$. This rule is coupled with the new rule \Rule{Let} that
infers the type $t_1$ of the argument of the \K{let}-expression and
generalizes this type by binding in the type of $x$ all variables
$\vec\alpha$ that are not free in $\Gamma$.  A final observation,
while by the rule \Rulem{\land} it is possible to deduce intersection
types for functions, it is not possible to deduce negation types (other
than by subsumption). Nevertheless the system is sound, but the proof
needs to deduce these negation types which, since the
$\lambda$-abstractions are not annotated, is not straightforward:
see~\citet[\S3.3]{Pet19phd}.

\subsubsection{Type reconstruction.}\label{sec:reconstruction}

The next problem is to define an algorithm of \emph{type
reconstruction}\footnote{We use this term in the sense
of \citet{Pierce2002}, that is, reconstructing the type information in
an implicitly-typed language. Sometimes the terms \emph{type inference}
or \emph{type assignment system} are equivalently used in the literature.}
for this implicitly-typed language with set-theoretic types (that we
dub Implicit
CDuce, for short).
The algorithm defined by \citet[Chapter 4]{Pet19phd} does not attempt
to infer intersection types:
that would complicate type inference
because we cannot easily know how many types we should infer and intersect
for a given expression, notably for a function.
Therefore, the algorithm of type reconstruction is sound
with respect to the type system in Figure~\ref{fig:implicit} and
complete with respect the same system without the
rule \Rulem{\land}. The algorithm follows a
pattern that is common with Hindley-Milner system and consists in producing
sets of structured constraints, that are then simplified into sets of
subtyping constraints to be solved by the tallying algorithm we hinted at in
Section~\ref{sec:polyduce}. For space reasons we just outline its main characteristics
and some specificities of type
reconstruction for set-theoretic types.

The main difference with respect to
Hindley-Milner systems is that we reduce type reconstruction to
solving sets of constraints that are subtyping constraints (rather
than type equality constraints) that we then solve by using tallying
(rather than unification).

A subtlety of Hindley-Milner type systems is the restriction used in generalization:
to type $ e_2 $ in $ \Let{x}{e_1}{e_2} $,
we assign to $ x $ the type scheme
obtained from the type of $ e_1 $
by quantifying over all type variables
\emph{except those that are free in the type environment}.
This restriction is needed to ensure soundness.  Therefore, whether
the binding for a variable $x$ of a \K{let}-expression is polymorphic or not (and if it is,
which type variables we can instantiate) depends on a comparison of
the type variables that appear syntactically in the type of the bound
expression and in the type environment.  This is problematic with
semantic subtyping: we want to see types up to the equivalence
relation $ \simeq $ (that is, to identify types with the same
set-theoretic interpretation), but two types can be equivalent while
having different type variables in them.  For instance,
$ \alpha \land \Empty $ and $ \alpha \setminus \alpha $ are both
equivalent to $ \Empty $, but $ \alpha $ occurs in them and not in
$ \Empty $. This means that we cannot see type environments up to
equivalence of their types, since the type schemes in them were
generalized according to the variables that syntactically occurred in the
environment. The absence of this property is problematic during
constraint resolution, in particular when
applying type substitutions.
The solution to this problem is to adopt
a technique akin to the \emph{reformulated typing rules}
of~\citet{Dolan2017}. First of all, note that our current type
environments $\Gamma$ bring two different kinds of hypotheses: they
map $\lambda$-abstracted variables into types (that cannot be
instantiated: the variables have monomorphic types)
and \K{let}-abstracted variables into type schemes (which can be
instantiated by replacing the quantified variables by types: the
variables have polymorphic types). As a first step, let us separate
these two kinds of hypotheses and replace our type environment $\Gamma$ by
a monomorphic type environment $M$ for $\lambda$-abstracted variables
and a polymorphic type environment $P$ for \K{let}-abstracted
variables: for clarity we distinguish the latter variables by
superposing a hat on them, such as in $\xh$ and $\yh$. The second step is to
notice that type schemes are obtained by generalizing the type
variables that do not occur in the \emph{monomorphic} part of the
type environment. This observation allows us to get rid of type schemes and generalization
by replacing them with \emph{typing schemes}~\cite{Dolan2017} that
record how a polymorphic type  depends on the current monomorphic
type environment. So while a monomorphic type environment $M$ still
maps a $\lambda$-abstracted variable $x$ into a type $t$, a
polymorphic environment $P$ maps a \K{let}-abstracted variable $\xh$
into a \emph{typing scheme} $\tschemer M t$ where $M$ is a monomorphic
type environment and $t$ a type. For instance, consider the expression
$ \lambda{x}. (\Let\xh{\lambda{y. (x, y)}} e)$. With type schemes we
would choose $ \alpha $ as the type of $ x $, type $ \lambda{y. (x,
y)} $ as $ \beta \to \alpha \times \beta $, and then, to type $ e $, we would
assign to $ \xh $ the type scheme
$ \tscheme{\beta}{\beta \to \alpha \times \beta} $ (we generalized
$\beta$ but not $\alpha$).  In the
reformulated system, in contrast, $ \xh $ is assigned the \emph{typing
scheme} $ \tschemer{x\colon \alpha}{(\beta \to \alpha \times \beta)} $
where all type variables are implicitly quantified (the reconstruction
algorithm will be allowed to instantiate all of them: \emph{cf.},
rule \Rulem{\xh} in Figure~\ref{fig:constrsimpl}) and can be
$\alpha$-renamed: we could equivalently choose for $ \xh $ the typing
scheme $ \tschemer{x\colon \gamma}{(\delta \to \gamma \times \delta)}
$, since we do not care which type variables we use, but only that the
dependency is recorded correctly.  Using this system, the previous difficulties
with generalization do not arise: we can give $\xh$ the type
$\alpha\lor(\gamma\setminus\gamma)$ equivalent to $\alpha$ as long as this does not
capture an implicitly quantified type variable of the typing scheme
(in the present case $\gamma$).

Once we have fixed this point, then reconstruction consists in 
constraint generation and constraint solving. On the lines
of~\citet{Pottier2005}, Petrucciani~\citep[Chapter~4]{Pet19phd} introduces two notions of
constraint.  The first, \emph{type constraints} $ (t_1 \ctsub t_2) $,
constrain a solution (a type substitution $ \sigma $) to satisfy
subtyping between two types (that is, to satisfy $ t_1 \sigma \leq
t_2 \sigma $). In the absence of let-polymorphism,
the type inference problem can be reduced to solving such type constraints,
as done by \citet{Wand1987} for unification.
In our setting, as for type inference for ML,
it would force us to mix constraint generation with constraint solving.
Therefore, we introduce \emph{structured constraints},
which allow us to keep the two phases
of constraint generation and constraint solving separate.
These constraints can mention expression variables
and include binders to introduce new variables.
Structured constraints are closely related to those
in the work of \citet{Pottier2005} on type inference for ML and are
defined as follows:
  \begin{align*}
    C \doublecoloneq {} &
      (t \ctsub t) \mid (x \cssub t) \mid (\xh \cssub t) \mid
      C \land C \mid C \lor C \mid \cex{\veca. C} 
      \mid \cdef x t C \mid
      \clet{\xh} \alpha  C  C
  \end{align*}
Structured constraints include type constraints $(t \ctsub t)$
but also several other forms.
The two forms $ (x \cssub t) $ and $ (\xh \cssub t) $
constrain the type or typing scheme of the variable.
Constraints include conjunction and disjunction.
The existential constraint $ \cex{\veca. C} $ introduces new type
variables (it simplifies freshness conditions).
Finally, the \K{def} and \K{let} constraints
introduce the two forms of expression variables
and are used to describe
the constraints for $\lambda$-abstractions and \K{let}-expressions,
respectively. Their meaning can be evinced from the definition of the \emph{constraint
generation} function $ \constrgen{\cdot:\cdot} $
that, given an expression $ e $ and a type $ t $,
yields a structured constraint $ \constrgen{e: t} $.
This constraint expresses the conditions under which $ e $ has type $ t \sigma $
for some type substitution $ \sigma $. It is defined in Figure~\ref{fig:constgen}
\begin{figure}\vspace{-8mm}
\begingroup
\addtolength{\jot}{-2pt}
\begin{align*}
  \constrgen{\xh: t} = {} &
    (\xh \cssub t)
  \\
  \constrgen{x: t} = {} &
    (x \cssub t)
  \\
  \constrgen{c: t} = {} &
    (\basic{c} \ctsub t)
  \\
  \constrgen{(\lambda{x. e}): t} = {} &
      \cex{ \alpha_1, \alpha_2.
        (\cdef{x}{\alpha_1}{\constrgen{e: \alpha_2}})
        \land (\alpha_1 \To \alpha_2 \ctsub t)}
  \\
  \constrgen{{e_1}{e_2}: t} = {} &
    \cex{ \alpha. \constrgen{e_1: \alpha \to t} \land \constrgen{e_2: t} }
  \\
  \constrgen{(e_1, e_2): t} = {} &
    \cex{ \alpha_1, \alpha_2.
      \constrgen{e_1: \alpha_1} \land \constrgen{e_2: \alpha_2}
      \land (\alpha_1 \Times \alpha_2 \ctsub t) }
  \\
  \constrgen{\pi_i{e}: t} = {} &
    \cex{ \alpha_1, \alpha_2.
      \constrgen{e: \alpha_1 \Times \alpha_2} \land (\alpha_i \ctsub t) }
  \\
  \constrgen{(\tcase{e_0}\tau{e_1}{e_2}): t} = {} &
      \cex{ \alpha.
        \constrgen{e_0: \alpha}
        \land
        \big( (\alpha \ctsub\!\lnot\tau) \lor \constrgen{e_1: t} \big) \land
        \big( (\alpha \ctsub \tau) \lor \constrgen{e_2: t} \big)
      }
  \\
  \constrgen{(\Let{\xh}{e_1}{e_2}): t} = {} &
    \clet{\xh}{\alpha}{\constrgen{e_1: \alpha}}{\constrgen{e_2: t}}
\end{align*}
\endgroup
\caption{Constraint generation}\label{fig:constgen}\vspace{-4mm}
\end{figure}
where $\alpha, \alpha_1, \alpha_2$ do not occur in $t$.
The constraints
for variables and constants are straightforward. To type $\lambda x.e$
with type $t$, the system generates two fresh variables $\alpha_1$
and $\alpha_2$, generates the constraint for $e$ to be of type
$\alpha_2$ under the hypothesis that $x$ is of type $\alpha_1$, and adds
the constraint that $t$ subsumes $\alpha_1\to\alpha_2$. Note
that the constraint generation associates  $\lambda x.e$ to a single
arrow  $\alpha_1\to\alpha_2$ since, as anticipated, it does not attempt
to infer intersection types. The
constraints for applications, pairs and projections are
self-explaining. For type-cases, the system generates the constraint
for the tested expression to be of type $\alpha$, for a fresh $\alpha$, and
then it types the two branches provided that they can be selected,
viz., either the constraint that $e_1$ is of type $t$ must be satisfied
or the first branch cannot be selected since $\alpha$ is a subtype of
$\neg\tau$, and similarly for the second branch. Finally,
for \K{let}-expressions it generates the constraints for $e_1$
remembering that the type $\alpha$ of $e_1$ can be generalized, and
under this hypothesis generates the constraints for $e_2$ to be of type
$t$.

Once the structured constraints are generated for a given expression
they are simplified to obtain a set of \emph{type constraints} whose
solution yields the type of the expression. This is done by an
algorithm that takes as input a polymorphic type environment $P$ and a
structured constraint $C$ and produces a set of type constraints $D$
(which is then solved by tallying), a monomorphic type-environment $M$
(which collects the constraints $x\cssub t$ in $C$) and a set of
variables $\veca$ (that collects the type variables introduced during
the simplification of $C$). This is written as $\constrsimpl PCDM\veca$ and defined by the deduction
rules in Figure~\ref{fig:constrsimpl}. 
\begin{figure*}\vspace{-4mm}
\begin{mathpar}
  \Infer[$\ctsub$]
    { }
    {
      \constrsimpl{\polyenv}{(t_1 \ctsub t_2)}{
        \Set{t_1 \ctsub t_2}}{\emptymonoenv}{\varnothing}
    }
    {}
  \and
  \Infer[$x$]
    { }
    {
      \constrsimpl{\polyenv}{(x \cssub t)}{\varnothing}{(x\colon t)}{\varnothing}
    }
    {}
  \and
  \Infer[$\xh$]
    { }
    {
      \constrsimpl{\polyenv}{(\xh \cssub t)}{
        \Set{t_1 \subst{\veca}{\vecb} \ctsub t}}{
        \monoenv_1 \subst{\veca}{\vecb}}{\vecb}
    }
    {{
      \sideconditionbracebegin
      \begin{array}{l}
        \polyenv(\xh) = \tschemer{\monoenv_1}{t_1} \\
        \veca = \mathsf{tvar}(\tschemer{\monoenv_1}{t_1}) \\
        \vecb \disjoint t
      \end{array}
      \sideconditionbraceend
    }}
  \and
  \Infer[$\land$]
    {
      \constrsimpl{\polyenv}{C_1}{D_1}{\monoenv_1}{\veca_1} \\
      \constrsimpl{\polyenv}{C_2}{D_2}{\monoenv_2}{\veca_2}
    }
    {
      \constrsimpl{\polyenv}{C_1 \land C_2}
        {D_1 \cup D_2}{\monoenv_1 \land \monoenv_2}{\veca_1 \cup \veca_2}
    }
    {{
      \sideconditionbracebegin
      \begin{array}{l}
        \veca_1 \disjoint \veca_2, C_2 \\
        \veca_2 \disjoint C_1
      \end{array}
      \sideconditionbraceend
    }}
  \and
  \Infer[$\lor$]
    {
      \constrsimpl{\polyenv}{C_i}{D}{\monoenv}{\veca}
    }
    {
      \constrsimpl{\polyenv}{C_1 \lor C_2}{D}{\monoenv}{\veca}
    }
    {}
  \and
  \Infer[$\exists$]
    {
      \constrsimpl{\polyenv}{C}{D}{\monoenv}{\veca'}
    }
    {
      \constrsimpl{\polyenv}{\cex{\veca. C}}{D}{\monoenv}{\veca' \cup \veca}
    }
    { \veca' \disjoint \veca }
  \and
  \Infer[Def]
    {
      \constrsimpl{\polyenv}{C}{D}{\monoenv}{\veca}
    }
    {
      \constrsimpl{\polyenv}{\cdef x t C} {D \cup \{t\ctsub \monoenv(x)\}}
        {\monoenv \setminus x}{\veca}
    }
  {{
      \veca \disjoint t
  }}
  \and
  \Infer[Let]
    {
      \constrsimpl{\polyenv}{C_1}{D_1}{\monoenv_1}{\veca_1} \\
      \constrsimpl{
        (\polyenv, \xh\colon \tschemer{\monoenv_1 \sigma_1: \alpha \sigma_1})
       }{C_2}{D_2}{\monoenv_2}{\veca_2}
    }
    {
      \constrsimpl{\polyenv}{\clet{\xh}{\alpha}{C_1}{C_2}}{
        D_2}{\monoenv_1 \sigma_1 \subst{\veca}{\vecb} \land \monoenv_2}{\veca_2 \cup \vecb}
    }
    {{
      \sideconditionbracebegin
      \begin{array}{l}
        \sigma_1 \in \mathsf{tally}(D_1) \\
        \veca = \mathsf{tvar}(\monoenv_1 \sigma_1) \\
        \veca_1 \disjoint \alpha \\
        \vecb \disjoint C_1, \veca_2
      \end{array}
      \sideconditionbraceend
    }}
\end{mathpar}
  \caption{Constraint simplification rules}\vspace{-4mm}
  \label{fig:constrsimpl}
\end{figure*}

A type constraint yields the singleton containing the type constraint
itself (rule \Rulem\ctsub) while a constraint for a
$\lambda$-abstracted variable returns the corresponding monomorphic
environment without any other constraint (rule \Rulem x). The first
interesting rule is the one for the constraint of a \K{let}-abstracted
variable (rule \Rulem\xh), since it performs the instantiation: if the
typing scheme of $\xh$ is $ \tschemer{\monoenv_1}{t_1}$, then the
simplification instantiates \emph{all} the type variables in the typing
scheme (i.e., $\mathsf{tvar}(\tschemer{\monoenv_1}{t_1})$) by some fresh variables $\vecb$ (precisely, some variables
$\vecb$ not
occurring in $t$, noted $\vecb\disjoint t$), and returns the constraint
that the type of $\xh$ so instantiated is subsumed by $t$, the
monomorphic environment $\monoenv_1$ of the constraint so instantiated,
and the set $\vecb$ of fresh variables used for this
instantiation. The rule for conjunction \Rulem{\land} requires all the
constraints to be satisfied and merges the monomorphic environments
(where $\monoenv_1 \land \monoenv_2$ denotes the pointwise
intersection of the environments\footnote{In this rule and in the
rule \Rule{Def} we suppose that $M(x)=\Any$ for $x\not\in\dom M$. Thus
if $x\not\in\dom M$, then $(M\land M')(x) = M'(x)$ and $(t\ctsub
M(x))= (t\ctsub\Any)$.}). Rule \Rulem{\lor} non-deterministically
chooses a constraint, while \Rulem{\exists} ensures that the
constraint uses fresh variables and records them. Rule \Rule{Def}
simplifies the constraint $C$ and adds a new type constraint $t\ctsub
M(x)$ (notice the contravariance, since $t$ is the type hypothesis of
a $\lambda$-abstracted variable it must be \emph{smaller} than the
type $M(x)$ needed to type the body of the function) to remove the binding of $x$ from
$M$, so that the domain of a monomorphic environment obtained by
simplifying a constraint $C$ is always the set of $\lambda$-abstracted
variables free in $C$. Finally, because of let-polymorphism, the
simplification algorithm uses the tallying algorithm internally to
simplify \K{let}-constraints. This is done in \Rule{Let} where
$\mathsf{tally}(D)$ denotes the set of type-substitutions that solve
the set of type constraints $D$. The rule first simplifies the
structured constraint $ C_1 $ and solves the resulting $ D_1 $ using tallying.
Then it non-deterministically chooses a solution $ \sigma_1 $ of $D_1$ to
obtain the typing scheme for $ \xh $, and simplifies $ C_2 $ in the
expanded environment.  The final monomorphic environment returned is
the intersection of $ \monoenv_2 $ and a fresh renaming of
$ \monoenv_1 \sigma_1 $.  In most rules, the side conditions force the
choice of fresh variables.

The type reconstruction algorithm is sound: let $e$ be a program
(i.e., a closed expressions) and $\alpha$ a type variable, if
$\constrsimpl\varnothing{\constrgen{e:\alpha}}D\varnothing\alpha$ and
$\sigma\in\textsf{tally}(D)$, then $\varnothing\vdash e:\alpha\sigma$
is derivable by the system in Figure~\ref{fig:implicit}.  The
algorithm is also complete with respect to the system without the
intersection rule, viz., if a type $t$ can be deduced for an
expression $e$ without using the rule \Rulem{\land}, then
$\constrsimpl\varnothing{\constrgen{e:\alpha}}D\varnothing\alpha$ for
some $D$ and there exists $\sigma\in\textsf{tally}(D)$, such that $t$
is an instance of $\alpha\sigma$.  The system we presented here is a
simplification of the one by~\citet{Pet19phd}. In particular we
glossed over how to handle non-determinism (disjunctive constraints
and multiple solutions of $\mathsf{tally}(D)$ are the two sources of
non-determinism for the algorithm) and how the introduction of fresh
variables during tallying is addressed. The reader can find these
details in~\citet[Chapter~4]{Pet19phd}.

In this system we can now write the \code{map} function defined
in \eqref{map} without specifying its type in the annotation: the type
reconstruction algorithm will deduce it for us. This same
type is deduced for the \code{map} function by any language of the
ML-family. But of course, the use of set-theoretic types goes beyond
what can be reconstructed in ML. We already gave an
example in Section~\ref{sec:motivations} that shows that, thanks to
set-theoretic types, pattern matching can be typed to ensure
exhaustivity. A second example is given by the function \code{f} below
which returns true for the pair of \emph{tags} (akin to
user-defined constants) \code{(\`{\!}A,\`{\!}B)}, and
false for the symmetric pair:
\begin{alltt}\color{darkblue}
  let f = function                   let g = function
    | (`A,`B) -> true                  | `A -> `B              \refstepcounter{equation}\text{\color{black}\rm(\theequation)}\label{eq:implicit}
    | (`B,`A) -> false                 |  x -> x
\end{alltt}
the type returned by the reconstruction algorithm for implicit CDuce
and the one by OCaml (where this kind of tags are
called \emph{polymorphic variants}) are given below.
\begin{center}
\begin{tabular}{rcc}
&Implicit CDuce & OCaml\\[1mm]
\code{f:}
   &\type{(\`{\!}A,\`{\!}B)$\lor$(\`{\!}B,\`{\!}A) $\to$ \Bool}
   & \type{(\`{\!}A$\lor$\`{\!}B\,,\,\`{\!}A$\lor$\`{\!}B) $\to$ \Bool}\\
\code{g:}
   &\hspace*{.5cm}\type{$\forall\alpha$.\`{\!}A$\lor$\`{\!}B$\lor$($\alpha\setminus$(\`{\!}A$\lor$\`{\!}B))
   $\to$ \`{\!}B$\lor\alpha$}\hspace*{1cm}
   & \type{$\forall(\alpha\geq$\`{\!}A$\lor$\`{\!}B$)\,.\,\alpha\to\alpha$}
\end{tabular}
\end{center}
While OCaml states that the function \code{f} can be applied to any
pair of tags \code{\`{\!}A} or \code{\`{\!}B} (but the type-checker warns that
pattern matching may not be exhaustive since it fails for, say,
the pair  \code{(\`{\!}A,\`{\!}A)}) the reconstruction in implicit
CDuce bars out all pairs that would make pattern matching fail. But even
when exhaustivity is not an issue, implicit CDuce can return more
precise types, as the function \code{g} defined in \eqref{eq:implicit} shows.
The type returned by OCaml states that the function \code{g} will return either
\code{\`{\!}A} or \code{\`{\!}B} or any other value that is passed to
the function.\footnote{In OCaml this value can only be another
polymorphic variant.}
The type inferred by implicit CDuce states that the function \code{g} will
return either \code{\`{\!}B} or any other value passed to the function
provided that it is neither \code{\`{\!}A} or \code{\`{\!}B}: contrary
to OCaml, it
correctly detects that \code{g} will never return a
tag \code{\`{\!}A}.

Finally, to understand how the reconstruction algorithm works in the presence
of subtyping, consider the following OCaml code snippet (that does not
involve any pattern matching or fancy data type: just products) that
OCaml fails to type:
\begin{alltt}\color{darkblue}
    fun x -> if (fst x) then (1 + snd x) else x
\end{alltt}
Our reconstruction algorithm deduces for this function the type\\[1mm] \centerline{
\type{(Bool$\times$Int) $\to$ (
    Int\,|\,(Bool$\times$Int) )}
}\\[1mm]
To that end, the constraint generation and simplification systems assign to the function
the type $\alpha\to\beta$ and, after simplification,
generates a set of four constraints:
$\{(\alpha\ctsub\Bool{\times}\Any),(\alpha \ctsub \Any{\times}\Int),(
\Int\ctsub\beta),(\alpha\ctsub\beta)\}$. The first constraint is
generated because \code{fst\,x} is used in a position where a Boolean
is expected; the second comes from the use of \code{snd\,x} in an
integer position; the last two constraints are produced to type the
result of an \code{if\_then\_else} expression (with a supertype of the
types of both branches). To compute the solution of two constraints of
the form $\alpha\ctsub t_1$ and $\alpha\ctsub t_2$, the tallying
algorithm must compute the greatest lower bound of $t_1$ and $t_2$ (or
an approximation thereof); likewise for two constraints of the form
$s_1\ctsub\beta $ and $s_2\ctsub\beta$ the best solution is the least
upper bound of $s_1$ and $s_2$. This yields \code{$\Bool\times\Int$}
for the domain ---i.e., the intersection of the upper bounds for
$\alpha$--- and \code{ (\Int\,|\,(\Bool$\times$\Int))} for the
codomain---i.e., the union of the lower bounds for $\beta$.

This last example further witnesses the interest of having set-theoretic
types exposed to the programmer rather than just as meta-operations
implemented by the type checker. To perform type reconstruction in the
presence of subtyping, one must be able to compute unions and
intersections of types. In some cases, as for the domain in the
example above, the solution of these operations is a type of ML (or of
the language at issue): then the operations can be meta-operators
computed by the type-checker but not exposed to the programmer. In
other cases, as for the codomain in the example, the solution is a
type which might not already exist in the language: therefore, the
only solution to type the expression precisely is to add the
corresponding set-theoretic operations to the types of the language.

\subsubsection{Adding type annotations.}\label{sec:annotations}

The type reconstruction algorithm we just described cannot infer
intersection types for functions. However it is possible to explicitly
annotate functions (actually, any expression) with an intersection
type and the system will \emph{check} whether the function has that
type~\citep[see][Chapter~5]{Pet19phd}. For instance, we can specify for the functions \code{f}
and \code{g} in \eqref{eq:implicit} the following annotations.

\code{f : }\type{((\`{\!}A,\`{\!}B)\,$\to$\,\true)\;$\land$\;((\`{\!}B,\`{\!}A)\,$\to$\,\false)}

\code{g : }\type{$\forall\alpha$.(\`{\!}A$\,\to\,$\`{\!}B)\;$\land$\;(($\alpha\setminus$\`{\!}A)\,$\to$\,($\alpha\setminus$\`{\!}A))}

\noindent
and the type system will accept both of them. With  these    explicit annotations
we almost recover all the expressiveness of the system in
Section~\ref{sec:polyduce} (it just lacks the possibility of testing function
types other than $\Empty\To\Any$). So for instance, the application of the
(implicitly-typed) \code{map} to the function \code{g} explicitly
annotated with the type above, will return in the system
of \citet[Chapter~5]{Pet19phd} exactly the same type as
the type of \code{map\,even} given in \eqref{mapeven} where \type{\`{\!}A} is
replaced for \Int{} and \type{\`{\!}B} is replaced for \Bool.

Adding annotations requires few modifications to the previous
system. First of all we have, of course, to add annotations to our syntax. Here we
present a simplified setting in which annotations are added only
to \K{let}-expressions (see~\citep[Chapter~5]{Pet19phd} for the system
where annotations can be added to any expressions anywhere in a program), which corresponds
to adding the
following production:
$$e ::= \Let{\xh:\forall\veca.t}{e}{e}$$
A \K{let}-abstracted variable can now be annotated with an annotation
$\forall\veca.t$ which specifies the type $t$ to check for the 
expression bound to the variable, as well as the type
variables $\veca$ that are polymorphic in $t$. Like
in the annotation given above to the function \code{g}, we can specify all the
variables occurring in $t$,  but we can also omit some, meaning that they
will be considered
monomorphic.  For instance, $
  \Let{\xh :\forall\alpha. \alpha \To \alpha}{\lambda{x. x}}{{\xh\,}{\K{3}}}
$ is well typed,
because $ \alpha $ is bound in the \K{let}
and can be instantiated in the body of the \K{let}.
Instead, $
  \Let{\xh :\alpha \To \alpha}{\lambda{x. x}}{{\xh\,}{\K{3}}}
$
is ill-typed,
because $ \alpha $ is not bound in the \K{let}
and cannot be instantiated when typing the body $ {\xh\,}{\K{3}}
$---in practice, this means that $\alpha$ is bound in some outer scope
and is polymorphic only outside that scope.

The addition of annotations has as a consequence that now expressions
may have some free \emph{type} variables (e.g.,
$\mathsf{tvar}(\Let{\xh:\forall\veca.t}{e_1}{e_2})=((\mathsf{tvar}(t)\cup\mathsf{tvar}(e_1))\setminus\veca)\cup\mathsf{tvar}(e_2)$)
which are monomorphic and, thus, cannot be instantiated. To cope with
this fact all the constructions we introduced previously in this
section must be enriched by a set $\Delta$ of monomorphic type
variables that cannot be instantiated. So for instance the typing rule
for \K{let}-expressions has $\Delta$ as extra hypothesis and becomes:
\begin{mathpar}
  \Infer[Let]
    {\Gamma;\Delta{\cup}\veca\vdash e_1:t_1\leq t'\and\Gamma,x:\forall\vec\alpha.t_1;\Delta\vdash
      e_2:t}
    {\Gamma;\Delta\vdash\Let{\xh:\forall\veca.t'} {e_1}{e_2}:t}
    {\vec\alpha\disjoint\Gamma,\Delta}
\end{mathpar}
The set $\Delta$ must also be added as a parameter of constraint
generation. Furthermore, constraint generation must be modified to
exploit type annotations. In particular, we want to generate different
constraints for an $ \xh $ variable or a function when we know the
type it should have.  For instance, if a function $\lambda x.e$ is
annotated by an intersection type $\tbigwedge_{i \in I} t_i' \to t_i$,
then we want to generate separate constraints from $ e $ for each
arrow: we break up the intersection into the set \hbox{$ \{ t_i' \to
t_i \mid i \in I \} $} and generate a \K{def}-constraint for each element in
the set.  If the type in the annotation is not syntactically an
intersection of arrows, we can still try to rewrite it to an equivalent
intersection (as a trivial example, we could treat the annotation $
(t' \to t) \lor \Empty $ like $ t' \to t $).  Formally, we need a
function $\decomposed(t)$ that given a type $t$ and a set of
monomorphic variables $\Delta$ returns a set of arrow types such that,
if it is not empty, then it satisfies $(i)$
$t\simeq \tbigwedge_{t'\in\decomposed(t)}t'$; $(ii)$
$\mathsf{tvar}(\tbigwedge_{t'\in\decomposed(t)}t')\subseteq\Delta$;
$(iii)$ for all $t_1\to t_2\in\decomposed(t)$,
$t_1\not\simeq\Empty$. Essentially,  $\decomposed(t)$ decomposes the
type $t$ into an equivalent intersection of arrow types such that these arrows are not of
the form $\Empty\to s$ (which not only would be redundant but also
problematic~\citep[see][Section 5.2.2]{Pet19phd}) and do not contain
monomorphic variables. If this decomposition is not possible
$\decomposed(t)$ returns the empty set.
Once we have a function satisfying these properties
(its definition is not important), then we can modify the constraint
generation function so that it takes into account the monomorphic
variables $\Delta$ and the annotations. The crucial modifications are
the following ones.
\[
\begin{array}{ll}
  \constrgen{\xh: t}^\Delta = {} 
    \tbigwedge_{i\in I}(\xh \cssub t) &
                \text{if }t\simeq \tbigwedge_{i\in I} t_i
  \\[1mm]
   \constrgen{(\lambda{x.e}): t}^\Delta  =
     \cex{ \alpha_1, \alpha_2.
        (\cdef{x}{\alpha_1}{\constrgen{e: \alpha_2}^\Delta})
        \land (\alpha_1 \To \alpha_2 \ctsub t)}
                & \text{if }\decomposed(t)=\varnothing
   \\[1mm]
   \constrgen{(\lambda{x.e}): t}^\Delta  =
      \bigwedge_{t_1\To t_2\in\decomposed(t) } (\cdef{x}{t_1}{\constrgen{e: t_2}^\Delta})&\text{otherwise}
   \\[1mm]
   \constrgen{(\Let{\xh:\forall\veca. t'}{e_1}{e_2}): t}^\Delta = {} 
    \clet{\xh}{\veca,\alpha}{\constrgen{e_1:t'}^{\Delta{\cup}\veca}\land(t'\ctsub\alpha)}{\constrgen{e_2: t}^\Delta} \hspace*{-12mm}
\end{array}
\]
with the conditions $\alpha_1,\alpha_2\disjoint t, e, \Delta$ in the
second line and $\alpha,\veca\disjoint e_1,\Delta$ in the last one. If
a \K{let}-abstracted variable is typed by an intersection, then we
generate the constraints for each type in the intersection separately
and take their conjunction. If the type of a $\lambda$-abstraction can
be decomposed into an intersection of arrows, then we generate the
constraints for each single arrow separately and take their
conjunction; otherwise we proceed as before (in this case  the type
$t$ is likely to be a type variable). For annotated \K{let}-expressions we
generate the constraint that expresses the conditions under which
$e_1$ has the type in the annotation, adding the variables $\veca$ to
those that cannot be instantiated when typing $e_1$.  Note that the
freshness conditions now regard both $\Delta$ and the subexpressions
of the program (since free type variables may occur in them). Similar
modifications must be done on the freshness conditions of the remaining
generation rules.

Finally, the constraint simplification rules must also take into
account the set of monomorphic variables. Thus, for instance, we have
to modify the simplification rule for \K{let}-abstracted variables, so
that the fresh instantiation does not use variables in $\Delta$, and
likewise for \K{let}-expressions:
\begin{mathpar}
  \Infer[$\xh$]
    { }
    {
      \constrsimpl{\polyenv;\Delta}{(\xh \cssub t)}{
        \Set{t_1 \subst{\veca}{\vecb} \ctsub t}}{
        \monoenv_1 \subst{\veca}{\vecb}}{\vecb}
    }
    {{
      \sideconditionbracebegin
      \begin{array}{l}
        \polyenv(\xh) = \tschemer{\monoenv_1}{t_1} \\
        \veca = \mathsf{tvar}(\tschemer{\monoenv_1}{t_1}) \\
        \vecb \disjoint t,\Delta
      \end{array}
      \sideconditionbraceend
    }}
\\
 \Infer[Let]
    {
      \constrsimpl{\polyenv;\Delta\cup\veca}{C_1}{D_1}{\monoenv_1}{\veca_1} \\
      \constrsimpl{
        (\polyenv, \xh\colon \tschemer{\monoenv_1 \sigma_1: \alpha \sigma_1});\Delta
       }{C_2}{D_2}{\monoenv_2}{\veca_2}
    }
    {
      \constrsimpl{\polyenv;\Delta}{\clet{\xh}{\veca,\alpha}{C_1}{C_2}}{
        D_2}{\monoenv_1 \sigma_1 \subst{\vecb}{\vecc} \land \monoenv_2}{\veca_2 \cup \vecb}
    }
    {{
      \sideconditionbracebegin
      \begin{array}{l}
        \sigma_1 \in \mathsf{tally}_{\Delta\cup\veca}(D_1) \\
        \veca\disjoint\Delta,M_1\sigma_1\\
        \vecb = \mathsf{tvar}(\monoenv_1 \sigma_1)\setminus\Delta \\
        \veca_1 \disjoint \alpha \\
        \vecc \disjoint C_1, \veca_2,\Delta
      \end{array}
      \sideconditionbraceend
    }}
\end{mathpar}
notice in the last rule that the appropriate set of monomorphic
variables is now passed to \textsf{tally}, so that it will not
instantiate them to solve the constraints (see~\citet{Pet19phd} for  details).

\subsubsection{A Remark on Occurrence Typing.}
As a final remark, notice that since the type-system in
Figure~\ref{fig:implicit} does not include any form of a union
elimination rule, this system cannot perform occurrence typing.
It is possible to proceed as in
Section~\ref{sec:polyduce} and change the syntax of type-cases so as
they specify a binding for the tested expression obtaining the same
limited form of occurrence typing present in the CDuce language.

\subsection{Occurrence Typing and Reconstruction of Intersections}\label{sec:occtyping}

\beppe{There are some redundancy in this section: it can be cut}

The two systems described in the preceding sections---i.e., the
explicitly-typed version and the implicitly-typed version of CDuce---present two
limitations with respect to the theoretical framework of Section~\ref{sec:theoretical}:
\begin{enumerate}[nosep]
\item \emph{No occurrence typing:} neither system includes the union
elimination rule \Rulem{\lor} of Figure~\ref{fig:declarative} which,
combined with the rules \Rulem{\in_i}, implements
occurrence typing. 
\item \emph{No reconstruction for intersection types:} in both systems
the only way to deduce an intersection type for a function is to
explicitly annotate it with the sought type.
\end{enumerate}
The approach we describe next, proposed by~\citet{CLNL22},
targets precisely these two problems but, for the time being, at the expense of
polymorphism. The work studies whether it is possible to define a
type-inference algorithm for the system of the theoretical framework,
\emph{as is}: we use the language
defined in~\eqref{slgrammar} with the type-system defined by rules in
Figure~\ref{fig:declarative} and the monomorphic types of
Definition~\ref{def:types}. The technical problems to solve in order to define a typing
algorithm for this system are those
evoked in Sections~\ref{sec:feasibility} and \ref{sec:problems},
namely, $(i)$ how to determine the arrows that form the
intersection type of a $\lambda$-abstraction that is not annotated, $(ii)$
how to deduce negation types for a function, $(iii)$ which expressions
and which occurrences of these expressions should the system choose
when it applies
an instance of the rule \Rulem{\lor}, and $(iv)$  how to determine the
union of types into which the system should
split the type of an expression
chosen for \Rulem{\lor}.

We have seen that the previous two systems simply avoided the
technical problems
$(iii)$ and $(iv)$ by excluding the rule \Rulem{\lor} and by typing
type-case expressions with custom rules (possibly adding an explicit
binding to the syntax of the type-cases so as to have a limited form of
occurrence typing). The system described in~\citep{CLNL22}, instead,
follows the opposite approach: it keeps the rule \Rulem{\lor} as is and
introduces specific sound (though, not complete) algorithmic solutions for
these two technical problems. For $(iii)$ it virtually applies the \Rulem{\lor}
rule to \emph{all} subexpressions of a program and for each such
subexpression it takes into account \emph{all} its occurrences in the program. For
$(iv)$ it uses the type-cases and the applications of overloaded
functions that occur in the program to determine the split in union types:
for instance, if $e_1:(\Int\to\Char)\land(\Bool{\lor}\Char\to\Bool)$,
$e_2:\Int{\lor}\Bool$ and there is in the program a type-case of the form
$\tcase{e_1e_2}{\Bool}{\cdots}{\cdots}$, then the system splits
the type of $e_1e_2$ (which is $\Char\vee\Bool$) into two separate
types, $\Char$ and $\Bool$, since they yield different results for the
type-case; but the system will also split the type of $e_2$ into $\Int$
and $\Bool$ since they yield two distinct result types for the
application of the overloaded function $e_1$ (and incidentally for the
type-case at issue). The same solution as for technical problem $(iv)$ is also used
for the technical problem $(i)$, viz., given a function with a certain
domain the system uses the type-cases and the applications of
overloaded functions occurring in the program to determine how to
split the function's domain into a union of types to be checked
separately and, thus, deduce an intersection type for the function:
for instance, if $e_1$ has the same type as above and it is applied to the
parameter $x$ of some function---e.g., $\lambda x\,.\, ...e_1(x)...$---,
then the system will deduce that the domain of the function is (a
subtype of) $\Int\vee\Bool\vee\Char$ and split this domain in two,
that is, it tries to type the body of the function under the
hypothesis $x:\Int$ and under the hypothesis $x:\Bool\vee\Char$ to
deduce for the function a type of the form
$(\Int\to...)\land(\Bool{\vee}\Char\to...)$. Finally, the system
in~\citet{CLNL22} avoids technical problem $(ii)$ in the same way as
implicit CDuce does: negation types
are not inferred, but type-cases cannot test functional types other
than $\Empty\to\Any$. This of course implies that
property \eqref{prop:values} in Sectionn~\ref{sec:negation}---i.e., that every value has a type or its
negation---does not hold. But this does not hinder the property of type
preservation since, as we explained in Section~\ref{sec:negation}, the
presence of the union elimination rule \Rulem{\lor} suffices for it
(even though it holds only for \emph{ad hoc} parallel reductions: cf.~\citet{BDD95} and~\citet{CLNL22}). 

To obtain a type-inference algorithm with the characteristics outlined
above, \citet{CLNL22} proceed in four steps, that we describe next.

First, we  introduce an intermediate language that adds to the
theoretical framework's
original language defined in~\eqref{slgrammar} (henceforth, the \emph{source language}) a ``bind'' construct that factors out common
subexpressions. The type system of this new intermediate language
limits the introduction of intersection and union types in the rules
for typing functions and bind forms, respectively. Typeability in
the source and the intermediate language coincide up to refactoring
with bind.

Second, we introduce a syntactic restriction on terms of the
intermediate language dubbed \emph{maximal-sharing canonical form}
(MSC-form), reminiscent of an aggressive A-normal form~\cite{SF92}. A
MSC-form is essentially a list of bindings from variables to
\emph{atoms}. An atom is either an expression of our source language
in which all subexpressions are variables, or it is a $\lambda$-abstraction whose
body is a MSC-form. These forms are called \emph{maximal-sharing} forms because
they must satisfy the property that there cannot be two distinct bindings for
the same atom. This is a crucial property because it ensures that  every
expression of the  source language $(i)$  is equivalent
to a unique  (modulo the order of bindings) MSC-form and $(ii)$  is well-typed if and only if its MSC-form
is. For instance, consider the expression
\begin{equation}\label{eq:slexp}
\tcase{a_1a_2}{\Int}{(a_2+1)}{((a_1a_2)\code{\,@\,}a_2)}
\end{equation}
where $a_1$ and $a_2$ are generic atoms of type
$t_1=(\Int\To\Int)\wedge(\String\To\String)$ and
$t_2=\Int\vee\String$, respectively, and \code{@} denotes string
concatenation. This expression is well-typed with type
$\Int{\lor}\String$. Its MSC-form will look like the term in
Table~\ref{tab:mscfun}.
\begin{table}\vspace{-2mm}
\begin{minipage}{.41\textwidth}
 \begin{alltt}\setlength{\baselineskip}{10pt}
bind \(x\sb1\) = \(a\sb1\) in
bind \(x\sb2\) = \(a\sb2\) in
bind \(x\sb3\) = \(x\sb1x\sb2\) in
bind \(x\sb4\) = \(x\sb2\,+\,1\) in
bind \(x\sb5\) = \(x\sb3\)@\(x\sb2\) in
bind \(x\sb6\) = \(\tcase{x\sb3}{\Int}{x\sb4}{x\sb5}\)
  in \(x\sb6\)
 \end{alltt}\vspace{-3mm}
 \captionof{table}{Pure MSC-form\hspace*{1cm}}\label{tab:mscfun}
\end{minipage}
\begin{minipage}{.58\textwidth}\setlength{\baselineskip}{10pt}\vspace{-1.6pt}
\begin{alltt}
bind \(x\sb1:\{t\sb1\}\) = \(a\sb1\) in
bind \(x\sb2:\{\Int\,,\,\String\}\) = \(a\sb2\) in
bind \(x\sb3:\{\ann{x\sb2{:}\Int}\Int\,,\,\ann{x\sb2{:}\String}\String\}\)\;=\;\(x\sb1x\sb2\)\;in
bind \(x\sb4:\{\Int\}\) = \(x\sb2\,+\,1\) in
bind \(x\sb5:\{\String\}\) = \(x\sb3\)@\(x\sb2\) in
bind \(x\sb6:\{t\sb2\}\) = \(\tcase{x\sb3}{\Int}{x\sb4}{x\sb5}\)
  in \(x\sb6\)
 \end{alltt}\vspace{-3mm}
 \captionof{table}{Annotated MSC-form}\label{tab:mscfty}
\end{minipage}\vspace{-8mm}
\end{table}
Notice that this term satisfies the maximal sharing property because the
two occurrences of the application $a_1a_2$ in the source language
expression~\eqref{eq:slexp} are bound by the same variable
$x_3$. Essentially MSC-forms are our solution to technical problem $(iii)$ we
evoked at the beginning of this section, namely, which subexpressions and
which occurrences of these subexpressions should the system choose for
applying \Rulem{\lor}: the fact that all proper subexpressions of an atom are variables means that the system chooses \emph{all}
subexpressions, while the maximal sharing property means that the system
chooses \emph{all} occurrences of each subexpression since it replaces
all of them by the same variable.

Third, we prove  that an MSC-form is
well-typed if and only if it is possible to explicitly annotate all
the bindings of variables so that the MSC-form type-checks. The annotations
essentially define how to split the type of the bound variables into a
union of types (when the variable is bound by a $\lambda$ this
corresponds to splitting the type of the $\lambda$-abstraction into an
intersection, when the variable is bound  by a \K{bind} this corresponds
to splitting the argument of the \K{bind}-expression into a union) and the
annotated MSC-form type-checks if the rest of the expression type-checks for
each of the splits specified in its annotations. Table~\ref{tab:mscfty} gives the annotations for
the MSC-form of Table~\ref{tab:mscfun}. The important annotations are those of the
variables $x_2$ and $x_3$. The first states that to type the
expression, the type $\Int{\vee}\String$ of $a_2$ must be split and the
expression must be checked separately for $x_2:\Int$ and
$x_2:\String$. The annotation of $x_3$ states that when $x_2$ has type
$\Int$ then $x_3$ must be assumed to be of type $\Int$ and when $x_2$
has type $\String$ so must have $x_3$.
Since we can effectively transform a source language expression into
its MSC-form, then we have a method to check the well-typedness of an
expression of the source language: transform it into its MSC-form and
infer all the annotations of its variables, if possible. Inferring
the annotations of a MSC-form boils down to deciding how to split the
types of its variables.

Fourth,  we define an algorithm which infers how to split the types of
atoms. It starts from a MSC-form in which all
variables are annotated with the top type $\ANY$ and performs several
passes to refine these annotations.  Each pass has three possible
outcomes: either $(a.)$ the MSC-form type-checks with its current
annotations and the algorithm stops with a success, or $(b.)$ the
MSC-form does not type-check, the pass proposes a new version of the
same MSC-form but
with refined annotations, and a new pass is started, or $(c.)$ the
MSC-form does not check and it is not possible to further refine the
annotations so that the form may become typable, then the algorithm
stops with a failure. The algorithm refines the annotations
differently for variables that are bound by lambdas and by binds. For
the variables in binds the algorithm produces a set of disjoint types
so that their union is the type of the atom in the bind; for lambdas
the algorithm splits the type of the parameter into a set of disjoint types
and rejects the types in this set
for which the function does not type-check, thus determining the domain of
the function.  The very last point that remains to explain is how to
determine the split of a type: as a matter of fact, in general there
are infinitely many different ways to split a type. The split of the
types is driven by the types tested in type-cases and the
operations applied to their components. For instance, the split of the
type of $a_2$ for the variable $x_2$ in Tables~\ref{tab:mscfun}
and~\ref{tab:mscfty} is determined by the test $x_3{\in}\Int$: the
algorithm will propose to split the type $t_3$ of $x_3$ into
$t_3\wedge\Int$ and $t_3\wedge\neg\Int$. Since $t_3$ is
$\Int{\vee}\String$, the split proposed for $x_3$ is actually $\Int$
or $\String$. This split triggers in the subsequent pass the split
for the type of $x_2$ since $x_3$ is defined as $x_1x_2$ and $x_3$ can be
of type $\Int$ only if $x_2$ is of type $\Int$ and it can be of type
$\String$ only if $x_2$ is of type \String. We just got the expected
annotations. Essentially, this fourth step is our solution to the
technical problems $(iv)$ and $(i)$ we
evoked at the beginning of this section, namely, how to split the type
of a subexpression chosen to apply \Rulem{\lor} into a union of types,
and how to split the type of an implicitly-typed function into an
intersection of arrows: we split these types
by analyzing  the type-cases and the overloaded function applications
occurring in the program.

Formally,~\cite{CLNL22} defines the following intermediate language
\begin{equation}
\hspace*{-3mm}  \begin{array}{lrclr}
    \textbf{Intermediate exp} &\eb\hspace*{-1mm} &::=& c\alt x\alt\lambda x.\eb 
    \alt \eb  \eb \alt (\eb ,\eb )\alt \pi_i \eb   \alt
    \tcase{\eb }{\tau}{\eb }{\eb } \alt \bindexp {x} \eb  \eb 
  \end{array}
\end{equation}
with the typing rules given in Figure~\ref{fig:intermediate} (where we
omitted the rules for constants, variables, and pairs 
since they are the same as in Figure~\ref{fig:declarative}).\footnote{%
Notice that we do not define a reduction semantics for the
intermediate language since the sole purpose of the intermediate expressions is to
encode typing derivations. But a call-by-need semantics for the new
bind-expressions would be appropriate~\cite[see Appendix A.6]{CLNL22}.}
\begin{figure}\vspace{-5mm}
\begin{mathpar}
  \Infer[$\to$I\IR]
    {{\small(\forall j\in J)} \\\Gamma,x:t_j\vdashI \eb:s_j}
    {\Gamma\vdashI\lambda
      x.\eb:
      \textstyle\bigwedge_{j\in J}\arrow{t_j}{s_j}}
    { J\neq\varnothing }
    \qquad
 \Infer[$\to$E\IR]
  {
    \Gamma \vdashI \eb_1: {t_1}\quad
    \Gamma \vdashI \eb_2: t_2
  }
  {\Gamma \vdashI {\eb_1}{\eb_2}: t_1\circ t_2 }
  {\begin{array}{l}t_1\leq\Empty\to\Any\\[-1mm]t_2\leq\dom{t_1}\end{array} }
  \vspace{-1.5mm}\\
  \Infer[$\times$E$_1$\IR]
  {\Gamma \vdashI \eb:t\leq(\Any\times\Any)}
  {\Gamma \vdashI \pi_1 \eb:\pi_1(t)}
  { } \ 
  \Infer[$\times$E$_2$\IR]
  {\Gamma \vdashI  \eb:t\leq(\Any\times\Any)}
  {\Gamma \vdashI \pi_2 \eb:\pi_2(t)}
  { }
 \quad
  \Infer[$\Empty$\IR]
  {
    \Gamma \vdashI \eb:\Empty
  }
  {\Gamma\vdashI \tcase {\eb} t {\eb_1}{\eb_2}: \Empty}
  { }
  \vspace{-1.5mm}\\
  \Infer[$\in_1$\IR]
  {
    \Gamma \vdash \eb:t_0\leq t \and \Gamma\vdash \eb_1: t_1
  }
  {\Gamma\vdash \tcase {\eb} t {\eb_1}{\eb_2}: t_1}
  {t_0\not\simeq\Empty}
  \quad
  \Infer[$\in_2$\IR]
  {
    \Gamma \vdashI \eb:t_0\leq\neg t \and \Gamma\vdash \eb_2: t_2
  }
  {\Gamma\vdashI \tcase {\eb} t {\eb_1}{\eb_2}: t_2}
  {t_0\not\simeq\Empty}
  \vspace{-1.5mm}\\
  \Infer[$\vee_1$\IR]
  {\Gamma \vdashI \eb_2:s}
  {
  \Gamma\vdashI\bindexp {x} {\eb_1} {\eb_2} : s
  }
  {x{\not\in}\dom\Gamma}
\ 
  \Infer[$\vee_2$\IR]
  {\Gamma\vdashI \eb_1:\textstyle\bigvee_{j\in J}t_j\quad
    {\scriptscriptstyle(\forall j{\in} J)}\ \Gamma, x\col t_j\vdashI \eb_2:s_j}
  {
  \Gamma\vdashI\bindexp {x} {\eb_1}
                      {\eb_2} :\textstyle \bigvee_{j\in J}s_j
  }
  { J{\neq}\varnothing }\vspace{-3mm}
\end{mathpar}
\caption{Intermediate typing rules\label{fig:intermediate}\vspace{-2mm}}
\end{figure}

A well-typed expression of the intermediate language is typed by 
derivations in which every instance of the \Rulem{\vee} rule (here
declined in two forms)
corresponds to a bind-expression. Any such  derivation corresponds to a
canonical derivation (Figure~\ref{fig:canonical}) for a particular
expression of the source 
language in~\eqref{slgrammar}. This expression can be obtained from the intermediate
language expression  by unfolding 
its bindings. Formally, this is obtained by the \emph{unwinding} operation, noted
$\eras .$ and defined for  binding expressions as  $\eras
{\bindexp{x}{\eb_1}{\eb_2}}\eqdef \eras{\eb_2}\subs{x}{\eras{\eb_1}}$,
as the identity for constants and variables,
and homomorphically for all the other expressions.
It is possible to prove that the problem of typing an expression of our
source language is
equivalent to the problem of finding a typable intermediate expression
whose unwinding is that declarative expression. In other terms, a
declarative expression is typable if and only if we can enrich it with
bindings so that it becomes a typable intermediate expression.

The definition of the intermediate expressions is a step forward in
solving the problem of typing a declarative expression, but it also
brings a new problem, since we now have to decide where to add the bindings
in a declarative expression so as to make it typable in the
intermediate system. We  get
rid of this problem by defining the \emph{maximal sharing
canonical forms} (\emph{MSC-form} for short). The idea is pretty simple, and consists in adding a
new binding for every \emph{distinct} (modulo $\alpha$-conversion)
sub-expressions of a declarative expression. Formally, this
transformation yields a MSC-form:
\begin{definition}[MSC Forms]\label{def:maximalsharing-short}
An intermediate expression $\eb$ is a \emph{maximal sharing canonical form} if it is produced by
the following grammar:
\begin{equation}
  \begin{array}{lrclr}
    \textbf{Atomic expressions} &\ab &::=&   c\alt\lambda x.\kappab\alt (x,x)\alt x x\alt \tcase{x}{\tau}{x}{x}\alt \pi_i x\\
    \textbf{MSC-forms} & \kappab &::=& x\alt \bindexp{x}{\ab}{\kappab}
  \end{array}
\end{equation}
and is  $\alpha$-equivalent to an expression $\kappab$ that
satisfies the following properties:
(1) if  $\bindexp{x_1}{\ab_1}{\kappab_1}$ and
      $\bindexp{x_2}{\ab_2}{\kappab_2}$ are distinct
      sub-expressions of $\kappab$, then $\eras{\ab_1}\not\equiv_\alpha \eras{\ab_2}$;
(2) if $\lambda x.\kappab_1$ is a sub-expression of
      $\kappab$ and
      $\bindexp{y}\ab{\kappab_2}$ a sub-expression of $\kappab_1$, then
      $\textsf{fv}(\ab)\not\subseteq\textsf{fv}(\lambda x.\kappab_1)$;
(3) if $\bindexp{x}\ab{\kappab'}$ is a sub-expression of
      $\kappab$, then $x\in\fv(\kappab')$.
\end{definition}
MSC-forms, ranged over by $\kappab$,  are variables preceded
by a list of bindings of variables to atoms. Atoms are either
$\lambda$-abstractions whose body is a MSC-form or any other
expression in which all proper sub-expressions are variables. Therefore, bindings can appear in a MSC-form
either at top-level or at the beginning of the body of a
function. Definition~\ref{def:maximalsharing-short} ensures that given an expression  $e$ of the source language~\eqref{slgrammar}
there exists a unique (modulo
$\alpha$-conversion and the order of bindings) MSC-form 
whose unwinding is $e$: we denote this MSC-form by $\MSCF e$ and it is
easy to effectively produce it from $e$ (roughly, visit
$e$ bottom up and generate a distinct binding for each distinct
sub-expression). Furthermore,  $e$
is typable if and only if $\MSCF e$ is: we reduced the problem of typing $e$
to the one of typing $\MSCF e$, a form that we can effectively
produce from $e$ and for which we have the
syntax-directed type system of Figure~\ref{fig:intermediate}.

The type system of Figure~\ref{fig:intermediate} is syntax directed,
but it still includes non-analytic rules for functions and
bind-expressions. Thus, the next step consists in adding annotations
to intermediate expressions, so as to make these rules analytic:  we consider expressions of the form  $\lambda
x{:}\anns.\eb$ and $\bindexp{x{:}\anns}{\eb}{\eb}$, where $\anns$ ranges over
annotations of the form  $\{\ann\Gamma t,\ldots{,}\ann\Gamma t\}$. Our
annotations are, thus, finite relations between type
environments and types.
An annotation of the form $x:\{\ann{\Gamma_i}{t_i}\}_{i\in I}$ indicates that under the hypothesis
$\Gamma_i $ the variable $x$ must be supposed to be of type $t_i$.
We write $\{t_1,\ldots{,} t_n\}$ for
$\{\ann\varnothing{t_1},\ldots{,}\ann\varnothing{t_n}\}$ and just $t$
for $\{\ann\varnothing{t}\}$. So for instance we write $\lambda
x{:}t.\eb$ for $\lambda x{:}\{\ann\varnothing{t}\}.\eb$ while, say, 
$\bindexp{x{:}\{t_1,\ldots{,} t_n\}}{\eb_1}{\eb_2}$ stands for
$\bindexp{x{:}\{\ann\varnothing{t_1},\ldots{,}\ann\varnothing{t_n}\}}{\eb_1}{\eb_2}$.

In this system terms encode derivations. Terms with simple annotations such as $\lambda x\col t.\eb$
represent  derivations as they can be found in the simply-typed
$\lambda$-calculus: in other terms, to type the function the system
must look for a type $s$ such that
$\lambda x\col t.\eb$ is of type $t\to s$.
When annotations are sets of
types, such as in $\lambda x\col\{t_1,\ldots,t_n\}.\eb$, then the term
represents a derivation for an intersection type, such as the
derivations that can be found in semantic subtyping calculi: in other
terms, to type the function the system look for a set of types $\{s_1,\ldots,s_n\}$ such that
$\lambda x\col\{t_1,\ldots,t_n\}.\eb$ has type $\textstyle\bigwedge_{i=1}^n\arrow{t_i}{s_i}$.
Finally, the reason why we need the more complex  annotations of the form
$\{\ann{\Gamma_1}{t_1},\ldots,\ann{\Gamma_1}{t_1}\}$ can be shown by an
example. Consider $\lambda x.((\lambda y.(x,y))x)$: in the declarative
system we can deduce for it the type
$(\Int\To\Int{\times}\Int)\wedge(\Bool\To\Bool{\times}\Bool)$. We must find the annotations
$\anns_1$ and $\anns_2$ such that  $\lambda x\col \anns_1.((\lambda y\col \anns_2.(x,y))x)$
has type $(\Int\To\Int{\times}\Int)\wedge(\Bool\To\Bool{\times}\Bool)$. Clearly
$\anns_1=\{\Int,\Bool\}$. However, the typing of the parameter $y$ depends on the typing
of $x$: when $x\col\Int$ then $y$ must have  type $\Int$ (the type of
$y$ must be larger than the one of $x$---the argument it will be bound
to---, but
also smaller than $\Int$ so as to deduce that $\lambda y.(x,y)$ returns
a pair in $\Int{\times}\Int$). Likewise when $x\col\Bool$, then $y$ must be of 
type \Bool, too. Therefore, we use as $\anns_2$ the annotation
$\{\ann{x\col\Int}\Int,\ann{x\col\Bool}\Bool\}$, which precisely states that
when $x\col\Int$, then we must suppose that $y$ (the variable annotated
by $\anns_2$) is of type $\Int$, and
likewise for $\Bool$.
\begin{figure}\vspace{-8mm}
\begin{mathpar}
  \Infer[$\to$I\AR]
    {{\small(\forall j\in J)} \\\Gamma,x:t_j\vdashA \kappa:s_j}
    {\Gamma\vdashA\lambda
      x{:}\{\ann{\Gamma_i}{t_i}\}_{i \in I}.\kappa:
      \textstyle\bigwedge_{j\in J}\arrow{t_j}{s_j}}
    { J=\{i\in I\alt \Gamma\leq\Gamma_i\}\neq\varnothing}
  \vspace{-1.5mm}\\
  \Infer[$\vee_1$\AR]
  {\Gamma \vdashA \kappa:s}
  {
  \Gamma\vdashA\bindexp {x{:}\{\ann{\Gamma_i}{t_i}\}_{i\in I}} {a} {\kappa} : s
  }
  {\begin{array}{c}x\not\in\dom\Gamma\\[-1mm]\{i\in I\alt \Gamma\leq\Gamma_i\}=\varnothing\end{array}}
  \vspace{-1.5mm}\\
  \Infer[$\vee_2$\AR]
  {\Gamma\vdashA a:\textstyle\bigvee_{j\in J}t_j\\{\small(\forall j\in J)} \\ \Gamma, x:t_j\vdashA \kappa:s_j}
  {
  \Gamma\vdashA\bindexp {x{:}\{\ann{\Gamma_i}{t_i}\}_{i\in I}} {a}
                      {\kappa} :\textstyle \bigvee_{j\in J}s_j
  }
  { J=\{i\in I\alt \Gamma\leq\Gamma_i\}\neq\varnothing}\vspace{-2mm}\
\end{mathpar}
\caption{Algorithmic typing rules\label{fig:algorithmic-short}\vspace{-4mm}}
\end{figure}

The type system for annotated terms is given by the rules for
abstractions and binding in Figure~\ref{fig:algorithmic-short} plus
all the other rules of the intermediate type system (specialized for MSC-forms,
i.e., where every subexpression is a variable). The system is
algorithmic since it is syntax-directed and uses only analytic rules.

The main interest of this algorithmic system is that a well-typed annotated
term univocally encodes a type derivation for a MSC-form and,
therefore, it  also encodes a particular canonical derivation for an
expression of the source language. All this gives us  a procedure to
check whether an expression $e$ of the source
language~\eqref{slgrammar} is well typed or not: produce $\MSCF e$ and look for a
way to annotate it so that it becomes a well-typed annotated
expression. If we find such annotations, then $e$ is well
typed. If such annotations do not exist, then $e$ is not well-typed.

The last final step is then to define an algorithm to find whether there
exists a way to annotate an MSC-form to make it well-typed. Different algorithms are
possible. \citet[Section 5]{CLNL22} describe an algorithm that starts
by annotating all bound variables with $\ANY$ and then performs several
passes in which it analyses the type-cases and overloaded
applications of the term to determine how to split the types of the
concerned expression and thus refining the annotations of their
bindings. The reader may refer to \cite[Section 5]{CLNL22} for the
details of this algorithm. Here we just stress that, contrary to the
previous systems presented here, the algorithm is able to deduce the
precise intersection types for the (non-annotated)
functions \code{not\_}, \code{and\_} (both versions), and \code{or\_}
we gave at the end of the introduction, as well as reconstruct the
type $ (\Int \To \Int) \land (\Bool \To \Bool) $ for the function
$ \lambda x. \tcase{x}\Int{(x + 1)}{\lnot x}$ given in
Section~\ref{sec:motivations}.

\subsection{Summary}
In this section we presented three practical variations of the
theoretical language we defined in
Section~\ref{sec:theoretical}. The ultimate goal of our research is to
have a unique language that covers the characteristics of the three of
them. The current implementation of polymorphic CDuce corresponds to
the language we presented in Section~\ref{sec:polyduce} but work is in
progress to merge it with the implicitly-typed language of
Petrucciani's
dissertation~\cite[Part I]{Pet19phd} that we presented in
Section~\ref{sec:implicit}. The idea is to completely move to the
constraint generation and resolution we surveyed in
Section~\ref{sec:reconstruction} and consider the current version of
polymorphic CDuce as the special case of the annotated expressions
presented in Section~\ref{sec:annotations}. This may require to change
the notation of explicitly-typed polymorphic functions, which is the
reason why a polymorphic version of CDuce was not released, yet. The
only problem to solve to obtain a conservative extension of (both
monomorphic and polymorphic) CDuce will then be how to deal in
Petrucciani's system with the type-cases of values with functional
components (since in CDuce you can test whether a function has a given
arrow type).  The final step will be then to integrate the resulting
system with the general usage of the union elimination rules on the
lines of the system we described in Section~\ref{sec:occtyping}. The
first results of this work, still in progress, are presented in~\cite{CLN24}.

\section{Further Features}
\label{sec:features}
In this section we briefly overview few extra features that were
developed in the context of the study of set-theoretic types.

\subsection{Pattern Matching}\label{sec:pm}

Several examples presented in this article use pattern matching. Furthermore,
in Section~\ref{sec:motivations} we cited the precise typing of
pattern matching expressions as one of the main motivations of using set-theoretic
types. However, the languages we formalized in
Section~\ref{sec:languages} do not include pattern matching
expressions:  they just have type-case expressions which, in
their binding variant of Section~\ref{typecase}, may be considered a very simplistic
version of pattern matching.  Here, we outline how  full-fledged
pattern matching can be added to the languages presented in Section~\ref{sec:languages}.
Similar formalizations of pattern matching for set-theoretic type systems
have been described by
\citet[Chapter 6]{Frisch04phd}, \citet[Appendix E]{polyduce2}, and \citet{CPN16}.

For simplicity, we only consider two-branch pattern matching.
We extend the syntax with the \K{match} construct and with patterns:
\begin{align*}
  e & \doublecoloneq \dots \mid \match{e}{p}e{p}e
  &
  p & \doublecoloneq
    \tau \mid x
    \mid (p, p) \mid \patand{p}{p} \mid \pator{p}{p}
  \: ,
\end{align*}
where $\tau$ are the test types defined in Section~\ref{sec:languages}
(ground types for CDuce and its explicit polymorphic variant, ground
non-functional types for implicitly-typed CDuce and the variant for
occurrence typing) and with some restrictions on the variables that can appear in patterns:
in $ (p_1, p_2) $ and $ \patand{p_1}{p_2} $,
$ p_1 $ and $ p_2 $ must have distinct variables;
in $ \pator{p_1}{p_2} $,
$ p_1 $ and $ p_2 $ must have the same variables.

A more familiar syntax for patterns is $
  p \doublecoloneq
    \patwild \mid c \mid x
    \mid (p, p) \mid \patas{p}{x} \mid \pator{p}{p}
$,
with wild-cards and constants instead of $ \tau $ types
and with as-patterns \enquote{$ \patas{p}{x} $}
(in OCaml syntax; $ x\textsf{@}p $ in Haskell) instead of conjunction.
We can encode
$ \patwild $ and $ c $ as $ \Any$ and $ \basic{c} $
(both are in the grammar for $ \tau $),
while \enquote{$ \patas{p}{x} $} is $ \patand{p}{x} $,
as will soon be clear.

\begin{figure}\vspace{-8mm}
  \begin{align*}
    v \matches x & = \subs{x}{v} \\
    v \matches \tau & = \{\,\} & \text{if $v\in \tau $} \\
    v \matches (p_1, p_2) & =
        \valsubst_1 \cup \valsubst_2
          & \text{if $ v = (v_1, v_2) $,
              $ v_1 \matches p_1 = \valsubst_1 $,
              and $ v_2 \matches p_2 = \valsubst_2 $} \\
    v \matches \patand{p_1}{p_2} & =
        \valsubst_1 \cup \valsubst_2
          & \text{if $ v \matches p_1 = \valsubst_1 $
              and $ v \matches p_2 = \valsubst_2 $} \\
    v \matches \pator{p_1}{p_2} & =
         v \matches p_1 & \text{if $ v \matches p_1 \neq \matchfail $} \\
    v \matches \pator{p_1}{p_2} & =
         v \matches p_2 & \text{if $ v \matches p_1 = \matchfail $}\\
    v \matches p & = \matchfail & \text{otherwise}
  \end{align*}\vspace{-6mm}
  \caption{Semantics of patterns}\vspace{-4mm}
  \label{fig:semantics-of-patterns}
\end{figure}

To describe the semantics of pattern matching,
we define a function $ (\cdot) \matches (\cdot) $
that, given a value $ v $ and a pattern $ p $,
yields a result $ v \matches p $ which is either $ \matchfail $
or a substitution $ \valsubst $
mapping the variables in $ p $ to values (subterms of $ v $).
This function is defined in Figure~\ref{fig:semantics-of-patterns}.
Then, we augment the reduction rules with
\begin{alignat*}{2}
  (\match v{p_1}{e_1}{p_2}{e_2})
  & \: \reduces \: e_1 \valsubst
  & \qquad & \text{if } v \matches p_1 = \valsubst
  \\
  (\match v{p_1}{e_1}{p_2}{e_2})
  & \: \reduces \: e_2 \valsubst
  & \qquad &
  \text{if } v \matches p_1 = \matchfail \text{ and } v \matches p_2 = \valsubst
\end{alignat*}
and add $ \match Epepe $
to the grammar of evaluation contexts.

Given each pattern $ p $,
we can define a type $ \patacc{p} $
that describes exactly the values that match the pattern:
\begin{align*}
  \patacc{ \tau } & = \tau
  &
  \patacc{ x } & = \Any
  \\
  \patacc{ (p_1, p_2) } & = \patacc{p_1} \times \patacc{p_2}
  &
  \patacc{ \patand{p_1}{p_2} } & = \patacc{p_1} \land \patacc{p_2}
  &
  \patacc{ \pator{p_1}{p_2} } & = \patacc{p_1} \lor \patacc{p_2}
\end{align*}
It can be shown that,
for every well-typed value $ v $ and every pattern $ p $,
we have $ v \matches p \neq \matchfail $
if and only if $ \varnothing \vdash v: \patacc{p} $.
This allows us to formalize purely at the level of types
the exhaustiveness and redundancy checks
that are often performed on pattern matching.
The typing rule for \K{match} is the following.
\[
  \hspace*{1cm}\Infer
  {
    \Gamma \vdash e_0: t_0
    \\\\
    \text{either $ t_0 \leq \lnot \patacc{p_1} $ or }
    {\Gamma, (t_0 \land \patacc{p_1}) / p_1 \vdash e_1: t}
    \\
    \text{either $ t_0 \leq \patacc{p_1} $ or }
    \Gamma, (t_0 \setminus \patacc{p_1}) / p_2 \vdash e_2: t
  }
  {
    \Gamma \vdash \match{e_0}{ p_1}{e_1}{p_2}{e_2}: t
  }
  {
      t_0 \leq \patacc{p_1} \lor \patacc{p_2} 
  }
\]
The \enquote{either \dots{} or \dots} conditions
have the same purpose as for type-case rule \Rule{Case} in
Figure~\ref{fig:implicit}, namely, they skip the typing of branches that
cannot be selected.
The side condition $ t_0 \leq \patacc{p_1} \lor \patacc{p_2} $
ensures that matching is exhaustive:
any value produced by $ e_0 $ has type $ t_0 $
and therefore matches either $ p_1 $ or $ p_2 $.
When a branch is selectable it is typed under the hypothesis $\Gamma$
extended with a type environment produced by applying the operator ${t}/{p }$, which
given a type $t$ and a pattern $p$ with $t\leq\patacc p$ produces
the type environment that can be assumed for the variables in $ p $
when a value of type $ t $ is matched against $ p $ and matching
succeeds. Thus $e_1$ is typed under the hypothesis obtained supposing
that $p_1$ was matched against a value produced by $e_0$ (i.e., in
$t_0$) and accepted by $p_1$ (i.e., in $\patacc{p_1}$), while the
hypotheses for $e_2$ are obtained supposing that $p_2$ was matched against a value produced by $e_0$ (i.e., in
$t_0$) and \emph{not} accepted by $p_1$ (i.e., in $\lnot\patacc{p_1}$:
remind that $t_0\setminus\patacc{p_1}=t_0\land\neg\patacc{p_1}$).
The operator is defined as follows
\begin{align*}
  t/\tau & =\varnothing \\
  t/x & =x:t\\
  t/(p_1,p_2) & = (\bpl t/p_1){\cup}(\bpr t/p_2)\\
  t/\patand{p_1}{p_2}&=(t/p_1){\cup}(t/p_2)\\
  t/\pator{p_1}{p_2} &=((t\land\patacc{p_1})/p_1){\cup}((t\setminus\patacc{p_1})/p_2)
\end{align*}
and satisfies the property that for every $ t $, $ p $, and $ v $,
if $ \varnothing\vdash v: t $ and $ v \matches p = \valsubst $,
then, for every variable $ x $ in $ p $, the judgment
$ \varnothing\vdash x \valsubst: (t/p)(x) $ holds.

Finally, we said that the condition $t_0 \leq \patacc{p_1} \lor
\patacc{p_2} $ in the typing rule for \K{match}-expressions ensures the exhaustiveness of pattern matching, but
what about redundancy? When  $t_0 \leq \neg\patacc{p_1}$ should not
the system return a warning that $e_1$ cannot be selected and likewise
for $e_2$ when  $t_0 \leq \patacc{p_1}$? In general it should not, since skipping
the typing of some branches is necessary for inferring intersection types for overloaded functions. For instance,
consider again the function we defined in
Section~\ref{sec:motivations} to explain overloading, that is,
$\lambda^{(\Int\To\Int)\land(\Bool\To\Bool)}  x. \tcase{x}\Int{(x +
  1)}{\lnot x}$ whose definition with pattern matching would be:\\[1mm]
\centerline{
$\lambda^{(\Int\To\Int)\land(\Bool\To\Bool)}  x\,.\, \match{x}\Int{(x +
    1)}{\Any}{\lnot x}$}
\\[1mm]
When typing the body of the function under the hypothesis $x:\Int$ it
is important not to check the type of the second branch (since $\neg x$ would be ill
typed) and under the hypothesis $x:\Bool$ it is important not to check
the type of the first branch (since $x+1$ would be ill typed). However, neither of
the branches is redundant because each of them is type-checked \emph{at least
once}. Redundancy corresponds to branches that are never type-checked,
as, for instance, the
second branch in the following definition\\[1mm]
\centerline{
$\lambda^{(\Int\To\Int)\land(\Bool\To\Bool)}
  x\,.\, \match{x}{(\Int{\lor}\Bool)}{x}{\Any}{\lnot x}$}
\\[1mm]
which is skipped both under the hypothesis $x:\Int$ and under the
hypothesis $x:\Bool$ and, therefore, must be fingered as redundant
(but the function is well-typed). In conclusion, as it is the case
for exhaustiveness, redundancy of pattern matching, too, can be
characterized in terms of a type system that includes
set-theoretic types.



\subsection{Gradual Typing}\label{sec:gradual}

Gradual typing is an approach proposed by~\citet{siek2006gradual} to
combine the safety guarantees of static typing with the programming
flexibility of dynamic typing. The idea is to introduce an \emph{unknown} 
(or \emph{dynamic}) type, denoted $\dyn$, used to inform the compiler that
some static type-checking can be omitted, at the cost of some additional
runtime checks. The use of both static typing and dynamic typing in a same
program creates a boundary between the two, where the compiler automatically
adds---often costly~\cite{takikawa2016sound}---dynamic type-checks to ensure 
that a value crossing the barrier is correctly typed.

Occurrence typing---that we discussed in Sections~\ref{sec:motivations}
and~\ref{sec:occtyping}---and gradual typing often have common use
cases. For instance the example we gave for occurrence typing
in Section~\ref{sec:motivations}, $ \lambda x. \tcase{x}\Int{(x +
  1)}{\lnot x}$, can also be typed by gradual typing as follows:
\begin{equation}\label{eq:gradual}
\lambda x:\textcolor{red}{\dyn}. \tcase{x}\Int{(x +
    1)}{\lnot x}
\end{equation}
``Standard'' or ``safe'' gradual typing inserts two dynamic checks
since it compiles the code above into $\lambda  x:{\dyn}. \tcase{x}\Int{(\textcolor{red}{\Cast{\type{\color{red}Int}}x} + 1)}{\lnot(\textcolor{red}{\Cast{\type{\color{red}Bool}}x})}$,
where {\Cast{$t$}{$e$}} is a type-cast that dynamically checks whether the value returned by $e$ has type $t$.\footnote{Intuitively, \code{\Cast{$t$}{$e$}} is
  syntactic sugar for, say, in JavaScript \code{(typeof($e$)==="$t$")\,?\,$e$\,:\,(throw "Type
    error")}. Not exactly though, since to implement compilation
  \emph{à la} sound gradual typing it is necessary to use casts on
  function types that need special handling.}
The type deduced for the function in \eqref{eq:gradual} is
\type{$\dyn\to\Int{\lor}\Bool$} meaning that it is a function that can
be applied to any argument (which may have its type dynamically
checked if needs to be)
and will return either an integer or a Boolean (or a cast exception if a
dynamic check fails). This
type is not very precise since it allows the function to be applied to
any argument, even if we already know that  it will fail with a cast
exception for arguments that are neither integers nor Booleans. Whence the
interest of having full-fledged set-theoretic types thanks to which
the programmer can shrink the domain of the function as follows:
\begin{equation}\label{eq:grad2}
\lambda x:\textcolor{red}{(\dyn\wedge(\Int{\lor}\Bool))}. \tcase{x}\Int{(x +
    1)}{\lnot x}
\end{equation}
Intuitively, this annotation means that the function above accepts for
$x$ a value of any type (which is indicated by \type{$\dyn$}), as long as
this value is also either an \Int{} or a \Bool. So the type-casts will
never fail. This was the initial motivation of our study of
integrating gradual and set-theoretic types~\cite{CL17,CLPS19}.  Of
course, the example above does not need gradual typing if the systems
provides occurence typing: this provides a better solution since, as
we showed before, it returns a more precise
type (\type{($\Int\To\Int$)$\wedge$(\Bool\To\Bool)}) and avoids the
insertion of superfluous run-time checks. But there are some cases in
which the occurrence typing analysis may fail to type-check, since either
they are too complex
or they are not covered by the formalism (e.g., when polymorphic
types are needed, which are not captured by the system presented in
Section~\ref{sec:occtyping}). In those cases gradual typing is a viable
alternative to no typing at all.  In a sense, occurrence typing is a
discipline designed to push forward the frontiers beyond which gradual
typing is necessary, thus reducing the amount of runtime checks needed
(see~\citet[Section 3.3]{CLNL22} for more a detailed treatment).

\vfill

But the interest of integrating gradual and set-theoretic types is not
limited to having a more precise typing of some applications like the
ones above. The main interest of this integration is that the
introduction of set-theoretic types allows us to give a semantic
foundation to gradual typing, and explain the type-theory of
gradual types only in terms of non-gradual ones. The core of the
type-systems for gradually-typed expressions such
as~\eqref{eq:gradual} or~\eqref{eq:grad2} is the definition of a
\emph{precision} relation $\mater$ on
types~\cite{Siek:2008kq,Garcia:2013fk} (called \emph{naive subtyping}
in~\cite{wadler2009well}). In the cited works the definition of this
relation is very simple: given two gradual types $\tau_1$, $\tau_2$,
the type
$\tau_1$ is less precise than $\tau_2$, written $\tau_1\mater\tau_2$,
if and only if $\tau_2$ is obtained from $\tau_1$ by replacing some
occurrences of $\dyn$ by some types (we use $\tau$ to range over
\emph{gradual types} to distinguish them from types in which
\type{$\dyn$} does not occur and that are called \emph{static
types}). So for instance
$\dyn\To\dyn\Times\dyn\mater\Int\To\dyn\Times\Bool\mater\Int\To\Int\Times\Bool$. Intuitively,
the precision relation indicates in which types a gradual type may
``materialize'' (i.e., turn out to be) at runtime. So the type $\dyn\To\dyn$ of a function
materializes into $\Int\To\dyn$ if this function happens at runtime
to be applied to an integer.
\ifcompact
\citet{CLPS19} demonstrate that to extend
with gradual typing an existing statically-typed language all is needed is $(i)$ to add
$\dyn$ to the types (as a new basic type), $(ii)$ define\begin{floatingfigure}[r]{3.5cm}
\hspace*{-3.0cm}%
\begin{minipage}{8.5cm}\vspace{-1.6mm}
\begin{mathpar}
      \Infer[$\mater$]
      {
        \Gamma\vdash e:\tau \\ \tau \mater \tau' 
       }{
       \Gamma\vdash e : \tau'
       }{}
\smallskip
\end{mathpar}
\end{minipage}
\end{floatingfigure}\noindent
the precision relation $\mater$ (and, if used,
the subtyping relation $\leq$) on the new types, and $(iii)$ add the
subsumption-like materialization rule here on the right to the
existing typing rules.
\else
\citet{CLPS19} demonstrate that to extend
with gradual typing an existing statically-typed language all is needed is $(i)$ to add
$\dyn$ to the types (as a new basic type), $(ii)$ define
the precision relation $\mater$ (and, if used,
the subtyping relation $\leq$) on the new types, and $(iii)$ add the following
subsumption-like materialization rule to the
existing typing rules:
\begin{mathpar}
      \Infer[$\mater$]
      {
        \Gamma\vdash e:\tau \\ \tau \mater \tau' 
       }{
       \Gamma\vdash e : \tau'
       }{}
\end{mathpar}
\fi
Of course, this does not immediately yield an
effective implementation (one has to find a type-inference algorithm,
define the language with the explicit casts and the compilation of
well-typed terms into it, compilation that, roughly, must insert a
dynamic type-cast wherever the typing algorithm had to use a
materialization rule) but conceptually this is all is needed. The
difficult point is to define the precision relation (but also to
extend an existing subtyping relation to gradual types). The simple
definition of the precision relation we gave above is 
syntax based and, as such, it shows its limits as soon as we add type
connectives. In semantic subtyping equivalence between types plays a
central role: two types are equivalent if and only if they represent
the same set of values, and this makes them to behave identically in
every context. So one would expect equivalent types to materialize in
the same set of types, but this is not the case: consider for example
the types \type{$\Int{\lor}\dyn$} and \type{$\dyn\lor\Int$}; although
they are equivalent, the former materializes into
\type{$\Int{\lor}\Bool$} while the latter does not. The latter does,
however, materialize into \type{$\Bool{\lor}\Int$} which is equivalent
to \type{$\Int{\lor}\Bool$}. A similar reasoning can be done for
\type{$\dyn$} and \type{$\lnot\dyn$} which intuitively behave in
exactly the same way. We thus need a more robust, syntax independent
characterization of the precision relation.

\vfill

In his
Ph.D.\ dissertation, \citet{Lanvin21phd} showed that this
characterization can be given just in terms of static types (i.e., the
types without any occurrence of $\dyn$ in them). Take as static types
either the set-theoretic types of Definition~\ref{def:types} or their
polymorphic extension given by grammar~\eqref{eq:polytypes}, together with
their respective subtyping relations. To obtain gradual types add to the grammars of these
types the production $t::=\dyn$ (still, we use $\tau$ to range over
gradual types and reserve $t$ for static types). Given a gradual type $\tau$, the set of all static
types it materializes to forms a complete lattice, with a maximum and
a minimum static type that we denote by $\tmax\tau$ and $\tmin\tau$,
respectively. So we have that for all $\tau$, if $\tau\mater t$, then
$\tmin\tau\leq t\leq\tmax\tau$, where $\leq$ is the subtyping relation
for the static types. It is very easy to derive the
materialization extrema $\tmax\tau$
and $\tmin\tau$ from $\tau$: you get $\tmax\tau$ by replacing in
$\tau$ every covariant occurrence of $\dyn$ by $\Any$ and every
contravariant occurrence of $\dyn$ by $\Empty$; $\tmin\tau$ is
obtained in the same way, by replacing in $\tau$ every covariant
occurrence of $\dyn$ by $\Empty$ and every contravariant one by
$\Any$. The definition of the minimal and maximal static materializations
together with the subtyping relation on static types is all is needed
to define the precision relation \emph{and} the subtyping relation on the
newly defined gradual types. \citet{Lanvin21phd} shows that it is possible to define
the precision relation $\mater$ and the subtyping relation $\dot\leq$ on
gradual types as follows:
\begin{eqnarray}
  \tau_1\mater\tau_2 & \quad\iffdef\quad &
  \tmin{\tau_1}\leq\tmin{\tau_2}\textsf{ and }\tmax{\tau_2}\leq\tmax{\tau_1}\label{eq:mater}\\
  \tau_1\;\dot\leq\;\tau_2 & \iffdef &
  \tmin{\tau_1}\leq\tmin{\tau_2}\textsf{ and }\tmax{\tau_1}\leq\tmax{\tau_2}\label{eq:gsub}
\end{eqnarray}
where $\leq$ denotes the subtyping relation given on static
types---e.g., the two subtyping relations
induced by interpretations of
types given in Sections~\ref{sec:semsub} and
\ref{sec:polysemsub}---.\footnote{Strictly speaking, it is necessary
slightly to modify these interpretations so that all the types of the
form $\Empty\To t$ are not all equivalent: see~\citet[Section~6.1.2]{Lanvin21phd}.}

Equation~\eqref{eq:mater} conveys a strong message: any gradual type
can be seen as an \emph{interval} of \emph{possible types}, where $\dyn$
denotes the interval of all types, and a type $\tau$ denotes the interval
ranging from $\tmin{\tau}$ to $\tmax{\tau}$ (or, more precisely, the sub-lattice of the types included
between the two). Semantic materialization then allows us
to reduce this interval, by going to any type $\tau'$ such that $\tmin{\tau} \leq \tmin{\tau'}$
and $\tmax{\tau'} \leq \tmax{\tau}$, possibly until reaching a static type
(that is, a type $\tau$ such that $\tmin{\tau} = \tmax{\tau}$).

Equation~\eqref{eq:gsub} extends this interval interpretation to
the subtyping relation of gradual types, stating that  a type $\tau_1$ is a subtype
of $\tau_2$ if the interval denoted by $\tau_1$ only contains subtypes of
elements of the interval denoted by $\tau_2$.

Notice that these two definitions also provide an effective way to
decide precision and subtyping for gradual types: generate the gradual
extrema and check on them the subtyping relations for static types
according to~\eqref{eq:mater} or~\eqref{eq:gsub}.

Lanvin justifies these definitions by giving a semantic interpretation
of all gradual types (i.e., not just of the static ones) and proving all the
needed properties. In particular, he proves a series of properties
that show the robustness of the relations defined in~\eqref{eq:mater}
and~\eqref{eq:gsub}. First, for all \emph{gradual} types $\tau$ and
$\tau'$ such that $\tau\mater\tau'$ we have
$\tmin\tau\mathrel{\dot\leq}\tau'\mathrel{\dot\leq}\tmax\tau$, that
is, all the materializations of a gradual type form a complete
sub-lattice (not just the static materializations). More surprisingly,
for every gradual type $\tau$ we have
$\tau\mathrel{\dot\simeq}\tmin\tau\lor(\dyn\land\tmax\tau)$, where
$\dot\simeq$ is the symmetric closure of the gradual subtyping
relation $\dot\leq$.  According to this last property, every gradual
type $\tau$ is equivalent to the $\dyn$ type as long as we bound it
with the two extrema $\tmin{\tau}$ and $\tmax{\tau}$, thus
strengthening the intuition of gradual types as intervals of
static types. Therefore, every gradual type can be represented by a
pair of static types, and to add gradual typing to a system, it
suffices to augment the types with a single constant $\dyn$ that only needs to
appear at top level, that is, under neither an arrow nor a product.  This
characterization can then be used to define the gradual counterparts
$\gdomType{.}$,  $\gresType{}{}$, and $\gprojType{i}{.}$ of the type
operators $\dom{.}$, $\resType{}{}$, and $\bpi{.}$ we defined at the end of
Section~\ref{sec:feasibility}, thus providing a further proof of the
robustness of the definitions in~\eqref{eq:mater} and~\eqref{eq:gsub}.
So we have:
\begin{eqnarray*}
      \gdomType{\tau} &\eqdef& \dom{\tmax{\tau}} \lor (\dyn \land \dom{\tmin{\tau}})\\
      \gresType{\tau}{\tau'} &\eqdef& (\resType{\tmin{\tau}}{\tmax{\tau'}}) \lor (\dyn \land (\resType{\tmax{\tau}}{\tmin{\tau'}}))\\
      \gprojType{i}{\tau} &\eqdef& (\bpi{\tmin{\tau}}) \lor (\dyn \land (\bpi{\tmax{\tau}}))  
\end{eqnarray*}
for which we can prove that $\gdomType\tau = \max \{ \tau' \mid \tau\mathrel{\dot\leq} \tau'\to \Any\}\),
\(\gresType{\tau_1}{\tau_2} = \min \{ \tau \mid \tau_1\mathrel{\dot\leq} \tau_2\to
\tau\}\),  $\gprojType{1}{\tau}=\min\{\tau'\mid \tau
\mathrel{\dot\leq}\pair{\tau'}\Any\}$, and $\gprojType{2}{\tau}=\min\{\tau'\mid \tau \mathrel{\dot\leq}\pair\Any{\tau'}\}$.

\subsection{Denotational Semantics}\label{sec:semantics}

We have seen in Section~\ref{sec:types} that the essence of semantic
subtyping is to interpret types as sets of values. However, for the
circularity problem described in Section~\ref{sec:closing} this cannot be done
directly on the values of some language, but must pass via an
interpretation in a domain $\Domain$ whose elements
represent these values. Furthermore, for cardinality problems,
functional values
cannot  be represented directly as elements of the domain, and one
has to interpret types as sets containing only functions with finite
graphs. Even if at the end one obtains the same subtyping relation as if we
had considered infinite functions (cf., Section~\ref{sec:semsub}) this
solution has been making readers uneasy. The fact of using
finite graph functions to define a relation for general function spaces
looked more as a technical trick than as a theoretical
breakthrough. Pierre-Louis Curien suggested that the construction was
a {\it pied de nez\/} to (it cocked a snook at) denotational
semantics, insofar as it used a semantic construction to define a
language for which a denotational semantics was not known to exist. The common belief
was that the solution worked because considering \emph{all} finite
functions in the interpretation of a function space was equivalent to give
the finite approximations of the non-finite functions in that space, in the same
way as, say, Scott domains are built by giving finite approximations of the
functions therein.

Very recently, Lanvin's
Ph.D.\ dissertation~\citep[Part~2]{Lanvin21phd} has formalized this intuition and defined a denotational
semantics for a language with semantic subtyping (actually, the
language of Sections~\ref{inferringintersection}--\ref{sec:choice}), in which functions
are interpreted as the infinite set of their finite
approximations. This yields a model with a simple \emph{inductive} definition, which does not need
isomorphisms or the solution of domain equation. The idea is to
interpret not only types but also terms in the domain $\Domain$ of
Definition~\ref{def:domain}. Unfortunately, the domain $\Domain$
cannot be used as is,\footnote{Actually, it can but it yields a weak
property of computational soundness: cf.~\citep[Chapter~9]{Lanvin21phd}.}
but it must be slightly modified to account for the fact that
functions map finite approximations (rather than single denotations) into
denotations. In practice, one has to modify the domain as follows
\begin{align*}
  d & ::=  c \mid (d, d) \mid \Set{(S, \domega), \dots, (S,
    \domega)} &
  S &::= \{d,\ldots,d\}& 
    \domega & ::= d \mid \Omega
\end{align*}
where, thus, the domain now includes finite maps from \emph{finite
and non-empty}
sets of elements (ranged over by $S$) into other elements or $\Omega$. The interpretation
of types must also be slightly modified to make the interpretation of 
arrows satisfy the following
equation:
\begin{align*}
    \Inter{t_1 \to t_2} & =
    \Set{R \in \PsetFin(\PsetFin(\Domain) \times \Domain_\Omega) \mid
      \forall (S, \domega) \in R\:.\: S\cap\Inter{t_1}\neq\varnothing \implies \domega \in \Inter{t_2}}\}
\end{align*}
modification that yields the same subtyping relation as the one produced by the
interpretation of Definition~\ref{def:interpretation-of-types}. In
this domain it is then not very difficult to interpret every term of the
language into the (possibly infinite) \emph{set} of its finite approximations. For
instance, a constant $c$ will be interpreted as the singleton
$\{c\}$. The only delicate part is the interpretation of
$\lambda$-abstractions, that is defined as follows:
\[
  \begin{array}{rclrl}
    \Inter{\lambda{x\col t.}{e}}_{\rho} &=&
      \{ R \in \PsetFin(\PsetFin(\Domain) \times \Domain_\Omega)  \mid \forall (S, \domega) \in R,
      & \text{either} & S \subseteq \Inter{t} \text{ and } \domega \in \Inter{e}_{\rho,\, x \mapsto S}\\
      &&&\text{or} &  S \subseteq \Inter{\lnot t} \text{ and } \domega = \Omega\}
  \end{array}
  \]
where $\rho$ is a semantic environment that maps variables into
approximations, that is, into sets in $\PsetFin(\Domain)$. The definition
above states that a $\lambda$-abstraction is interpreted as the set
of all finite approximations that map any approximation in the domain of
the function to the interpretation of the body where the parameter is
associated to that approximation, and  any approximation that is
outside this domain to the failure $\Omega$. Notice that, as we
anticipated in Section~\ref{inferringintersection}, the denotation of
a function depends on its type annotation.

For this
interpretation it is possible
to prove three fundamental properties:
\begin{enumerate}
  \item Type soundness: if $\Gamma\vdash e:t$, then
    $\Inter{e}_\rho\subseteq\Inter t$, for every
    $\rho\in\Inter\Gamma$.\footnote{Where
  $\Inter\Gamma=\{\rho\mid\forall
  x\in\dom\Gamma\;.\;\rho(x)\subseteq\Inter{\Gamma(x)}\}$.}
  \item Computational soundness: if $\Gamma\vdash e:t$, and $e\reduces
    e'$, then $\Inter{e}_\rho=\Inter{e'}_\rho$, for all
    $\rho\in\Inter\Gamma$.
\item Computational adequacy: $\Inter e_\rho=\varnothing$, for every
  well-typed closed \emph{diverging} term $e$. 
\end{enumerate}
This interpretation works only for a language without type-cases and
overloaded functions (notice the syntax of annotations which are just on the parameter
of $\lambda$-abstractions and not on the full term). The
reader can refer to~\citet[Part~2]{Lanvin21phd} for all details and a
denotational semantics of the whole Core CDuce language.

\section{Conclusion}
\label{sec:conclusion}
In this essay I tried to survey the multiple advantages and
usages of
set-theoretic types in programming. Set-theoretic types are sometimes
the only way to type some particular functions, sometimes as simple as the
\code{flatten} function of the introduction. This is so because
set-theoretic types provide a suitable language to describe many
non-conventional, but not uncommon, programming patterns. This is
demonstrated by the fact that the need of set-theoretic types
naturally arises when trying to fit type-systems on dynamic languages:
union and negations become necessary to capture the nature of
branching and of pattern matching, intersections are often the only
way to describe the polymorphic use of some functions whose definition
lacks the uniformity required by parametric polymorphism.  The
development of languages such as Flow, TypeScript, and Typed Racket is
the latest witness of this fact. I also showed that even when
set-theoretic types are not exposed to the programmer, they are often
present at meta level since they provide the basic tools to precisely
type some program constructions such as type-cases and pattern
matching. Finally, set-theoretic types provide a powerful theoretic
toolbox to explore, understand, and formalize existing type disciplines:
I demonstrated this with gradual types which, thanks to set-theoretic
types, can be understood as intervals of static types, an analogy
that we can use to rethink both their theory (see Lanvin's
dissertation~\cite{Lanvin21phd}) and their practice (e.g., the
implementation of gradual virtual machines as in~\cite{CDLS19}).

This survey is necessarily incomplete. For instance, I barely spoke of
XML types and XML programming even though they were the first
motivation for developing the theory of semantic subtyping and to
design and implement programming languages such as XDuce and
CDuce. Also, I completely swept under the carpet how to handle features that
are common in modern programming languages such as the use of abstract
types---whose integration with structural subtyping and polymorphism
may result delicate---and the presence of
side-effects. The latter is particular sensitive for the language presented
in Section~\ref{sec:occtyping}, insofar as the use of MSC-forms is sound
only for pure expressions.
Nevertheless, I hope I gave a good idea of the potentiality of
having set-theoretic types in a programming language and how the
addition of these types can be done.

It is not all a bed of roses though. From a formal point of view we
did not succeed, yet, to define a unique formalism that mixes
implicitly and explicitly typed functions, reconstruction of
intersection types, and an advanced use of occurrence typing. But we
are not far from it: see~\cite{CLN24}. From a practical viewpoint even more work is
needed. We have seen that parametric polymorphism with set-theoretic
types implies constraint generation and constraint resolution (i.e.,
structured and (sub-)typing constraints in the implicitly-typed language and only
the latter in the explicitly-typed one). This
has several drawbacks.  Foremost, because of the presence of unions
and of subtyping, constraint solving is a potential source of
computational explosion that we do not master well, yet. Furthermore,
constraint solving makes the generation of informative error messages
very difficult for the case when it fails, but even pretty printing
the deduced types in a form easily understandable by the programmer may
sometimes happen to be challenging. So the positive message with which
I want to conclude this presentation is that, all in all, the research
of set-theoretic types still is a very nice playground that reserves
us several interesting and challenging problems yet to be solved.

\subsubsection*{Acknowledgements:}
The work presented here is the result of many collaborations with many
coauthors that I listed in the first page of this article. The ``we''
I used all the presentation long and abandoned from the conclusion must be intended as
inclusive of all of them. Not only the results exposed here were first presented in
articles I co-authored or dissertations I supervised but, in some
cases, I also adapted or reused verbatim the text of these works. Thus
several parts of this presentation are borrowed from~\citet{CLNL22}.
Section~\ref{sec:motivations} and the beginning of Section~\ref{sec:types} faithfully reproduce and extend
Section 1.1 of Tommaso Petrucciani PhD
thesis~\cite{Pet19phd} whose reading I warmly recommend. Sections~\ref{sec:semsub} and~\ref{sec:polysemsub} follow a
presentation that can be found in several papers I co-authored. Most of
Section~\ref{sec:cduce} comes from some unpublished notes that Victor
Lanvin and I wrote on the denotational semantics of
CDuce, while Section~\ref{sec:gradual} reuses parts of~\citet[Chapter
6]{Lanvin21phd}. James Clark, Guillaume Duboc, Victor Lanvin, Micka\"el Laurent,
Matt Lutze, Kim Nguyen,  and José Valim provided useful feedback on
an early version of this manuscript.

%
%
%

\bibliography{main}

\begin{thebibliography}{64}
\providecommand{\natexlab}[1]{#1}
\providecommand{\url}[1]{\texttt{#1}}
\expandafter\ifx\csname urlstyle\endcsname\relax
  \providecommand{\doi}[1]{doi: #1}\else
  \providecommand{\doi}{doi: \begingroup \urlstyle{rm}\Url}\fi

\bibitem[Barbanera et~al.(1995)Barbanera, Dezani-Ciancaglini, and de'Liguoro]{BDD95}
Franco Barbanera, Mariangiola Dezani-Ciancaglini, and Ugo de'Liguoro.
\newblock Intersection and union types.
\newblock \emph{Inf. Comput.}, 119\penalty0 (2):\penalty0 202–230, June 1995.
\newblock ISSN 0890-5401.
\newblock \doi{10.1006/inco.1995.1086}.

\bibitem[Benzaken et~al.(2003)Benzaken, Castagna, and Frisch]{BCF03}
V\'eronique Benzaken, Giuseppe Castagna, and Alain Frisch.
\newblock {CD}uce: an {XML}-centric general-purpose language.
\newblock In \emph{ICFP~'03, 8th ACM International Conference on Functional Programming}, pages 51--63, Uppsala, Sweden, 2003. ACM Press.
\newblock \doi{10.1145/944705.944711}.

\bibitem[Cardelli and Wegner(1985)]{Fun}
Luca Cardelli and Peter Wegner.
\newblock On understanding types, data abstraction, and polymorphism.
\newblock \emph{{ACM} Comput. Surv.}, 17\penalty0 (4):\penalty0 471--522, 1985.
\newblock \doi{10.1145/6041.6042}.

\bibitem[Cardelli et~al.(1994)Cardelli, Martini, Mitchell, and Scedrov]{Fsub}
Luca Cardelli, Simone Martini, John~C. Mitchell, and Andre Scedrov.
\newblock An extension of system {F} with subtyping.
\newblock \emph{Inf. Comput.}, 109\penalty0 (1/2):\penalty0 4--56, 1994.
\newblock \doi{10.1006/inco.1994.1013}.

\bibitem[Castagna(2020)]{Cas20}
Giuseppe Castagna.
\newblock Covariance and controvariance: a fresh look at an old issue (a primer in advanced type systems for learning functional programmers).
\newblock \emph{Logical Methods in Computer Science}, 16\penalty0 (1):\penalty0 15:1--15:58, 2020.
\newblock \doi{10.23638/LMCS-16(1:15)2020}.
\newblock New and extended version available at the author's web page: \url{https://www.irif.fr/~gc/papers/covcon-again.pdf}.

\bibitem[Castagna and Frisch(2005)]{CF05}
Giuseppe Castagna and Alain Frisch.
\newblock A gentle introduction to semantic subtyping.
\newblock In \emph{\emph{Proceedings of} PPDP '05, the 7th ACM SIGPLAN International Symposium on Principles and Practice of Declarative Programming, \emph{pages 198-208, ACM Press (full version) and} ICALP '05, 32nd International Colloquium on Automata, Languages and Programming, \emph{Lecture Notes in Computer Science n.\ 3580, pages 30-34, Springer (summary)}}, Lisboa, Portugal, July 2005.
\newblock \doi{10.1145/1069774.1069793}.
\newblock Joint ICALP-PPDP keynote talk.

\bibitem[Castagna and Lanvin(2017)]{CL17}
Giuseppe Castagna and Victor Lanvin.
\newblock Gradual typing with union and intersection types.
\newblock \emph{Proc. ACM Program. Lang.}, 1, Article 41\penalty0 (ICFP\,'17), September 2017.
\newblock \doi{10.1145/3110285}.

\bibitem[Castagna and Xu(2011)]{CX11}
Giuseppe Castagna and Zhiwu Xu.
\newblock Set-theoretic foundation of parametric polymorphism and subtyping.
\newblock In \emph{ICFP\,'11: 16th {ACM-SIGPLAN} International Conference on Functional Programming}, pages 94--106, 2011.
\newblock \doi{10.1145/2034773.2034788}.

\bibitem[Castagna et~al.(2014)Castagna, Nguyen, Xu, Im, Lenglet, and Padovani]{polyduce1}
Giuseppe Castagna, Kim Nguyen, Zhiwu Xu, Hyeonseung Im, Sergue\"{\i} Lenglet, and Luca Padovani.
\newblock Polymorphic functions with set-theoretic types. {Part 1}: Syntax, semantics, and evaluation.
\newblock In \emph{Proceedings of the 41st Annual {ACM} {SIGPLAN}-{SIGACT} Symposium on Principles of Programming Languages}, {POPL}\,'14, pages 5--17, January 2014.
\newblock \doi{10.1145/2676726.2676991}.

\bibitem[Castagna et~al.(2015)Castagna, Nguyen, Xu, and Abate]{polyduce2}
Giuseppe Castagna, Kim Nguyen, Zhiwu Xu, and Pietro Abate.
\newblock Polymorphic functions with set-theoretic types. {Part 2}: local type inference and type reconstruction.
\newblock In \emph{Proceedings of the 42nd Annual \acro{ACM} \acro{SIGPLAN}-\acro{SIGACT} Symposium on Principles of Programming Languages}, \acro{POPL}\,'15, pages 289--302, January 2015.
\newblock \doi{10.1145/2676726.2676991}.

\bibitem[Castagna et~al.(2016)Castagna, Petrucciani, and Nguyen]{CPN16}
Giuseppe Castagna, Tommaso Petrucciani, and Kim Nguyen.
\newblock Set-theoretic types for polymorphic variants.
\newblock In \emph{ICFP\,'16, 21st ACM SIGPLAN International Conference on Functional Programming}, pages 378--391, September 2016.
\newblock \doi{10.1145/2951913.2951928}.

\bibitem[Castagna et~al.(2019{\natexlab{a}})Castagna, Duboc, Lanvin, and Siek]{CDLS19}
Giuseppe Castagna, Guillaume Duboc, Victor Lanvin, and Jeremy~G. Siek.
\newblock A space-efficient call-by-value virtual machine for gradual set-theoretic types.
\newblock In \emph{Proceedings of the 31st Symposium on Implementation and Application of Functional Languages}, IFL '19, New York, NY, USA, 2019{\natexlab{a}}. Association for Computing Machinery.
\newblock \doi{10.1145/3412932.3412940}.

\bibitem[Castagna et~al.(2019{\natexlab{b}})Castagna, Lanvin, Petrucciani, and Siek]{CLPS19}
Giuseppe Castagna, Victor Lanvin, Tommaso Petrucciani, and Jeremy~G. Siek.
\newblock Gradual typing: a new perspective.
\newblock \emph{Proc. ACM Program. Lang.}, 3\penalty0 ({POPL}), January 2019{\natexlab{b}}.
\newblock \doi{10.1145/3290329}.

\bibitem[Castagna et~al.(2022{\natexlab{a}})Castagna, Laurent, Nguyen, and Lutze]{CLNL22}
Giuseppe Castagna, Micka\"el Laurent, Kim Nguyen, and Matthew Lutze.
\newblock On type-cases, union elimination, and occurrence-typing.
\newblock \emph{Proc. {ACM} Program. Lang.}, 6\penalty0 ({POPL}), 2022{\natexlab{a}}.
\newblock \doi{10.1145/3498674}.

\bibitem[Castagna et~al.(2022{\natexlab{b}})Castagna, Laurent, Lanvin, and Nguyen]{CLLN20}
Giuseppe Castagna, Mickaël Laurent, Victor Lanvin, and Kim Nguyen.
\newblock Revisiting occurrence typing.
\newblock \emph{Science of Computer Programming}, 217:\penalty0 102781, mar 2022{\natexlab{b}}.
\newblock ISSN 0167-6423.
\newblock \doi{10.1016/j.scico.2022.102781}.
\newblock Preprint available at \url{https://arxiv.org/abs/1907.05590}.

\bibitem[Castagna et~al.(2024)Castagna, Laurent, and Nguyen]{CLN24}
Giuseppe Castagna, Mickaël Laurent, and Kim Nguyen.
\newblock Polymorphic type inference for dynamic languages.
\newblock \emph{Proc. ACM Program. Lang.}, 8\penalty0 (POPL), January 2024.
\newblock \doi{10.1145/3632882}.

\bibitem[{CDuce}()]{cduce}
{CDuce}.
\newblock {The CDuce Compiler}.
\newblock \url{https://www.cduce.org}.

\bibitem[Chaudhuri et~al.(2017)Chaudhuri, Vekris, Goldman, Roch, and Levi]{Chaudhuri2017}
Avik Chaudhuri, Panagiotis Vekris, Sam Goldman, Marshall Roch, and Gabriel Levi.
\newblock Fast and precise type checking for javascript.
\newblock \emph{Proceedings of the {ACM} on Programming Languages}, 1\penalty0 ({OOPSLA}):\penalty0 48:1--48:30, October 2017.
\newblock ISSN 2475-1421.
\newblock \doi{10.1145/3133872}.

\bibitem[Dezani-Ciancaglini et~al.(2003)Dezani-Ciancaglini, Frisch, Giovannetti, and Motohama]{itrs02}
Mariangiola Dezani-Ciancaglini, Alain Frisch, Elio Giovannetti, and Yoko Motohama.
\newblock The relevance of semantic subtyping.
\newblock \emph{Electronic Notes in Theoretical Computer Science}, 70\penalty0 (1):\penalty0 88 -- 105, 2003.
\newblock ISSN 1571-0661.
\newblock \doi{10.1016/S1571-0661(04)80492-4}.
\newblock ITRS '02, Intersection Types and Related Systems.

\bibitem[Dolan and Mycroft(2017)]{Dolan2017}
Stephen Dolan and Alan Mycroft.
\newblock Polymorphism, subtyping, and type inference in mlsub.
\newblock In \emph{Proceedings of the 44th {ACM} {SIGPLAN} Symposium on Principles of Programming Languages}, {POPL} 2017, pages 60--72. {ACM}, 2017.
\newblock ISBN 978-1-4503-4660-3.
\newblock \doi{10.1145/3009837.3009882}.

\bibitem[{Facebook}()]{Flow}
{Facebook}.
\newblock {Flow}.
\newblock \url{https://flow.org/}.

\bibitem[Frisch(2004)]{Frisch04phd}
Alain Frisch.
\newblock \emph{Th{\'e}orie, conception et r{\'e}alisation d'un langage de programmation adapt{\'e} {\`a} {XML}}.
\newblock PhD thesis, Universit{\'e} Paris 7 -- Denis Diderot, 12 2004.
\newblock \url{http://www.cduce.org/papers/frisch_phd.pdf}.

\bibitem[Frisch et~al.(2002)Frisch, Castagna, and Benzaken]{FCB02}
Alain Frisch, Giuseppe Castagna, and V\'eronique Benzaken.
\newblock Semantic {S}ubtyping.
\newblock In \emph{LICS '02, 17th Annual IEEE Symposium on Logic in Computer Science}, pages 137--146. IEEE Computer Society Press, 2002.
\newblock \doi{10.1109/LICS.2002.1029823}.

\bibitem[Frisch et~al.(2008)Frisch, Castagna, and Benzaken]{FCB08}
Alain Frisch, Giuseppe Castagna, and V{\'e}ronique Benzaken.
\newblock Semantic subtyping: dealing set-theoretically with function, union, intersection, and negation types.
\newblock \emph{Journal of the \acro{ACM}}, 55\penalty0 (4):\penalty0 19:1--19:64, September 2008.
\newblock ISSN 0004-5411.
\newblock \doi{10.1145/1391289.1391293}.

\bibitem[Garcia(2013)]{Garcia:2013fk}
Ronald Garcia.
\newblock Calculating threesomes, with blame.
\newblock In \emph{{ICFP} '13: Proceedings of the International Conference on Functional Programming}, 2013.
\newblock \doi{10.1145/2500365.2500603}.

\bibitem[Gesbert et~al.(2011)Gesbert, Genevès, and Layaïda]{Gesbert2011}
Nils Gesbert, Pierre Genevès, and Nabil Layaïda.
\newblock Parametric polymorphism and semantic subtyping: the logical connection.
\newblock In \emph{Proceedings of the 16th \acro{ACM} \acro{SIGPLAN} International Conference on Functional Programming}, \acro{ICFP} '11, pages 107--116, New York, NY, \acro{USA}, 2011. \acro{ACM}.
\newblock ISBN 978-1-4503-0865-6.
\newblock \doi{10.1145/2034773.2034789}.

\bibitem[Gesbert et~al.(2015)Gesbert, Genev{\`e}s, and Layaïda]{Gesbert2015}
Nils Gesbert, Pierre Genev{\`e}s, and Nabil Layaïda.
\newblock A logical approach to deciding semantic subtyping.
\newblock \emph{\acro{ACM} Transactions on Programming Languages and Systems}, 38\penalty0 (1):\penalty0 3, 2015.
\newblock \doi{10.1145/2812805}.

\bibitem[Google()]{googledart}
Google.
\newblock Dart programming language specification.
\newblock \url{https://dart.dev/guides/language/spec}.

\bibitem[Greenberg(2019)]{gre19}
Michael Greenberg.
\newblock {The Dynamic Practice and Static Theory of Gradual Typing}.
\newblock In \emph{3rd Summit on Advances in Programming Languages (SNAPL 2019)}, volume 136 of \emph{Leibniz International Proceedings in Informatics (LIPIcs)}, pages 6:1--6:20, 2019.
\newblock ISBN 978-3-95977-113-9.
\newblock \doi{10.4230/LIPIcs.SNAPL.2019.6}.

\bibitem[Guibas and Sedgewick(1978)]{GS78}
Leo~J. Guibas and Robert Sedgewick.
\newblock A dichromatic framework for balanced trees.
\newblock In \emph{19th Annual Symposium on Foundations of Computer Science (sfcs 1978)}, pages 8--21, 1978.
\newblock \doi{10.1109/SFCS.1978.3}.

\bibitem[Harper(2006)]{harperbook06}
Robert Harper.
\newblock \emph{Programming Languages: Theory and Practice}.
\newblock Carnegie Mellon University, 2006.
\newblock Available on the web: \url{http://fpl.cs.depaul.edu/jriely/547/extras/online.pdf}.

\bibitem[Hindley and Seldin(2008)]{HS08}
J.~Roger Hindley and Jonathan~P. Seldin.
\newblock \emph{Lambda-Calculus and Combinators An Introduction}.
\newblock Cambridge University Press, 2008.

\bibitem[Hosoya(2001)]{hosoya01thesis}
Haruo Hosoya.
\newblock \emph{Regular Expression Types for {XML}}.
\newblock PhD thesis, The University of Tokyo, 2001.

\bibitem[Hosoya and Pierce(2001)]{hosoya01patter}
Haruo Hosoya and Benjamin~C. Pierce.
\newblock Regular expression pattern matching for {XML}.
\newblock In \emph{POPL\,'01, 25th {ACM} Symposium on Principles of Programming Languages}, 2001.
\newblock \doi{10.1145/360204.360209}.

\bibitem[Hosoya and Pierce(2003)]{xduce_toit}
Haruo Hosoya and Benjamin~C. Pierce.
\newblock {XD}uce: A statically typed {XML} processing language.
\newblock \emph{ACM Trans. Internet Techn.}, 3\penalty0 (2):\penalty0 117--148, 2003.
\newblock \doi{10.1145/767193.767195}.

\bibitem[Hosoya et~al.(2000)Hosoya, Vouillon, and Pierce]{hosoya00regular}
Haruo Hosoya, J\'er\^ome Vouillon, and Benjamin~C. Pierce.
\newblock Regular expression types for {XML}.
\newblock In \emph{ICFP\,'00}, volume 35(9) of \emph{SIGPLAN Notices}, 2000.
\newblock \doi{10.1145/351240.351242}.

\bibitem[Hosoya et~al.(2009)Hosoya, Frisch, and Castagna]{HFC09}
Haruo Hosoya, Alain Frisch, and Giuseppe Castagna.
\newblock Parametric polymorphism for {XML}.
\newblock \emph{ACM Transactions on Programming Languages and Systems}, 32\penalty0 (1):\penalty0 1--56, 2009.
\newblock \doi{10.1145/1596527.1596529}.

\bibitem[JetBrains(2018)]{Kotlin}
JetBrains.
\newblock Kotlin documentation.
\newblock Available at \url{http://kotlinlang.org/docs/reference}, 2018.

\bibitem[King(2017)]{Ceylon}
Gavin King.
\newblock The {Ceylon} language specification, version 1.3.
\newblock Available at \url{https://ceylon-lang.org/documentation/1.3/spec}, 2017.

\bibitem[Lanvin(2021)]{Lanvin21phd}
Victor Lanvin.
\newblock \emph{A Semantic Foundation for Gradual Set-Theoretic Types}.
\newblock PhD thesis, Université de Paris, November 2021.
\newblock URL \url{https://theses.hal.science/tel-03853222}.

\bibitem[MacQueen et~al.(1986)MacQueen, Plotkin, and Sethi]{MacQueen1986}
David MacQueen, Gordon Plotkin, and Ravi Sethi.
\newblock An ideal model for recursive polymorphic types.
\newblock \emph{Information and Control}, 71\penalty0 (1):\penalty0 95--130, 1986.
\newblock ISSN 0019-9958.
\newblock \doi{10.1016/S0019-9958(86)80019-5}.

\bibitem[Martin-L{\"o}f(1994)]{ML1994}
Per Martin-L{\"o}f.
\newblock \emph{Analytic and Synthetic Judgements in Type Theory}, pages 87--99.
\newblock Springer Netherlands, Dordrecht, 1994.
\newblock ISBN 978-94-011-0834-8.
\newblock \doi{10.1007/978-94-011-0834-8\_5}.

\bibitem[{Microsoft}()]{TypeScript}
{Microsoft}.
\newblock {TypeScript}.
\newblock \url{https://www.typescriptlang.org/}.

\bibitem[Muehlboeck and Tate(2018)]{Muehlboeck2018}
Fabian Muehlboeck and Ross Tate.
\newblock Empowering union and intersection types with integrated subtyping.
\newblock \emph{Proceedings of the {ACM} on Programming Languages}, 2\penalty0 ({OOPSLA}):\penalty0 112:1--112:29, October 2018.
\newblock ISSN 2475-1421.
\newblock \doi{10.1145/3276482}.

\bibitem[Okasaki(1998)]{okasaki_1998}
Chris Okasaki.
\newblock \emph{Purely Functional Data Structures}.
\newblock Cambridge University Press, 1998.
\newblock \doi{10.1017/CBO9780511530104}.

\bibitem[Okasaki(1999)]{okasaki99}
Chris Okasaki.
\newblock Red-black trees in a functional setting.
\newblock \emph{J. Funct. Program.}, 9\penalty0 (4):\penalty0 471--477, 1999.

\bibitem[Pearce(2013)]{Pearce2013}
David~J. Pearce.
\newblock Sound and complete flow typing with unions, intersections and negations.
\newblock In \emph{Verification, Model Checking, and Abstract Interpretation}, pages 335--354. Springer, 2013.

\bibitem[Pearce and Groves(2013)]{Pearce2013whiley}
David~J. Pearce and Lindsay Groves.
\newblock Whiley: a platform for research in software verification.
\newblock In \emph{Software Language Engineering}, pages 238--248, Cham, 2013. Springer International Publishing.
\newblock ISBN 978-3-319-02654-1.

\bibitem[Petrucciani(2019)]{Pet19phd}
Tommaso Petrucciani.
\newblock \emph{Polymorphic Set-Theoretic Types for Functional Languages}.
\newblock PhD thesis, Joint Ph.D.\ Thesis, Università di Genova and Université Paris Diderot, March 2019.
\newblock URL \url{https://tel.archives-ouvertes.fr/tel-02119930}.

\bibitem[Pierce(2002)]{Pierce2002}
Benjamin~C. Pierce.
\newblock \emph{Types and programming languages}.
\newblock {MIT} Press, 2002.

\bibitem[Pierce(1992)]{Pierce92phd}
Benjamin~Crawford Pierce.
\newblock \emph{Programming with Intersection Types and Bounded Polymorphism}.
\newblock PhD thesis, Carnegie Mellon University, USA, 1992.
\newblock URL \url{https://www.cis.upenn.edu/~bcpierce/papers/thesis.pdf}.

\bibitem[Pottier and Rémy(2005)]{Pottier2005}
François Pottier and Didier Rémy.
\newblock \emph{The essence of ML type inference}, chapter~10, pages 389--489.
\newblock {MIT} Press, 2005.

\bibitem[Reynolds(1996)]{forsythe}
John~C. Reynolds.
\newblock Design of the programming language {F}orsythe.
\newblock Technical Report CMU-CS-96-146, Carnegie Mellon University, 1996.

\bibitem[Reynolds(2003)]{Rey03}
John~C. Reynolds.
\newblock \emph{Programming Methodology}, chapter What do types mean? -- From intrinsic to extrinsic semantics.
\newblock Monographs in Computer Science. Springer, 2003.

\bibitem[Sabry and Felleisen(1992)]{SF92}
Amr Sabry and Matthias Felleisen.
\newblock Reasoning about programs in continuation-passing style.
\newblock In \emph{Proceedings of the 1992 ACM Conference on LISP and Functional Programming}, page 288–298, New York, NY, USA, 1992. Association for Computing Machinery.
\newblock \doi{10.1145/141471.141563}.

\bibitem[Siek and Taha(2006)]{siek2006gradual}
Jeremy~G. Siek and Walid Taha.
\newblock Gradual typing for functional languages.
\newblock In \emph{Scheme and Functional Programming Workshop}, volume~6, pages 81--92, 2006.

\bibitem[Siek and Vachharajani(2008)]{Siek:2008kq}
Jeremy~G. Siek and Manish Vachharajani.
\newblock Gradual typing with unification-based inference.
\newblock Technical Report CU-CS-1039-08, University of Colorado at Boulder, January 2008.

\bibitem[Takikawa et~al.(2016)Takikawa, Feltey, Greenman, New, Vitek, and Felleisen]{takikawa2016sound}
Asumu Takikawa, Daniel Feltey, Ben Greenman, Max~S. New, Jan Vitek, and Matthias Felleisen.
\newblock Is sound gradual typing dead?
\newblock In \emph{Proceedings of the 43rd Annual ACM SIGPLAN Symposium on Principles of Programming Languages}, POPL '16, pages 456--468. ACM, 2016.
\newblock ISBN 978-1-4503-3549-2.
\newblock \doi{10.1145/2914770.2837630}.
\newblock URL \url{http://doi.acm.org/10.1145/2914770.2837630}.

\bibitem[Tobin-Hochstadt and Felleisen(2008)]{THF08}
Sam Tobin-Hochstadt and Matthias Felleisen.
\newblock The design and implementation of typed scheme.
\newblock In \emph{Proceedings of the 35th Annual ACM SIGPLAN-SIGACT Symposium on Principles of Programming Languages}, POPL\,'08, pages 395--406, New York, NY, USA, 2008. ACM.
\newblock ISBN 978-1-59593-689-9.
\newblock \doi{10.1145/1328438.1328486}.
\newblock URL \url{http://doi.acm.org/10.1145/1328438.1328486}.

\bibitem[Tobin-Hochstadt and Felleisen(2010)]{THF10}
Sam Tobin-Hochstadt and Matthias Felleisen.
\newblock Logical types for untyped languages.
\newblock In \emph{Proceedings of the 15th {ACM} {SIGPLAN} International Conference on Functional Programming}, {ICFP} '10, pages 117--128, New York, NY, {USA}, 2010. {ACM}.
\newblock ISBN 978-1-60558-794-3.
\newblock \doi{10.1145/1863543.1863561}.
\newblock URL \url{http://doi.acm.org/10.1145/1863543.1863561}.

\bibitem[Types()]{types2019}
Types.
\newblock What exactly should we call syntax-directed inference rules?
\newblock Discussion on the Types mailing list, jun 2019.
\newblock \url{http://lists.seas.upenn.edu/pipermail/types-list/2019/002138.html}.

\bibitem[Wadler and Findler(2009)]{wadler2009well}
Philip Wadler and Robert~Bruce Findler.
\newblock Well-typed programs can’t be blamed.
\newblock In \emph{European Symposium on Programming}, ESOP '09, pages 1--16. Springer, 2009.
\newblock \doi{10.1007/978-3-642-00590-9\_1}.

\bibitem[Wand(1987)]{Wand1987}
Mitchell Wand.
\newblock A simple algorithm and proof for type inference.
\newblock \emph{Fundamenta Informatic\ae}, 10:\penalty0 115--122, 1987.
\newblock \doi{10.3233/FI-1987-10202}.

\bibitem[Wright(1995)]{wright95}
Andrew~K. Wright.
\newblock Simple imperative polymorphism.
\newblock \emph{{LISP} Symb. Comput.}, 8\penalty0 (4):\penalty0 343--355, 1995.
\newblock \doi{10.1007/BF01018828}.

\end{thebibliography}

\newpage

\appendix

\section{Core Calculus of CDuce~\cite{FCB08}}\label{app:cducecore}
\subsection*{Syntax}
\[
\begin{array}{lrcl}
  \textbf{Types}& t & \doublecoloneq & b\mid t\times
  t\mid t\to t\mid t\lor t \mid \lnot t\mid \Empty\\
  \textbf{Expressions}& e & \doublecoloneq & c\mid
x \mid  \lambda^{\wedge_{i{\in}I}t_i\to t_i} x.e \mid ee\mid \pi_i
e\mid (e,e)\mid \ctcase x e t e e\mid \choice(e,e)\\
\textbf{Values}& v & \doublecoloneq & c\mid x \mid  \lambda^{\wedge_{i{\in}I}s_i\to t_i} x.e \mid (v,v)
\end{array}
\]
\subsection*{Reduction semantics}
\begin{align*}
 (\lambda^{\wedge_{i{\in}I}s_i\to t_i} x.e)v & \reduces e\subs x v\\[-1mm]
 \pi_i(v_1,v_2) & \reduces v_i & i=1,2 \\[-1mm]
 \choice(e_1,e_2)  & \reduces e_i & i=1,2 \\[-1mm]
  \ctcase x v t {e_1}{e_2} & \reduces e_1\subs x v &\text{if }  v\in t\\[-1mm]
  \qquad\ctcase x v t {e_1}{e_2} & \reduces e_2\subs x v &\text{if } v\not\in
  t
 \end{align*}
where $v\in t\iffdef \exists s{\in}\typeof(v).s\leq t$ with\\
$$\begin{array}{rcl}
   \typeof(c) &\eqdef& \{\BasicTypeOfConstant{c}\}\\
   \typeof(\lambda^{\bigwedge_{i{\in}I}s_i\to t_i}
   x.e) &\eqdef& \{t\mid t\simeq(\bigwedge_{i{\in}I}s_i\to
  t_i)\land(\bigwedge_{j{\in}J}\lnot(s'_j\to
  t'_j)), t\not\leq\Empty\}\\
   \typeof((v_1,v_2))&\eqdef&\typeof(v_1)\times\typeof(v_2)
 \end{array}$$
plus the  standard context rule implementing a leftmost outermost
strategy, namely, $\ctx{e} \reduces \ctx{e'}$ if $e \reduces e'$, where 
$$E ::=
[\,]\mid E e\mid vE  \mid (E ,e)\mid (v,E)\mid \pi_i{E}\mid\ctcase x {E} t e e$$

\subsection*{Type-system}
\begin{mathpar}
    \Infer[Const]
    {{~} }
    {\Gamma\vdash c:\basic{c}}
    { }
    \and
    \Infer[Var]
  { }
  {\Gamma \vdash x: \Gamma(x)}
  {x\in\dom\Gamma}
\\  
\Infer[$\to$I]
{\forall i\in I\quad \Gamma, x:s_i\vdash e:t_i}
{\Gamma \vdash \lambda^{\wedge_{i{\in}I}s_i\to t_i} x. e: t\wedge t'}
{\begin{array}{l}
t=\wedge_{i{\in}I}(s_i\to t_i)\\[-1mm]
t'=\wedge_{j{\in}J}\neg(s'_j \to t'_j)\\[-1mm]
t\wedge t'\not\simeq\Empty
\end{array}
}
\quad
\Infer[$\to$E]
  {
    \Gamma \vdash e_1: {t_1}\quad
    \Gamma \vdash e_2: t_2
  }
  {\Gamma \vdash {e_1}{e_2}: t_1\circ t_2 }
  {\begin{array}{l}t_1\leq\Empty\to\Any\\[-1mm]t_2\leq\dom{t_1}\end{array}
  }
\\
  \Infer[$\times$I]
  {\Gamma \vdash e_1:t_1 \and \Gamma \vdash e_2:t_2}
  {\Gamma \vdash (e_1,e_2):\pair {t_1} {t_2}}
  { }
\\
  \Infer[$\times$E$_1$]
  {\Gamma \vdash e:t}
  {\Gamma \vdash \pi_1 e:\bpl t}
  {t\leq\Any\times\Any}
  \and
  \Infer[$\times$E$_2$]
  {\Gamma \vdash e:t}
  {\Gamma \vdash \pi_2 e:\bpr t}
  {t\leq\Any\times\Any}
  \\
  \Infer[Case]
        {\Gamma\vdash e:t' \quad \Gamma, x:t{\wedge} t'\vdash e_1:s\quad  \Gamma, x:\neg t{\wedge} t'\vdash e_2:s}
        {\Gamma\vdash\ctcase x e t {e_1}{e_2}:s}{}
\qquad
  \Infer[Efq]
        {\quad}
        {\Gamma,x:\Empty\vdash e:t}{}
\\        
  \Infer[Choice]
        {\Gamma\vdash e_1:t_1\and\Gamma\vdash e_2:t_2}
        {\Gamma\vdash\choice(e_1,e_2):t_1\vee t_2}
        {}
\end{mathpar}

\end{document}